\documentclass[preprint2]{aastex}

\begin{document}



\title{Diffuse X-ray-emitting Gas in Major Mergers}

\author{Beverly J. Smith}
\affil{Department of Physics and Astronomy, East Tennessee State University, Johnson City TN  37614}

\author{Kristen Campbell}
\affil{Department of Physics and Astronomy, East Tennessee State University, Johnson City TN  37614}

\author{Curtis Struck}
\affil{Department of Physics and Astronomy, Iowa State University,
Ames, IA 50011
}

\author{Roberto Soria}
\affil{National Astronomical Observatories, Chinese Academy of Sciences, Beijing 100012, China;
\\Radio Astronomy Research, Curtin University, GPO Box U1987, Perth, WA 6845, Australia}

\author{Douglas Swartz}
\affil{Astrophysics Office, NASA Marshall Space Flight Center, ZP12, Huntsville, AL 35812
}

\author{Macon Magno}
\affil{Department of Physics and Astronomy, East Tennessee State University, Johnson City TN  37614}

\author{Brianne Dunn}
\affil{Department of Physics and Astronomy, East Tennessee State University, Johnson City TN  37614}

\author{Mark L. Giroux}
\affil{Department of Physics and Astronomy, East Tennessee State University, Johnson City TN  37614}


\begin{abstract}

Using archived data from the Chandra X-ray telescope, we have 
extracted the diffuse X-ray emission from 49 equal-mass
interacting/merging galaxy pairs in a merger sequence, 
from widely separated pairs to merger remnants. 
After removal of contributions from unresolved point sources,
we compared the diffuse thermal X-ray luminosity 
from hot gas
(L$_{\rm X}$(gas)) 
with the global star formation rate (SFR).
After correction for absorption within the target galaxy,
we do not see strong trend of L$_{\rm X}$(gas)/SFR 
with SFR or merger stage
for galaxies with SFR $>$ 1 M$_{\sun}$~yr$^{-1}$.
For these galaxies, the 
median L$_{\rm X}$(gas)/SFR is 5.5 $\times$ 10$^{39}$ 
((erg~s$^{-1}$)/M$_{\sun}$~yr$^{-1}$)), similar to that of normal
spiral galaxies.
These results suggest that stellar feedback in star forming galaxies
reaches an approximately 
steady state condition, in which a relatively 
constant fraction of about 2\% 
of the total 
energy output from supernovae and stellar winds is converted into
X-ray flux.
Three late-stage merger remnants with low SFRs 
and 
high K band luminosities (L$_{\rm K}$)
have 
enhanced 
L$_{\rm X}$(gas)/SFR;
their UV/IR/optical colors suggest that they are post-starburst galaxies,
perhaps in the process of becoming ellipticals.
Systems with L$_{\rm K}$ $<$ 10$^{10}$~L$_{\sun}$
have lower L$_{\rm X}$(gas)/SFR ratios than the other
galaxies in our sample, perhaps due to 
lower gravitational fields 
or 
lower metallicities.
We see no relation between 
L$_{\rm X}$(gas)/SFR and Seyfert activity in this sample,
suggesting that feedback from active galactic nuclei is not a major
contributor to the hot gas in our sample galaxies.

\end{abstract}



\vfill
\eject

\section{Introduction }

The idea 
that mergers of equal-mass spiral galaxies 
produce elliptical galaxies 
was first suggested by \citet{toomre72} and 
\citet{toomre77}.
Early computer simulations found that major mergers (i.e., equal mass pairs)
may destroy the disks of spirals, creating spherical systems resembling ellipticals
\citep{barnes91a, barnes96, hernquist92, bekki98, naab99,
bournaud05,
cox06a}. 
More recent models show that if sufficient cold interstellar gas 
survives the merger, it may settle into
a plane, re-forming a disk \citep{barnes02, springel05, robertson06}.
This means that 
whether the final merger remnant is an elliptical or a spiral depends upon the 
fate of the interstellar gas.
The final quantity of cold gas remaining after the merger depends upon the initial gas mass,
the star formation rate (SFR),
the amount of heating of the gas, and the amount of infall back onto the galaxies late in the merger.
Galaxies that were gas-rich before the merger are more likely to produce remnants with disks,
since more gas survives
\citep{springel05, hopkins09a, robertson06, athanassoula16}.
Tidal forces can pull gas out of the galaxy into extended tails, 
which may then fall back into the main galaxies \citep{hibbard95, hancock09}.
In a major merger, gas can be driven 
into the center of the system \citep{negroponte83,
barnes91b, barnes96, mihos94, mihos96},
triggering bursts of star formation,
which may deplete the gas \citep{dimatteo07, dimatteo08, cox08}.    
However, 
simulations predict that only some major mergers will have strong starbursts, and the 
timing of the burst varies from system to system depending upon the parameters of the 
system
\citep{dimatteo07, dimatteo08, lotz08, sparre16}.

Other processes that affect the 
interstellar gas in a major merger, and therefore 
the final morphological
type, 
are winds
driven by supernovae and active galactic nuclei (AGN).
Stellar and supernova feedback heats the gas, potentially
lowering the star formation efficiency \citep{cox06a}.
Supernova-driven winds may drive gas out into the halo; this 
hot halo material may then cool and re-form a disk, triggering
delayed star formation \citep{hopkins13a}.
Winds due to supernovae may remove gas from the galaxy entirely;
in some simulations 
the mass loss rate from supernovae-driven winds 
is greater than the SFR \citep{hopkins12a, hopkins13a}.
AGN feedback also heats the gas, potentially quenching star formation 
\citep{dimatteo05, springel05, choi15}.  This may
allow more gas to 
survive to form a new disk 
\citep{hayward14, karman15}.
Alternatively, AGN feedback may remove gas from a merger 
remnant, allowing the formation of an elliptical
\citep{springel05, khalatyan08}.
Some simulations indicate that, without AGN feedback, 
star formation will continue long after a major 
merger is complete \citep{springel05, lotz08}.
In some models, 
producing an elliptical-like remnant with an old stellar population 
requires a powerful
AGN to clear out
the left-over gas \citep{springel05, hopkins06, sparre17}.
When pre-existing hot gaseous halos are included in merger simulations, infall from the halo
helps sustain star formation, but the efficiency of star formation is decreased
because of heating of halo gas
by shocks and by the conversion of satellite orbital energy into heat
\citep{sinha09,
moster11, karman15, hwang15}.
Model predictions depend upon the resolution of the simulation and the details of the calculations,
with higher resolution models including multi-phase interstellar gas
producing more efficient 
star formation \citep{teyssier10, hopkins13a, hayward14, sparre16}.
The duration and intensity of the starburst 
as well as the
final morphology of the merger remnant 
depend 
strongly upon the 
prescription for 
stellar feedback assumed in the model \citep{hopkins12a, fensch17}.
How effective AGN feedback is in quenching star formation also depends upon
the details of the model \citep{choi15}.

To test these feedback models, X-ray observations are required.
With high resolution X-ray imaging, the distribution, temperature, and mass
of the hot gas within galaxies can be studied, and compared to
other properties of the galaxies.
In spiral galaxies, the bulk of the hot gas is attributed to feedback
from Type II supernova and young stars
\citep{strickland00, grimes05, owen09, li13, mineo12b}, while the origin of the hot gas
in ellipticals is still under debate.
Elliptical galaxies typically have 
hot gas masses in excess of that expected from star formation
feedback alone \citep{osullivan01a, su15, goulding16}.
As summarized by \citet{mathews03}, traditionally several processes 
were thought to contribute
to the hot gas in ellipticals:
1) a component associated with the older stellar population of the
elliptical, 
including thermalization of the gas lost by red giants and 
AGB stars, and
heating by type Ia supernovae,
2) re-acquisition of hot gas previously ejected into the halo
by type II supernovae during an earlier star-forming phase,
3) accretion of left-over primordial gas, 
and 
4) feedback from AGN.  
In the first models of hot gas production in ellipticals, an early starburst
used up or cleared out most of the interstellar gas in the system, leaving
the galaxy gas-deficient until the hot halo was slowly replenished by mass loss
from the older stellar population and Type Ia supernovae 
\citep{ciotti91, pellegrini98}.   AGN feedback was then added to the
scenario to prevent
over-cooling (see \citealp{mathews03} and \citealp{ciotti17}, and references
therein).
In the paradigm of ellipticals being formed by major mergers, \citet{cox04}
and \citet{cox06b}
suggested another possible source of the hot gas in ellipticals:
5) shocks from the direct collision between two gas disks.
Yet another possible source
of hot gas in merger remnants was suggested by
\citet{hibbard95} 
and \citet{read98}: 
6) gas shocking 
during infall from tidal features.  However, based on hydrodynamical
simulations, \citet{cox06b} 
conclude that such infall is not 
a major contributor to the hot gas.

Observational studies 
show that the total X-ray luminosity of ellipticals 
scales with L$_{\rm B}$$^{2.2}$
\citep{osullivan01a},
while the hot gas L$_{\rm X}$
scales as L$_{\rm K}$$^{2.3}$ \citep{su15} or L$_{\rm K}$$^{\rm 2.5~to~2.8}$
\citep{goulding16}.
The hot gas luminosity in ellipticals scales steeply with 
gas temperature, as T$_{gas}^{4.5}$ \citep{goulding16}.  
These relations indicate that simple models of
X-ray production due to virialization of gas from stellar mass loss are 
insufficient, and other factors contribute \citep{goulding16}.
There is an anti-correlation between the global L$_{\rm X}$/L$_{\rm B}$
ratio of ellipticals
and fine structure parameters indicative of a past merger, suggesting that
the amount of hot gas increases late in a merger
\citep{mackie97, sansom00, osullivan01b}.
The global L$_{\rm X}$/L$_{\rm B}$ of ellipticals also increases with stellar
population age \citep{osullivan01b}, supporting
a picture in which the hot gas halo is built by stellar mass loss from older stars.

In the current study, 
we aim to better understand 
the spiral-to-elliptical transformation process and the origin of the hot gas 
in ellipticals
by using 
archival X-ray imaging data from the Chandra telescope to 
measure the hot ionized interstellar gas in a sample of 49 major mergers.
We will compare these to 
models of 
hot gas 
production during mergers.
These models include a range of processes, including 
shock heating of gas due to the collision itself
\citep{cox04, cox06b, sinha09}, 
shock-heated infalling tidal gas \citep{hibbard95, cox06b}, 
winds from Type II supernovae and stellar winds associated with a young stellar population \citep{hopkins13a},
and AGN feedback \citep{cox06b}.
The goal of the current study is to test models of 
stellar and AGN feedback in major mergers
by determining how the diffuse X-ray-emitting gas varies with SFR, 
merger stage, starburst age, and AGN activity, and comparing with normal spiral and elliptical galaxies.

Since the timescale for mergers is long (greater than 1 Gyrs),
testing models of hot gas production in mergers 
observationally
requires
a large sample of major mergers in a range of merger stages.
The first X-ray vs. merger stage study 
was conducted by
\citet{read98}
using low 
resolution X-ray images of eight systems
from the ROSAT 
satellite.
Their ROSAT measurements included flux from 
both hot interstellar gas and point sources. 
They found that, near the middle of the sequence, where 
L$_{\rm FIR}$ and the 
far-infrared-to-blue luminosity
ratio L$_{\rm FIR}$/L$_{\rm B}$ increased, 
the ratio of the total X-ray luminosity to that in the far-infrared
L$_{\rm X}$/L$_{\rm FIR}$ decreased.
\citet{lehmer10} also 
found a deficiency of total X-ray luminosity
L$_{\rm X}$ compared to L$_{\rm FIR}$ for 
high FIR luminosity galaxies.
The cause of this deficiency is uncertain;  it may be due to
absorption of the X-rays, or to contributions from AGN to powering
the FIR light. 

High resolution X-ray observations with Chandra 
are able to resolve the brightest of the 
point sources, and separate their fluxes from
that of the hot gas.
In an earlier Chandra study, we found a 
deficiency of ultra-luminous X-ray point sources (ULXs;
L$_{\rm X}$ $>$ 10$^{39}$ erg~s$^{-1}$) relative to L$_{\rm FIR}$ for 
ultra-luminous infrared galaxies (ULIRGs)
\citep{smith12}.
This deficiency in ULXs in high SFR galaxies was 
later confirmed by \citet{luangtip15}.
In the current study, we investigate
whether there is a corresponding depression in the diffuse
X-ray luminosity relative to the SFR at high L$_{\rm FIR}$, or if the global deficiency is due to
point sources alone.

The first Chandra study of the diffuse X-ray emission from hot gas in a merger sequence was done by 
\citet{brassington07},
using nine systems.
They found that in mid-sequence the X-ray dropped relative to L$_{\rm FIR}$, then late in the merger
there is an increase in L$_{\rm X}$ to the level found in ellipticals.
They concluded that the mid-merger drop
in the X-ray flux 
was because hot gas was escaping from the system, while in the early stages, 
the gas is confined. Freely-flowing hot gas produces little X-ray emission in contrast to 
hot gas confined by surrounding cooler gas \citep{hopkins12a}.
\citet{brassington07} suggest
that L$_{\rm X}$ drops off before the FIR because 
of the initiation of large-scale outflows from starburst-driven winds
before the starburst reaches its peak.   
They suggest that, once extended winds 
form, L$_{\rm X}$ drops due to a rapid decrease in the 
density of the gas, and therefore its X-ray emissivity.

The \citet{brassington07} results, although intriguing,
are tentative due to the small sample size.
Due to the unique interaction parameters associated with each system,
there are likely galaxy-to-galaxy variations between
galaxies in the same evolutionary stage as well as
variations in the gas content and mass of individual galaxies in
their sample.   
In order to better understand the evolution of the hot gas in galaxy mergers, a 
larger sample size is needed.
More than one system in each stage of the merging sequence must be observed
for a reliable test of the models.

In the current study, we use Chandra data
to investigate the hot X-ray-emitting gas in 49 major mergers
in a range of merger stages
spanning the full merger sequence.
In addition to X-ray observations,
observations at a wide range of other wavelengths are
needed to track the evolution of
various components of the galaxy.
Broadband optical
and near-infrared images
trace morphological transformations in the underlying stellar
component of the galaxy, for example,
the development of the characteristic r$^{1/4}$ law radial light
profile of ellipticals \citep{schweizer82, wright90, stanford91, scoville00,
chitre02} or other morphological signatures of ellipticals
\citep{cox06a, naab09}.
UV, IR, and H$\alpha$ observations help quantify the SFR,
and show that the highest rates are found 
in systems in the middle of the merger sequence \citep{casoli91, keel95, read98, brassington07, larson16}.
However, there is a lot of system-to-system variation in the SFR and mass-normalized SFR along the 
merger sequence, with some mid-merger systems being fairly
quiescent
\citep{keel95, ellison13, larson16}, and pre-merger interacting galaxies on average having elevated
SFRs compared to more isolated systems \citep{bushouse87, kennicutt87, smith07}.

Our goal in the current study is to track the evolution of hot gas in mergers compared
to other components of the galaxies.  
For our sample of merging galaxies we compare the 
X-ray luminosity from hot gas 
with other properties of the galaxies, and 
with normal spirals and ellipticals.  We compare the X-ray luminosity of the hot gas 
with the SFR as determined from UV and IR data, the 
stellar mass as traced by near-IR observations, and the merger
stage as indicated by optical and near-IR images.

In Section 2 of this paper, we describe our merger sample,
while the data at other wavelengths is
described in Section 3.  Section 4 describes our comparison samples of
ellipticals and spirals.   In Section 5, we outline the processing
and analysis of the Chandra data for the mergers.
We compare L$_{\rm X}$(gas) with various other properties of the system in Sections 6.
In Section 7, we discuss the results.
Conclusions are presented in Section 8.
The Appendix of this paper includes a detailed discussion of each system in the sample,
including morphology.

\section{The Merger Sample }

\subsection{Sample Selection}

From the \citet{arp66}
Atlas of Peculiar Galaxies, we selected the subset of systems that are approximately equal-mass spiral pairs or merger remnants, 
eliminating triples, groups, unequal-mass pairs/mergers, 
radio galaxies, and pairs containing ellipticals. 
We found 32 Arp galaxies that fit these criteria, and have archival Chandra data with sufficient sensitivity to 
detect the most luminous point sources (i.e.,  0.3 - 8 keV X-ray luminosity (L$_{\rm X}$) $\ge$ $10^{40}$ erg~s$^{-1}$). 
We excluded Arp 245 from the sample, due to significant pile-up 
in the Chandra data because of a powerful AGN.
We supplemented this sample by adding nearby non-Arp major mergers 
from the literature that have archival Chandra data to the same 
limiting L$_{\rm X}$.   
These additional systems were obtained from the surveys of 
\citet{keel95}, \citet{gao99}, \citet{rothberg04}, \citet{taylor07}, 
\citet{brassington07}, and \citet{ellison13}.
We also added the nearby pre-merger pair NGC 2207/IC 2163 
\citep{mineo14}.
This brings our final sample to 49 systems. 

The distances to each system are listed in Table 1.
These were obtained from the NASA Extragalactic Database 
(NED)\footnote{http://ned.ipac.caltech.edu} assuming a Hubble constant of 
73 km/s/Mpc, correcting for peculiar velocities 
due to the 
Virgo Cluster, the Great Attractor, and the Shapley Supercluster. 
All of the galaxies are within 180 Mpc, with
a median distance of 51.5 Mpc.
Figure 1 provides a histogram of the distances to the sample galaxies.

In the Appendix to this paper, we provide detailed descriptions of 
the sample galaxies.
Our sample includes both pre-merger pairs and post-merger remnants, as well as system in mid-merger.
For mid-merger systems and post-merger remnants, 
we selected systems with morphological signatures characteristic 
of major mergers, including two tidal tails.  
We have also included
galaxies with shell structures, although there is 
some uncertainty about the origin of such shells.
According to numerical models,
major mergers can produce shells \citep{barnes92,
hernquist92}, but so can minor mergers 
\citep{quinn84, dupraz86, hernquist87a, hernquist87b} and weaker
interactions \citep{thomson90, thomson91}.
We note that the original mass ratio of the 
progenitor galaxies is generally difficult to determine 
post-merger, thus it is possible that some of
our systems may be the product of minor mergers or mergers of multiple galaxies.

Since our sample was selected based on the existence of 
suitable data in the Chandra archives,
it may be biased towards X-ray-bright objects.  Furthermore,
since most of our galaxies were chosen from an 
optical-morphology-selected
catalog (the Arp Atlas), 
the galaxies, on average, may have lower SFRs than other 
merger samples that are IR flux and luminosity selected 
(e.g., \citealp{larson16}).
However, this bias is partially compensated for by the addition
of galaxies from the \citet{gao99} far-infrared-selected sample.

\subsection{Merger Stages}

We did a rough classification of the 49 systems in our sample into seven merger stages based on 
morphology (Table 1).  These stages are: 
1: separated but interacting pair with small tails or no tails. 
2: separated pair with moderate to long tails. 
3: pair with disks in contact. 
4: common envelope, two nuclei, and tails. 
5: single nucleus and two strong tails. 
6: single nucleus but weak tails. 
7: disturbed elliptical with little or no tails.
When possible, Hubble Space 
Telescope near-IR and optical images were used to discern double nuclei (e.g., 
\citealp{haan11, kim13}; see Appendix).
In Figure 1, we provide a histogram showing the number of galaxies in each of these stages.
For each of these stages, we have at least five systems, which will provide a measure of the scatter in the X-ray properties along the sequence. 
We emphasize that these stages are quite uncertain (as much as $\pm$ 1 stage in some cases), due to observational
limitations, viewing angle, projection effects,
and system-to-system variations in the parameters of the
interaction and the progenitor galaxies.   

The relationship between
these 
merger stages and the absolute timescale of the merger varies from system
to system,
as it depends upon the masses of the two galaxies,
the orbital parameters, and other properties of the system 
(e.g., \citealp{lotz08}).
Furthermore, the timescale of a merger is not necessarily
correlated with 
the age of a starburst triggered by that merger; if and when a starburst
is triggered by a merger depends in a complicated way on the properties
of the system (e.g., \citealp{dimatteo08}).
In spite of these limitations, however, these stages are helpful in 
searching for general trends in the X-ray properties of galaxies with
merger morphology.

In general, the latest stage merger remnants in our sample tend to 
be more nearby than the other galaxies in the sample.  This is shown
in Figure 2, where we plot stage vs.\ distance.   
Very late-stage merger
remnants are difficult to identify at large distances.
In contrast, several of the systems in the middle of the merger sequence
are relatively far away.   These include three IRAS-discovered mergers 
from the \citet{gao99}
survey.

\subsection{Active Galactic Nuclei}

Since the presence of an active galactic nucleus (AGN) may affect the diffuse
X-ray emission in a galaxy, in Table 1 we identify
which galaxies in the sample are listed as AGN in NED.
Six of the galaxies in the sample are classified in NED as Seyfert 2,
two as Seyfert 1, 
one as an unspecified Seyfert, and one as ``Sy2/HII".
Two were classified as ``Low Ionization Nuclear Emission Region" (LINER)
galaxies, one as ``Sy2/Liner", and one as ``Liner/HII".
More information 
about the spectral types of these galaxies, including references, is available
in the Appendix to this paper.
In all plots in this paper, the Seyfert galaxies are 
marked as open red circles.
We note that none of the merger stage 6 or 7 systems in our sample
are classified as Seyfert in NED.
The majority of the Seyferts in the sample are in the middle
merger stages (stages 3, 4, and 5).

Obscured AGN can sometimes be identified by their mid-IR neon line ratios.
To search for additional AGN in our sample, we scoured the 
literature for published neon line fluxes for our sample galaxies.
At least some neon data was available for 31 of our sample systems
\citep{verma03, bernard09, inami13, pereira10}.
A Ne~V $\lambda$14.32 $\mu$m/Ne~II $\lambda$ 12.8 $\mu$m ratio
greater than 0.1 is a good indicator of an AGN \citep{inami13}.
Only two of our systems meet this criteria, both of which are otherwise
identified as AGN (NGC 5256 and Mrk 273).

An alternative way to identify AGN is via X-ray observations, which in
some cases can reveal excess high energy photons above that expected from
a starburst alone.   In the Appendix of this paper, 
we summarize earlier analyses of the Chandra data for each galaxy in our
sample, as well as observations from other X-ray telescopes. 
In most cases, AGN identified by the X-ray spectra had already been 
identified as AGN by optical or IR observations.  In two cases,
Arp 293 and NGC 5018, 
the X-ray data revealed possible low luminosity AGNs not found at other 
wavelengths.  Based on published analyses, however, these sources contribute little to the total
bolometric luminosity of the galaxies (see Appendix).

\section{UV/IR Data and SFRs}

Our primary goal in this study is to compare the diffuse X-ray light from these galaxies
with other properties of the systems, including SFR, stellar mass, and stellar population age.
To this end, we have extracted UV and IR fluxes 
for these galaxies from archival Spitzer and 
Galaxy Evolution Explorer (GALEX) images
using the method described in Smith et al. (2007, 2010).   
In this study we use 
GALEX far-ultraviolet (FUV; $\lambda$$_{eff}$ = 
1516 \AA) and near-ultraviolet (NUV; $\lambda$$_{eff}$ = 2267 \AA) data, 
as well as 
Spitzer 3.6 $\mu$m and 24 $\mu$m fluxes. 
For the one galaxy without a Spitzer 24 $\mu$m image, NGC 2865, we used the WISE 22 $\mu$m 
elliptical aperture photometry from the AllWISE
Source Catalog \citep{cutri14} instead.
For two of the galaxies without GALEX UV images, Arp 186 and Markarian 231, we used
archival Hubble Space Telescope (HST) images in similar filters.
For NGC 1700, which had neither GALEX nor HST UV images, we used an archival 
SWIFT UV image in the uvm2 filter (central
wavelength 2246 \AA) in place of the GALEX NUV flux; 
SWIFT fluxes in this filter agree well with the GALEX NUV fluxes
\citep{hoversten09}.
Luminosities ($\nu$L$_{\nu}$) in these bands are provided in Table 1.

In Table 1, we provide estimates of the global SFRs of these galaxies.
For most of the galaxies, to calculate the SFR we used the 
\citet{hao11} prescription in terms of the FUV and the 24 $\mu$m 
luminosities.\footnote{SFR(M$_{\sun}$~yr$^{-1}$) = 4.5 $\times$ 10$^{-44}$(L$_{\rm FUV}$ + 3.89L$_{24}$),
where L$_{\rm FUV}$ and L$_{24}$ are $\nu$L$_{\nu}$ at FUV and 24 $\mu$m, respectively.  This assumes
a \citet{kroupa02} initial mass function (IMF).
Our SFRs are a factor of two times lower in the median than the SFRs quoted
by \citet{howell10} 
for the 16 galaxies in common with their sample of luminous infrared galaxies;  the difference is due to 
a different IMF and a different prescription for the SFR (FUV plus total IR rather than FUV+24 $\mu$m).
}

For the four galaxies with NUV data 
but no FUV measurement, we used an alternative relation from \citet{hao11}
for SFR as a function of the NUV plus the
24 $\mu$m luminosities.\footnote{SFR(M$_{\sun}$~yr$^{-1}$) = 6.76 $\times$ 10$^{-44}$(L$_{\rm NUV}$ + 2.26L$_{24}$),
where L$_{\rm NUV}$ and L$_{24}$ are $\nu$L$_{\nu}$ at FUV and 24 $\mu$m, respectively.} 
For the one galaxy with no UV observations, Arp 235,
we calculated the SFR from the 24 $\mu$m luminosity 
alone using the relationship from \citet{rieke09}.
The \citet{hao11} relation for SFR is thought to be reliable for a range of SFRs \citep{catalan15},
however, it may over-estimate the SFR for galaxies with powerful AGN or significant contributions
to the UV or mid-IR flux from older stars.
Contributions from AGN to powering the global UV and IR fluxes of our galaxies are
discussed further in Sections 6.1 and 6.3.

Table 1 also includes total
far-infrared luminosities (L$_{\rm FIR}$, from 42.5 $-$ 122.5 $\mu$m; \citealp{helou85}) for these systems, 
calculated from
Infrared Astronomical Satellite (IRAS) 60 $\mu$m and 100 $\mu$m
flux densities. We used the total IRAS fluxes quoted in NED, except in a 
few cases where we used the on-line {\it xscanpi }
software\footnote{http://irsa.caltech.edu/applications/scanpi/}
to extract
total fluxes.  
One of the galaxies in the sample, NGC 1700, is marginally detected or not 
detected by IRAS in these bands.\footnote{As indicated by 
the scanpi results; also
see G. Knapp 1994, priv. communication quoted in NED.}
For NGC 1700, we used Spitzer 24, 70, and 160 $\mu$m fluxes
from 
\citet{temi09}
to calculate a total IR flux (TIR) using the relation
in 
\citet{dale02}.
We then used a ratio of L$_{\rm FIR}$/L$_{\rm TIR}$ = 0.51 
\citep{dale02}
to estimate the FIR luminosity of NGC 1700.

Figure 1 displays a histogram of the L$_{\rm FIR}$ for the sample galaxies.
The median L$_{\rm FIR}$ for the sample galaxies is $4.5\times10^{10}$ L$_{\sun}$. 
Far-infrared luminosities are also sometimes used as a measure of the SFR
in galaxies
(e.g., \citealp{kennicutt98}), 
but in low SFR systems older stars may help
power the far-infrared (e.g., \citealp{smith91, smith94, sauvage92}), 
and in post-starburst
systems the far-infrared 
may over-estimate the current SFR (e.g., \citealp{hayward14}).
The FIR may also over-estimate the SFR in systems with AGN.

Table 1 also includes the near-infrared K 
luminosities\footnote{Calculated using the equation given in 
\citet{brassington07}
(from \citealp{seigar05}):
log(L$_{\rm K}$) = 11.364 - 0.4K$_{\rm T}$ + log(1+z) + 2log(D),
where K$_{\rm T}$ is the total K magnitude, z is the redshift, L$_{\rm K}$ is the K band luminosity in solar luminosities, 
and D is the distance (in Mpc).}
of the systems, obtained from the 2MASS database using 
the K$_{\rm S}$ total magnitudes from the 2MASS Extended Source Catalog 
\citep{cutri06}.
The values for L$_{\rm K}$ were coadded in cases where there were magnitudes for the individual galaxies in the 
pair.\footnote{We note that for three of our systems, 
Arp 242, Arp 270, and NGC 5256, 
our K-band luminosities differ significantly from those of 
\citet{brassington07};
for these systems, they used the smaller 2MASS point source 
catalog fluxes (N. Brassington, 2014, private communication) 
rather than the total fluxes. }
In the top panel of Figure 3, 
we plot L$_{\rm K}$ vs. merger stage. 
There is a lot of scatter in this plot.  
Note that there are a few galaxies at the beginning and end of
the merger sequence with considerably lower 
L$_{\rm K}$ than the majority of the galaxies.
The system with the highest L$_{\rm K}$ is the stage 5 Seyfert galaxy Markarian 231; the 2nd highest luminosity system
is the stage 4 system NGC 6240.
L$_{\rm K}$ is an approximate tracer of the stellar mass of the galaxy,
however, 
it is not perfect as 
the mass-to-L$_{\rm K}$ ratio can vary with the age of 
the stellar population due to contributions
from asymptotic giant branch stars and red supergiants
\citep{maraston98, bell01, into2013}.
Hot dust heated by an AGN can also contribute to the K band flux
(e.g., \citealp{oyabu11}).

In the middle panel of Figure 3, we display L$_{\rm FIR}$ vs.\ merger stage.
Note that some of the mid-merger systems have very high FIR luminosities,
while most of the late-stage mergers have lower 
L$_{\rm FIR}$.
The bottom panel of Figure 3 shows a plot 
of L$_{\rm FIR}$/L$_{\rm K}$ vs.\ merger stage. 
An enhanced L$_{\rm FIR}$/L$_{\rm K}$ ratio
is an indicator of an increase in star formation relative to 
the stellar mass of the 
galaxy. 
No strong trend 
of L$_{\rm FIR}$/L$_{\rm K}$ with merger stage
is visible, however a few systems in stage 3-6 have 
higher L$_{\rm FIR}$/L$_{\rm K}$ 
than the rest, while some of the stage 7 systems and one stage 6 system have low L$_{\rm FIR}$/L$_{\rm K}$ indicating a quiescent stellar population.

There is a large scatter in L$_{\rm FIR}$/L$_{\rm K}$ as a function
of merger stage.  This 
is consistent with earlier studies that found 
considerable scatter in the stellar-mass-normalized SFR along the merger sequence
\citep{keel95, ellison13}.
It is also consistent with numerical simulations, which predict that the 
duration of a starburst is short compared to the merger timescale, and 
powerful bursts are infrequent and depend on the parameters of the interaction 
\citep{dimatteo07, dimatteo08, lanz14}.

\section{Comparison Samples: Normal Spirals and Ellipticals}

As a comparison sample of normal spiral galaxies, 
we use the 
nine nearby galaxies in the \citet{mineo12b} Chandra study
that are classified in NED as spiral galaxies, and have no companion in
NED with an optical luminosity greater than about one quarter that of the 
target galaxy lying within four galaxian diameters.   
These spiral galaxies have Hubble types between Sb and Sd.

In addition to providing diffuse X-ray fluxes from Chandra,
\citet{mineo12b} also provide SFRs for their galaxies
derived from UV and IR data,
however, they use a different formula for the SFR than we do.  Therefore, 
for consistency with our mergers, 
we obtained published total
Spitzer and GALEX fluxes for the spirals from 
NED\footnote{Spitzer: \citet{smith07, dale07, dale09, engelbracht08},
GALEX: \citet{lee11, brown14, munoz09, hao11}},
from archival images, or from WISE, from either the AllWISE Catalog
\citep{cutri14} or \citet{ciesla14}.
We calculated SFRs for these galaxies
using the same formulae as for the mergers.
As we did for the mergers, for these spirals we used NED to look up 
IRAS 60 and 100 $\mu$m fluxes and 
2MASS K band magnitudes, and calculated 
L$_{\rm FIR}$ and L$_{\rm K}$.

We also compare our merging galaxies with two samples of early-type galaxies.
The first is
a set of 42 elliptical and S0 galaxies from the ATLAS$^{\rm 3D}$ survey
for which
\citet{su15} extracted the soft
diffuse X-ray fluxes.   
The second set consists of 33 more massive elliptical galaxies 
(M$_{\rm K}$ $<$ 25.3 magnitudes)
from \citet{goulding16}, who provide measurements of 
the diffuse X-ray fluxes from the hot gas.
The parent samples for both the \citet{su15} and the \citet{goulding16} surveys are
near-IR selected and volume-limited, however, like the current merger study, their
X-ray studies are Chandra-archive-selected, which may bias them towards
X-ray-bright objects.

As with the spirals, we obtained
total 2MASS, Spitzer, GALEX, and IRAS fluxes for these early-type galaxies
from NED, and calculated our own L$_{\rm K}$ and SFR estimates with the same formulae
as with the mergers.  In the discussion below, these SFR values
are treated as upper limits,
since the observed UV and MIR light may be powered in part by older stars.
\citet{su15} provide alternative estimates of the SFRs 
of their galaxies from UV-to-IR spectral energy
distribution (SED) fitting; these are less than ours by a factor of 1.9 in the median.
This difference may be due in part to contributions to the UV/IR light from older stars.
The E/S0 galaxies will be discussed further in Section 6.2. 

\section{Chandra Data and Reductions}

\subsection{Extraction of X-ray Spectra}

All of the Chandra data used in this study came from the 
Advanced CCD Imaging Spectrometer (ACIS) S-array.
We only used data from the S3 chip on this array
since it has the highest sensitivity.
The data was reprocessed using the 
chandra$\_$repro script, 
and the data were filtered, 
retaining grades 0, 2, 3, 4, and 6.
When multiple data sets were available for a system, we combined the datasets to improve the sensitivities. 

The basic data reduction was done using the Chandra Interactive Analysis of 
Observations (CIAO) software version 4.7. 
For each observation, blank regions of the sky off of the galaxy were 
identified and the light curve extracted.
All of our targets are small enough on the sky 
that they fit within the 8\farcm3 $\times$ 8\farcm3 field of view of the ACIS-S3 chip,
and nearby  
contemporaneous background can be defined on the same CCD.
Background regions used for the flux extraction
included most of the CCD excluding bright point sources and the target region.

The data were deflared and the good time intervals (GTIs) were identified. 
The GTIs were extracted by using the ciao command {\it deflare} with the sigma clipping routine with
nsigma = 3 and a bin size of 259.28 seconds, restricting the energy range to 0.3 $-$ 8 keV.
The final Chandra exposure times range from 2.6 ks to 456 ks, with a median time of 31.5 ks. 
These quantities are provided in Table 2, along with the datasets used.

Table 2 also includes 
0.3 $-$ 8 keV 
point source sensitivities, calculated using the 
PIMMS\footnote{Portable Interactive Multi-Mission Simulator;
http://asc.harvard.edu/toolkit/pimms.jsp} 
software
version 4.8
assuming 5 counts for a lower limit detection on 
axis and a $\Gamma$ = 1.8 power law spectrum,
including Galactic absorption but not absorption within the galaxy.

For each observation, 
the Chandra point spread function (PSF) as a function of the location on the sky was determined
using the CIAO command {\it mkpsfmap}.  With {\it mkpsfmap}, we used an effective energy of 2 keV and included
regions encompassing 30\% of the source counts of the point source.
We then used the PSF map and the 
CIAO routine 
{\it wavdetect} to identify point sources in or near the galaxy.
For {\it wavdetect},
we used an energy range of 1.5 $-$ 7 keV and 
wavelet scales of 1 and 2.
The point source regions were then excluded in
determining the diffuse X-ray light from the galaxies. 

For all the systems with SDSS images available,
after excluding the point sources
we extracted the X-ray spectrum for the diffuse light 
within the
SDSS g band surface brightness isophote of 24.58 mag~arcsec$^{-2}$.
This corresponds to a B surface brightness of 25.0 mag~arcsec$^{-2}$ (i.e., the D$_{25}$ isophote), assuming
the \citet{jester05} g to B conversion with
the median g $-$ r color for tidal tails from 
\citet{smith10}.
When SDSS images were not available we used the GALEX 
NUV image, measuring the X-ray flux within a NUV surface brightness isophote of 
26.99 mag~arcsec$^{-2}$, 
using the median NUV $-$ g color for tidal tails of 2.4 
\citep{smith10}.

We carefully inspected the Chandra maps for emission outside of this region.
In most cases the diffuse 
X-ray emission does not 
extend beyond the optical/UV isophotes given above. 
The exceptions are Arp 220, NGC 6240, and Mrk 273 (see Appendix
for details).
All of these are known AGNs.
As noted in the Appendix, most of the observed flux for these galaxies lies within our isophotal 
limits, thus our total flux determinations are reasonably accurate in spite of not including this
very extended emission.

The CIAO command {\it specextract} was used to extract the spectrum from the
selected regions.  When multiple datasets are available, 
the ``combine=yes" option was used, which calibrates each dataset individually,
then co-adds the weighted spectra.

\subsection{Correction for Absorption }

In fitting our X-ray spectra to spectral models, we need to take into account
two separate sources of absorption:
foreground gas in 
the Milky Way and gas within the target galaxy
itself.  
For absorption by Galactic material,
we used the 
Galactic hydrogen column densities N$_{\rm H}$ from
\citet{kalberla05}.\footnote{https://heasarc.gsfc.nasa.gov/cgi-bin/Tools/w3nh/w3nh.pl}
These values of Galactic N$_{\rm H}$ are provided in Table 3.

For galaxies 
with 
high S/N X-ray data, it is sometimes possible to 
constrain the total (Galactic plus internal)
absorbing column from the X-ray
spectrum itself.  
However, these results are often very uncertain (see Appendix and Section 5.3).

We therefore make an alternative
estimate of the  
internal
hydrogen column density
using data at other wavelengths,
and extrapolate to the X-ray.   
For all of the galaxies that have FUV data as well as 
24 $\mu$m photometry, we 
use the 
FUV/24 $\mu$m ratio 
as in \citet{hao11}
to estimate the attenuation
in the FUV band,
A$_{\rm FUV}$.
We then obtain the 
attenuation 
in the optical V band,  
A$_{\rm V}$, assuming 
the \citet{calzetti01} dust attenuation law.
For the four galaxies without FUV images but with NUV data,
we use the NUV/24 $\mu$m ratio 
as in 
\citet{hao11}
to estimate 
A$_{\rm V}$.
From A$_{\rm V}$, we infer the color excess E(B$-$V) using the 
relation A$_{\rm V}$/E(B$-$V) = 3.1 \citep{savage79}, 
and calculate the
hydrogen column density N$_{\rm H}$ using the \citet{bohlin78} equation
N$_{\rm H}$(cm$^{-2}$) = 5.8 $\times$ 10$^{21}$ E(B$-$V). 
We did not calculate
the internal absorption for the one galaxy in our sample that does 
not have any UV data available.

These estimates of the internal N$_{\rm H}$ in the
target
galaxies are given in Table 4.
We note that this method makes assumptions about the geometry 
of the X-ray-emitting gas
compared to the UV-emitting stars which may not always be reliable.
However, they provide a first estimate, when more direct measurements
from the X-ray spectra are not possible.

\subsection{Spectral Fitting }

We used the
xspec\footnote{https://heasarc.gsfc.nasa.gov/xanadu/xspec/}
software version 12.9.0 to
fit the extracted X-ray spectra using a two-component
spectral model: a thermal optically-thin plasma modeled by a MEKAL function 
\citep{mewe85, liedahl95}
combined with a power law 
component.   As discussed at length in Section 5.4 below,
the MEKAL component is likely dominated by emission from hot interstellar gas,
while the power law component is assumed to be mainly from unresolved point sources.
Throughout this paper we assume that the MEKAL component is
due to hot gas, thus we will refer to it as L$_{\rm X}$(gas) throughout.
The unresolved point sources making up the power law component
are discussed in detail in Section 5.4 below.

For our initial fits, we fixed the temperature
of the thermal component to 0.3 keV and the photon index
of the power law component to $\Gamma$ = 1.8.   
A temperature of 0.3 keV is characteristic of the hot gas in spiral galaxies
\citep{strickland04, grimes05, mineo12b}, while $\Gamma$ = 1.8
is typical of high mass X-ray binaries (HMXBs) 
with
$L_{\rm X}$
$<$ 10$^{39}$ erg~s$^{-1}$
in nearby 
galaxies \citep{kong02, swartz04}. 
The contributions from unresolved HMXBs to the diffuse
X-ray emission is discussed further in Section 5.4.1.
We repeated the fitting 
process twice, once with the absorbing
column fixed to the Galactic value, and second, with N$_{\rm H}$
set to the total attenuation, including the internal attenuation obtained from the
UV/IR data as described above.

Results from these fits are included in Tables 3 and 4, where we
provide the 
absorption-corrected 
0.3 $-$ 8 keV 
X-ray luminosities for these two
components, along with the best-fit reduced chi-squared $\chi$$_{\nu}^2$,
the chi-squared $\chi$$^2$, the degrees of freedom,
and the total observed X-ray flux.
Table 3 gives the results including only Galactic absorption, while Table 4 gives
the luminosities including both Galactic and internal absorption.

As can be seen by comparing Tables 3 and 4, 
the fits are generally better when internal absorption is
included, however, for some systems good fits are also found with only Galactic 
absorption.
For the galaxies with poor fits in Table 4, we experimented
with fitting additional parameters to improve the fit.   
For systems with long exposure times and high S/N spectra, we ran a 
total of 14 spectral models of increasing complexity.
These models included:
1) varying the temperature of the MEKAL component, 2) adding a second MEKAL component
and fitting for the second temperature as well, 
3) fitting for the absorbing column, 4) fitting for the photon index of the power law
component, 5) fitting for the absorbing column of the MEKAL component separately
from the absorbing component of the power law component, 6) varying the abundances of 
the $\alpha$ elements relative to iron by using the VMEKAL function instead of the
MEKAL function, 7) adding a Gaussian at 6.4 keV to match the Fe-K$\alpha$ line
(e.g., \citealp{grimes05})
and 8) various combinations of the above.
For the VMEKAL models, as in \citet{grimes05} we tied the abundances of the $\alpha$-elements together,
and tied the Fe abundance to that of Ni, Ca, Al, and Na.  We then fit for the $\alpha$/Fe ratio.

We then rejected models with non-physical results (i.e., extreme power law 
photon indices or very low temperatures ($<$0.2 keV)). We also rejected more
complex models in which one or both MEKAL components was detected at 
$<$3$\sigma$
levels, or models in which the column density was not well-determined.
We then sorted the
remaining models, and selected the model 
with the most degrees of freedom (least number of parameters) 
that had a 
null hypothesis probability $\ge$ 0.10 (i.e., if the null hypothesis is that the model
is a good fit to the data, we fail to reject the null hypothesis at the 10\% level).
If none of the models met this
criteria, we selected the model with the highest null hypothesis probability.
In cases with multiple good models with different parameters, we inspected the fits to the X-ray spectra
by eye, especially the low energy regime, to select the best model.
In choosing between a good-fit model with a high gas temperature and a low absorbing column,
and a second good-fit model with a lower gas temperature but higher absorbing column, we selected the
former.

The final best models selected by this procedure are included in Table 5, along with the parameters
of the model.
Some example spectra with best-fit models overlaid are presented in the Appendix.
In no case did adding the 6.4 keV Fe$-$K$\alpha$ Gaussian component significantly improve the fit,
so those models were not used.
In no case were we able to strongly constrain the photon index of the power law component,
and 
in no case did varying the photon index of the power law provide the best model.
Table 5 also includes the absorption-corrected 0.3 $-$ 8 keV luminosities of the MEKAL
and power law components
of these new fits. 
In most cases, our new final absorption-corrected MEKAL luminosities agreed well with the results obtained with 
fixed 0.3 keV temperature and fixed UV/IR-derived absorption (Table 4).
For the 23 galaxies in Table 5,
the mean and median ratio of the MEKAL luminosity in Table 5 to that in Table 4 is 0.99, and the
root mean square deviation in this ratio is 0.68.  The fluxes in the two tables
generally agree to within a factor of three (0.5 dex) or better.
This gives us an estimate of the uncertainty in the luminosities obtained by our simple fixed temperature,
fixed column density estimates for the lower S/N systems.
For the systems for which we can estimate the absorbing column from the X-ray spectra,
these values generally reasonably well 
with our UV/IR estimates of N$_{\rm H}$ 
(Tables 4 and 5).

To improve the fits in Table 5 further would require even more complex models.
For example,
one might divide the data into
different radial regimes, and fit different spatial regimes separately.
Such analyses have been done using the 
Chandra archival data for some of our galaxies by other research groups.
In the Appendix to this paper, the methods and 
results of these earlier studies
are described in detail, and compared with our results.
The assumptions and methods used in these earlier studies differed widely.
However, 
for most of these systems
our total absorption-corrected MEKAL luminosities agree within a factor of a few with
these earlier studies.  This gives another estimate of the uncertainties in our luminosities.

Yet another method of estimating the uncertainties in the derived absorption-correction MEKAL luminosities is
to compare the best-fit absorption-corrected MEKAL luminosities for different `good' models for the same galaxy.
There were 15 galaxies in our sample which had at least three different `good fit' models,
with null hypothesis probability greater than 0.10 and meeting the other criteria listed above.   
For each of these galaxies,
we calculated the rms spread in the absorption-corrected
MEKAL luminosities for the different models.  The median rms dispersion for the sample was 0.31 dex, or a factor
of two in the luminosity.  This means that the choice of model can cause the derived luminosity to differ
by about a factor of two on average.  

For the subsequent analysis in this paper, 
when a better fit is available (Table 5) we used that rather than the luminosities from
Table 4.   For the models with two MEKAL components, we used the sum of the two MEKAL luminosities 
in Table 5
as the X-ray luminosity of the
hot gas rather than the MEKAL luminosity from Table 4.
In Table 6, 
we provide the ratios of these estimates of L$_{\rm X}$(gas) with L$_{\rm FIR}$, L$_{\rm K}$,
and SFR, for both Galactic and total absorption.

\subsection{Contributions to the Observed Diffuse X-Ray Emission}

The observed diffuse X-ray emission from galaxies is made up of at least two 
distinct spectral
components (e.g., \citealp{long14, kuntz16}).  
First, there is hot gas with a thermal spectrum.
In spiral galaxies, this hot gas is mostly
due to supernova shocks and feedback from star formation (e.g., \citealp{grimes05}),
while 
in ellipticals and spheroidal bulges the hot gas is thought to be
mainly virialized gas ejected from old stars (e.g., \citealp{ciotti91, pellegrini98}).
Second, the observed diffuse X-ray emission from galaxies
also includes
the light from faint unresolved point sources below the point source detection 
threshold, most of which have a power law spectrum.
Unresolved sources include HMXBs, low mass X-ray binaries
(LMXBs), 
cataclysmic variables (CVs), 
coronally active binaries (ABs) and young supernovae
remnants, as well as objects associated
with a young stellar population, including protostars and young stars.
Unresolved young supernova remnants below the point source sensitivity limit
are expected to have
a thermal X-ray spectrum \citep{long10} thus are included in our hot
gas component.
The other components are discussed further below.
Another possible contributor 
to the measured diffuse X-ray emission is
contamination by 
incompletely-removed bright point sources, i.e.,
spillage outside the aperture in the wings of the point spread
function.
This light is also expected to have a power law X-ray spectrum.
This component is also discussed below.

\subsubsection{Contributions from Unresolved HMXBs}

The contribution from unresolved HMXBs is expected
to have a power law X-ray spectrum and 
to
scale with the SFR
(\citealp{grimm03, mineo12b}).
The luminosity of the unresolved sources will depend upon the point source sensitivity of the
observations.

To estimate the expected contributions to the observed diffuse X-ray emission
from HMXBs, we scale from the SFR using the 
\citet{mineo12a}
best-fit 
X-ray luminosity 
function for HMXBs in a sample of nearby star-forming galaxies:

$dN/dL_{\rm 38} = \xi ~ SFR ~ L_{\rm 38}^{-\gamma} $

Where L$_{\rm 38}$ is the 0.5-8 keV luminosity in units of 10$^{38}$ 
erg~s$^{-1}$. 

Integrating over this luminosity function gives the total
unresolved X-ray luminosity due to HMXB:

$L_{\rm X}({\rm unresolved~HMXBs}) = 
\int_{L_{lower}}^{L_{upper}} \xi ~ SFR ~ L_{\rm 38}^{-\gamma} L dL $

$L_{\rm X}({\rm unresolved~HMXBs}) 
= [(\xi ~ SFR)/(2 - \gamma)] [ L_{upper}^{(2-\gamma)} 
                - L_{lower}^{(2-\gamma)}] $

Using $L_{lower}$ = 10$^{34}$ erg~s$^{-1}$ = 10$^{-4}$ L$_{38}$ as
in \citet{mineo12b}, and the \citet{mineo12a} best-fit values 
of 
$\gamma$ = 1.6 and $\xi$ = 1.49 (also see \citealp{grimm03}) gives:

$L_{\rm X}({\rm unresolved~HMXBs}) 
= 3.725 ~ SFR [ L_{upper}^{0.4} - 0.025 ] $
   
A power law X-ray spectrum with a photon index $\Gamma$
is assumed (i.e., a photon flux A $\propto$ E$^{-\Gamma}$, where E is
the energy).
We assumed
$\Gamma$ = 1.8, the average value for 
resolved
$L_{\rm X}$
$<$ 10$^{39}$ erg~s$^{-1}$
point sources in nearby star forming
galaxies \citep{kong02, swartz04}. 
Using L$_{upper}$ equal to the 5-count PIMMS
sensitivity limit (Table 2), 
we integrated this formula to determine the total X-ray luminosity
expected due to unresolved HMXBs.  
These values are also included in Table 2.

In Figure 4, 
the 
ratio of the 
predicted $L_{\rm X}({\rm unresolved~HMXBs}$) to 
the best-fit 
power law X-ray luminosity is plotted against the SFR.
In the top panel,
the results assuming only Galactic absorption are shown,
while the bottom panel shows the results using the correction
for internal absorption.  
Red circles mark AGN.  The red dotted line indicates the expected result
if all of the measured power law flux is due to HMXBs.

These plots show a strong
trend.
At the high SFR end, the ratios are consistent with the power law flux being dominated by light from unresolved HMXBs.  
However, at the low SFR end, another source of hard X-ray light must be contributing.
The most likely contribution is from LMXBs associated with the older stellar population.
This is discussed in the next section.

\subsubsection{Contributions from LMXBs }

The light from unresolved LMXBs is expected to have a power law spectrum with a
photon index of $\Gamma$ $\sim$ 1.6 \citep{irwin03}.
To first approximation, the total X-ray luminosity of LMXBs in
a galaxy is expected to scale
with the stellar mass M$_*$ of the galaxy \citep{gilfanov2004}.
This ratio
also depends upon the 
average stellar population
age, with about a factor of 1.5 times higher total L$_{\rm X}$ per M$_*$ for older ages \citep{zhang12}.
For each galaxy in our sample, we calculated the expected total X-ray luminosity
from LMXBs, L$_{\rm X}$(LMXBs), scaling from the K band luminosity using the average value
of 6.1 $\times$ 10$^{39}$ erg~s$^{-1}$ per 10$^{11}$ L$_{\sun}$ 
for LMXBs from \citet{gilfanov2004}. 	
These estimates are also included in Table 2.

In Figure 5, 
the ratio of the
predicted 
best-fit
L$_{\rm X}$(LMXBs) to the 
absorption-corrected power law luminosity
is plotted against the SFR.
The top panel assumes only Galactic absorption, 
while the bottom includes absorption internal to the galaxy.
These plots show that in most cases, LMXBs alone are insufficient
to account for the observed power law flux.   However, for a few of the lower
SFR systems, the expected contribution from LMXBs could account for all of the
power law emission.

In Figure 6, we plot SFR against the sum of the expected HMXB plus LMXB flux,
divided by the absorption-corrected power law luminosity.
Given that there is considerable scatter in the HMXB$-$SFR relation 
(rms dispersion $\sim$ 0.4 dex; \citealp{mineo12a})
and the LMXB$-$L$_{\rm K}$ and LMXB$-$M$_{*}$ relations 
(rms dispersion $\sim$ 0.2 dex; 
\citealp{zhang12}), 
Figure 6 shows that in most cases, HMXBs and LMXBs together can plausibly account for most of the
observed power law flux.   
However, there are a few systems which appear to have
an excess of power law flux compared to these predictions.
In order from largest excess, these three galaxies are UGC 5189, Arp 263,
and Arp 295.
Our diffuse fluxes from these systems may include
some imperfectly-removed light from an AGN, although none of them
are classified as Seyfert galaxies.

\subsubsection{Contributions from CVs and ABs}

Another possible contributor to the diffuse X-ray emission from galaxies
is cataclysmic variables (CVs) and
coronally active binaries (ABs).
This component is not likely to be the cause of the possible excess power law component,
as it 
is expected to have an approximately thermal component, rather than
a power law \citep{revnivtsev08}.
Because they have similar X-ray spectra, the flux from CVs/ABs cannot
be separated spectroscopically from hot gas.
However, contributions from CVs/ABs are expected to be small.
\citet{revnivtsev08} find that
CVs and ABs with L$_{\rm X}$ $<$ 10$^{36}$ erg~s$^{-1}$ contribute a 
constant ratio log(L$_{\rm X}$/L$_{\rm K}$) $\sim$ $-$5.8 to the global light of galaxies.
This ratio is significantly smaller than the observed MEKAL-to-K-band ratios for almost
all of our
galaxies (see Table 6), thus we conclude that CVs and ABs contribute a negligible 
part of the observed diffuse emission for our galaxies.

\subsubsection{Residuals or Deficiencies Due to Point Source Removal}

To estimate whether spillage from incompletely-removed bright
sources is a significant source
of contamination, 
we compare the total point source counts for each galaxy with the
measured diffuse emission.  
Such spillage is typically expected to be 2$-$4 percent of the 
point source counts,
but can be up to 10\% \citep{mineo12b}.
The median ratio of the point source counts to the counts 
in the diffuse emission
is only 0.39, with the largest ratio, 4.6, being found for Arp 261.
In the median case, only 1.6\% of the observed diffuse counts are due to imperfectly-removed point sources.
For Arp 261, the residual counts due to contamination is at most 18\% of the measured diffuse flux.
Thus this factor is likely unimportant for our sample.
If such residuals are present in our data, they will contribute to the power law component,
possibly accounting for some of the excess power law flux above that expected from HMXBs and LMXBs.

A related issue is that, 
when removing the point sources from the image, one may inadvertently
leave `holes'
in the map of the diffuse emission, causing the total diffuse light to be 
under-estimated.  
To estimate how important this missing flux is, for each system
we calculated the 
total area covered by the point sources, and compared with the area
used to measure the diffuse emission, assuming that the diffuse gas uniformly covers the measured area.
The galaxy with the largest fractional area covered by point sources was IRAS 17208-0014, with 5.1\%.
The median fraction for the sample was only 0.3\%, thus little flux was missed.

A detailed analysis of the spatial extent of the hot gas and the locations of the point sources
relative to the hot gas is beyond the scope of this paper.   However, we can make 
a more exact estimate of the missed flux by assuming a centrally-peaked
Gaussian distribution for the hot gas rather than a uniform
distribution.  For each point source, we calculated
the positional offset of the source from the center of the galaxy, and scaled relative to the
optical extent of the galaxy as measured by the B band 25 mag~arcsec$^{-2}$ radius.  
The median offset was found to be 0.35 times the $\mu$$_{\rm B}$ = 25 mag~arcsec$^{-2}$ radius.  
At this distance from the center, the expected hot gas flux assuming a Gaussian
is about twice that calculated assuming a uniform surface density, assuming the 
full width tenth maximum intensity of the hot gas is coincident with the 
$\mu$$_{\rm B}$ = 25 mag~arcsec$^{-2}$ isophote.
With these assumptions, in the median only 0.6\% of the diffuse emission is lost due to point source
removal.  Thus this factor is also likely unimportant, and can be neglected.

\section{The X-Ray Luminosity from Hot Gas }

In the rest of this paper, we focus mainly on the X-ray luminosity from the hot gas L$_{\rm X}$(gas), 
which
we assume is equal to the MEKAL component from our best fits to the diffuse X-ray spectrum.
As noted above, in Table  6
we provide ratios of this component to the FIR and K band luminosities
and to the SFR.

\subsection{Comparison with Merger Stage}

The top panel of
Figure 7 shows a plot of L$_{\rm X}$(gas) vs. the interaction stage, where the X-ray luminosity
has only been corrected for Galactic extinction.
In the bottom panel, 
L$_{\rm X}$(gas) after correction for internal absorption is plotted
against interaction stage.
In this plot, and all subsequent plots, the upper limits plotted for
the X-ray luminosities are 3$\sigma$.
In the lower panel of this plot and subsequent
plots involving internal absorption, the stage 7 remnant
Arp 235 is not plotted because it
lacks UV data and is undetected in the MEKAL component.
The black open squares in all panels of
Figure 7 are the data for the individual galaxies; the blue filled diamonds that are offset slightly
to the left of the stage show the median value for that stage.  The errorbars on the blue diamonds show 
the semi-interquartile range, equal to half the difference between
the 75th percentile and the 25th percentile.
Upper limits above the median were not included in calculating the median.
In Figure 7, there is a decrease in the value of L$_{\rm X}$ at stage 7, 
and a possible increase in the middle of the sequence, but there is not a 
strong overall correlation. 

In Figure 8, we normalize L$_{\rm X}$(gas)
by the
stellar mass as traced by the K band luminosity, and plot against merger stage. 
The black open squares are the data for the galaxies, while the blue filled diamonds
are the median values for that stage.
The top panel gives the results assuming only Galactic absorption, while the
bottom plot includes internal absorption.
The L$_{\rm X}$(gas)/L$_{\rm K}$ ratio shows 
a lot of scatter without clear trends.
However, in the middle stages the median
L$_{\rm X}$(gas)/L$_{\rm K}$ ratio is higher than the later stages.
This result is consistent with a picture in which some of the mid-merger systems
have enhanced SFRs, increasing the amount of hot gas relative to the older stellar
population.

In the top panel of 
Figure 9, 
we plot L$_{\rm X}$(gas)/L$_{\rm FIR}$ vs.\ stage, where L$_{\rm X}$(gas) 
has only been corrected
for Galactic extinction.
The black open squares are the data for the galaxies, while the blue filled diamonds
are the median values for that stage.
In the middle panel of Figure 9, 
L$_{\rm X}$(gas)/SFR is plotted vs.\ stage, 
where L$_{\rm X}$(gas) has again only been corrected
for Galactic absorption.
In the bottom panel of Figure 9,
the 
L$_{\rm X}$(gas)/SFR corrected for internal absorption vs.\ merger stage is shown, 
where the X-ray luminosity has been corrected for
internal absorption attenuation.

In Figure 9,
considerable scatter is present in 
these quantities
from system to system for each stage. 
In the top and middle panels of Figure 9, when only Galactic absorption is
included, merger stages 4 and 5 show lower median 
L$_{\rm X}$(gas)/L$_{\rm FIR}$
and 
L$_{\rm X}$(gas)/SFR values.
However, when internal absorption is included, this effect disappears
(bottom panel Figure 9).
After correction for internal absorption,
no overall trends are visible in these plots, 
except that the median value for stage 7 is higher than 
the other stages.
The systems with the highest 
L$_{\rm X}$(gas)/SFR 
ratios
are NGC 1700, NGC 5018, and NGC 2865, all of which
are classified as stage 7.

The AGN do not stand out in Figure 9, except that none of our
stage 6 and stage 7 merger remnants are classified as Seyferts.
In addition, in stages 4 and 5 the AGN have somewhat lower
L$_{\rm X}$(gas)/FIR
and
L$_{\rm X}$(gas)/SFR 
ratios
on average.   This may be because of AGN contributions to powering
the UV and IR fluxes, causing us to slightly over-estimate our
SFRs in some cases.
The system with the 
lowest 
L$_{\rm X}$(gas)/SFR ratio,
the stage 5 merger Mrk 231, 
hosts a powerful
Seyfert 1 nucleus.  This AGN may contribute significantly to powering
the UV and IR fluxes (see Appendix for a detailed discussion
of the SFR in Mrk 231).  This means that our nominal SFR calculation
may over-estimate the SFR, which
will artifically lower the 
L$_{\rm X}$(gas)/SFR ratio.  
The stage 4 mergers Mrk 273 and Arp 220 are other possible examples.

\subsection{Comparison with Normal Spiral Galaxies and Ellipticals}

For the nine normal spiral galaxies in our comparison sample, we use the 0.5 $-$ 2 keV luminosities of the diffuse X-ray emission provided by \citet{mineo12b}.  
They have already subtracted an estimated contribution from unresolved point
sources, 
corrected for Galactic absorption, and fit to a MEKAL plus power law function. 
For consistency with our fluxes, we converted the \citet{mineo12b} MEKAL X-ray fluxes to the 0.3 $-$ 8 keV energy range.
In some cases, they were able to fit for internal absorption using the X-ray spectra, while in other cases, Galactic absorption alone provided a good
fit.   They found gas temperatures of 0.2 $-$ 0.3 keV, with some systems having a second 0.5 $-$ 0.9 keV component.

For their sample of ellipticals and S0 galaxies, \citet{su15} provide 0.1 $-$ 2 keV fluxes for the hot gas, 
after subtraction for contributions from CVs, ABs, and LMXBs, and correction for attenuation by Galactic absorption.  
For consistency with our galaxies, we approximately converted their fluxes into the 0.3 $-$ 8 keV range using PIMMS.
For the hot gas, they used an APEC model instead of a MEKAL function.
For many of their galaxies, \citet{su15} were able to derive a gas temperature from the X-ray spectra; 
their best-fit temperatures range from 0.15 $-$ 1.05 keV, with a median of 0.6 keV.

For their sample of massive elliptical galaxies,
\citet{goulding16} extracted 
measurements of the diffuse X-ray luminosity from the hot gas, after removal of the light from
LMXBs and AB/CVs.   They model the hot gas using an APEC model, and find a median gas temperature of 0.85 keV, with a range from
0.44 $-$ 2.9 keV.   
\citet{goulding16} provide both total hot
gas fluxes and fluxes within the effective radius of the galaxy.  
For the comparison to our mergers, we use the values within the effective radius, thus they are a lower limit to the total
X-ray flux of the galaxy.

In Figure 10, 
we plot  L$_{\rm X}$(gas)/SFR against L$_{\rm X}$(gas)/L$_{\rm K}$
for the mergers, the normal spirals, and the E/S0 galaxies.   
For the mergers and spirals, larger symbols represent data corrected
for internal absorption, while the small dots represent data assuming
only Galactic absorption.
In this Figure, the mergers are color-coded against merger stage.
After correction for internal absorption,
merger stages 1 and 2
are identified by open green triangles,
3, 4, and 5
are plotted as open cyan diamonds, and stages 6 and 7 are marked by
blue open squares.
The same colors are used as small dots for the data corrected only
for Galactic absorption.
Merging galaxies containing a Seyfert nucleus are circled in red.
Black crosses or dots mark the normal spirals from 
\citet{mineo12b}.

The red open squares in Figure 10 represent the \citet{goulding16} massive ellipticals,
while the \citet{su15} E/S0 galaxies are marked by magenta open circles.
The X-ray luminosities for the E/S0 galaxies have only been corrected for 
Galactic
absorption.   
For these E/S0 galaxies, the points marked by the red open squares and the 
magenta open circles
were calculated using SFRs derived from the UV/IR data in the same manner as for the mergers.
Since the E/S0 SFRs are likely
upper limits, due to contributions to the observed UV/IR fluxes 
from older stars, the 
L$_{\rm X}$(gas)/SFR values for the E/S0 galaxies are plotted 
as lower limits.
The magenta upside down filled
triangles mark the \citet{su15} E/S0 galaxies using SFRs from SED fitting.	

Figure 10 shows an apparent
trend for the E/S0 galaxies, in that
L$_{\rm X}$(gas)/SFR appears correlated with L$_{\rm X}$(gas)/L$_{\rm K}$.
This is an artifact, caused by the fact that the formula used for SFR is not valid for
these galaxies.   The 24 $\mu$m flux from most E/S0 galaxies may be dominated by 
circumstellar dust, which produces an approximately constant 
L$_{\rm K}$/L$_{\rm 24}$ \citep{temi09}.  In the formula used for
the SFR, 
the 24 $\mu$m component dominates over the FUV for the majority
of the ellipticals, thus a constant  
L$_{\rm K}$/SFR ratio is produced. 
In Figure 10, the black line 
is the relation expected using the
\citet{temi09} 
L$_{\rm K}$/L$_{\rm 24}$ ratio for an older stellar population and 
the \citet{hao11} SFR formula.
This line thus marks the dividing line between systems with reliable estimates of the SFR
and those that do not; points that lie above this line are lower limits.
Interestingly, when the SED-determined SFRs are used (magenta 
upside-down filled triangles) there is also a trend in this plot,
which argues that they are also unreliable.

All but one of the ellipticals/S0s lie above the black line;
the exception is the SB0 galaxy NGC 1266, which lies in the same
regime as most of the mergers. 
Three of the late-stage mergers (NGC 1700, NGC 5018, and NGC 2865)
also lie above the line, in the same regime as the ellipticals.
A fourth late-stage merger (Arp 222) is close to the line.   
This suggest that the SFR estimates may also be unreliable for these mergers.

In Figure 10, the spirals and the majority of the mergers are clearly separated from
the E/S0 galaxies, having lower L$_{\rm X}$(gas)/SFR values
for their L$_{\rm X}$(gas)/L$_{\rm K}$ ratios.
In the galaxies in the lower right of the plot, the 
hot gas is related to star formation.
It is not related to star formation for the galaxies above the line.
Only one of the E/S0 galaxies, NGC 1266, 
lie in the hot-gas-from-star-formation regime.

\subsection{Diffuse X-Ray Light from Hot Gas vs.\ SFR}

In the top panel of Figure 11, we compare 
L$_{\rm X}$(gas)/L$_{\rm FIR}$ with SFR, where the X-ray luminosity
has only been corrected for Galactic extinction.
The middle panel shows 
L$_{\rm X}$(gas)/SFR vs.\ SFR, where again the X-ray luminosity
has only been corrected for Galactic absorption.
The bottom panel again plots
L$_{\rm X}$(gas)/SFR vs.\ SFR, but this time 
the X-ray luminosity
has been corrected for absorption within the target galaxy.
In Figure 11, we include data for the spirals along with the mergers;
the E/S0 galaxies are not included.

In the top and middle panels of Figure 11,
there are weak anti-correlations, in that 
the galaxies with the highest SFRs
tend to have 
lower L$_{\rm X}$(gas)/L$_{\rm FIR}$ and
L$_{\rm X}$(gas)/SFR ratios.
However, when the diffuse X-ray luminosity is corrected for 
absorption within the galaxy itself,
no strong trend with 
SFR 
is present (bottom), except that a few of the low SFR systems have high
L$_{\rm X}$(gas)/SFR ratios.

To quantify these relationships, we calculated the Spearman rank correlation
coefficient for each plot, excluding galaxies with SFR $<$ 1 M$_{\rm sun}$~yr${-1}$
and only including the galaxies in the merger sample (ignoring the spiral control sample).
For the top, middle, and bottom panels of Figure 11, these correlation coefficients are
-0.74, -0.67, and -0.35, respectively, ignoring upper limits.
The likelihood of the correlations in the top two plots to have occurred by chance
is $<$0.1\% (i.e., 99.9\% confidence for the correlation).
In the bottom panel, however, the likelihood of the correlation occurring by
chance is $>$5\%, meaning that we must reject the hypothesis of a correlation.

The AGN do not stand out in these plots, except that
most of the galaxies classified as Seyferts have moderately high SFRs.
Also, the system with the highest SFR, the Seyfert 1 Mrk 231, has the lowest
L$_{\rm X}$(gas)/SFR ratio.  
As noted earlier, 
the AGN may contribute significantly to powering
the global UV and IR fluxes of this galaxy, causing our SFR value to be over-estimated
and so
L$_{\rm X}$(gas)/SFR artificially low.
Arp 220 and Mrk 273 are other examples of high SFR AGNs with 
moderately low 
L$_{\rm X}$(gas)/SFR ratios (Table 6); as with Mrk 231, this ratio maybe somewhat 
under-estimated if the AGN contributes significantly to powering the UV and IR fluxes.

The results presented in Figure 11 are displayed in a different way in
Figure 12, where we plot 
L$_{\rm X}$(gas) vs.\ SFR using only Galactic absorption (top panel) and internal absorption
(bottom panel).  Above SFR = 1 M$_{\sun}$~yr$^{-1}$, the hot gas luminosity is clearly
correlated with the SFR (Spearman correlation coefficients of 0.82 and 0.88, respectively).
When internal absorption is included and AGNs omitted, the best-fit slope 
is 1.00 $\pm$ 0.10, consistent with a constant  
L$_{\rm X}$(gas)/SFR ratio on average for star-forming galaxies.   
When AGNs are included, the slope is less than
one (0.74 $\pm$ 0.09 when internal absorption is included), perhaps due to an artificial depression of 
L$_{\rm X}$(gas)/SFR at high SFRs because of contributions to powering the UV/IR fluxes by the AGN.

\subsection{L$_{\rm X}$(gas)/SFR vs.\ Stellar Mass }

As noted earlier, the near-IR K band is an approximate
tracer of stellar mass in galaxies, with some possible
contamination from red supergiants, AGB stars, and AGN. 
In Figure 13, we plot 
L$_{\rm X}$(gas)/SFR 
corrected for internal absorption
against
L$_{\rm K}$ for the mergers and the spirals.  The mergers
are color-coded according to merger stage (see caption).  Figure 13
also contains the values for the E/S0 galaxies, corrected only
for 
Galactic absorption (magenta open circles and red open squares,
for the \citet{su15} and \citet{goulding16} samples, respectively).
The L$_{\rm X}$(gas)/SFR ratios for the E/S0 galaxies are plotted
as 
lower limits.

There is no strong correlation between 
L$_{\rm X}$(gas)/SFR and
L$_{\rm K}$ for the mergers in Figure 13.  The Spearman rank correlation
coefficient is only 0.23 ignoring upper limits.
The three stage 7 post-mergers with high diffuse L$_{\rm X}$(gas)/SFR all have high
K band luminosities.  
However, there are other mergers with equally high 
L$_{\rm K}$ which have substantially lower 
L$_{\rm X}$(gas)/SFR values.
Thus high 
L$_{\rm K}$ does not guarantee
high L$_{\rm X}$(gas)/SFR.
The mergers overlap with some of the lower mass \citet{su15}
E/S0 galaxies, although this is uncertain due to the fact that the 
L$_{\rm X}$(gas)/SFR 
values for the 
E/S0 galaxies are lower limits.
Many of the mergers have K band luminosities in the same range
as the \citet{su15} E/S0 galaxies, but only a fraction of the mergers
are as luminous in K as the massive E galaxies from \citet{goulding16}.

However, 
the median L$_{\rm X}$(gas)/SFR for the mergers with L$_{\rm K}$ $<$ 10$^{10}$
L$_{\sun}$ is lower than that for mergers with higher 
L$_{\rm K}$.   
Most of the systems with L$_{\rm K}$ $<$ 10$^{10}$ L$_{\sun}$ are undetected in
the MEKAL component of the diffuse emission (Figure 13) or were detected
at low S/N (Table 4).
However, our sample includes only a handful of galaxies with
L$_{\rm K}$ $<$ 10$^{10}$
L$_{\sun}$, so this result is uncertain.

The Seyfert galaxies do not stand out in Figure 13, except that
they all have moderately high K band luminosities, and the system
with the highest L$_{\rm K}$ by far, Mrk 231, is a Seyfert.
None of the galaxies with low K band luminosity are classified as Seyferts.

\subsection{L$_{\rm X}$(gas)/SFR vs. Stellar Population Age}

In Figures 14 and 15, we plot the 
L$_{\rm X}$(gas)/SFR ratio against two different measures of
the ratio of young-to-old stars in galaxies:
the L$_{\rm FIR}$/L$_{\rm K}$ ratio
(Figure 14)
and the [3.6 $\mu$m magnitude] $-$ [24 $\mu$m magnitude] color
(Figure 15).   
For both plots, for the mergers and the spirals the X-ray fluxes
have been corrected for internal absorption.
For the mergers, the data points are color-coded according
to merger stage.  The spirals are plotted as black crosses.
In Figures 18 and 19, the E/S0 galaxies have also been plotted, with the
understanding that their 
L$_{\rm X}$(gas)/SFR ratios are lower limits.

The near-infrared K and Spitzer 3.6 $\mu$m bands are both approximate
tracers of the older stellar population, while the total
FIR and the 24 $\mu$m bands are approximate tracers of young stars.
This means that 
galaxies with low 
L$_{\rm FIR}$/L$_{\rm K}$ ratios
and small
[3.6] $-$ [24]
in Figures 14 and 15
tend to be 
dominated by older stellar populations, while
galaxies with 
high L$_{\rm FIR}$/L$_{\rm K}$ ratios and 
large [3.6] $-$ [24] have large young stellar populations, relative
to the older stars.

In Figures 14 and 15, systems with large relative amounts of
older stars 
(log L$_{\rm FIR}$/L$_{\rm K}$ $<$ $-$1.2
and [3.6] $-$ [24] $<$ 4.0)
are found to have large 
L$_{\rm X}$(gas)/SFR ratios.  
Four of the late-stage mergers (NGC 1700, NGC 2865, NGC 5018, and Arp 222)
stand out in these plots as having 
high L$_{\rm X}$(gas)/SFR ratios,
low L$_{\rm FIR}$/L$_{\rm K}$ values, and low [3.6] $-$ [24]. 
These merger remnants lie in the same place on these plots
as the ellipticals.
Among the mergers, 
NGC 1700 has the smallest
young/old stellar population and the largest 
L$_{\rm X}$(gas)/SFR ratio.

For systems with 
high 
log L$_{\rm FIR}$/L$_{\rm K}$ 
ratios (younger stellar populations), 
above
log L$_{\rm FIR}$/L$_{\rm K}$ 
$>$ $-$1.2,
there is
no trend of 
L$_{\rm X}$(gas)/SFR with  
L$_{\rm FIR}$/L$_{\rm K}$ (Spearman rank correlation coefficient of -0.24 ignoring
upper limits, corresponding to greater than 5\% likelihood
these trends happened by chance).
For the mergers with 
[3.6] $-$ [24] $>$ 4.0, there is a weak correlation between
L$_{\rm X}$(gas)/SFR and
[3.6] $-$ [24] (Spearman rank correlation coefficient of -0.49 ignoring upper limits,
corresponding to about 1\% likelihood of happening
by chance).
Thus, 
there is a weak trend of enhanced X-ray emission
for younger average stellar populations.

The FUV $-$ NUV color is 
another tracer of average stellar age, although it is also affected by dust.
In the left panel of Figure 16, we compare FUV $-$ NUV against FUV $-$ [24],
another tracers of dust attenuation.
In the right panel, we plot
FUV $-$ NUV vs.\ 
the NUV $-$ [3.6] color.\footnote{For the GALEX FUV and NUV magnitudes, 
we use a zero magnitude 
flux density of 3631 Jy (i.e., the AB magnitude system).
For the Spitzer 3.6 $\mu$m and 24 $\mu$m magnitudes, 
we use zero magnitude flux densities
of
277.5 Jy 
and 7.3 Jy, respectively.  }
The panels in the top row show the mergers, while the panels
in the bottom row mark the spirals and E/S0 galaxies.
The mergers are color-coded into four classes according 
to their 
internal-absorption-corrected
L$_{\rm X}$(gas)/SFR ratios (see caption).

In Figure 16, we have overlaid stellar population
synthesis model colors from the Starburst99 code \citep{leitherer99, vazquez05}.
These are solar metallicity instantaneous burst models with a 
\citet{kroupa02}
initial mass function.  
We have plotted curves of constant age (green solid curve: 3 Myrs;
red dotted line: 20 Myrs; cyan short dashed line: 100 Myrs;
blue long dash: 250 Myrs; magenta dot-dash: 500 Myrs).
In these plots, model extinction increases left to right (towards larger
FUV $-$ [24] and larger NUV $-$ [3.6]), while age increases bottom
to top (i.e., towards increasing FUV $-$ NUV).   
The model FUV $-$ [24] colors in the left panel were
calculated assuming the \citet{hao11} dust attenuation relation.
Although these galaxies likely host a range of stellar ages rather
than a single stellar population, this comparison
provides a rough indication of average stellar age and attenuation in the
galaxies.

The two galaxies with the largest
FUV $-$ NUV colors 
in Figure 16
stand out as 
having the highest 
L$_{\rm X}$(gas)/SFR ratios. 
These are two of the late-stage mergers 
discussed earlier, NGC 5018 and NGC 2865. 
Comparison to the models indicates that these two systems have
luminosity-weighted average stellar ages of $\sim$500 Myrs.
The galaxy with the highest 
L$_{\rm X}$(gas)/SFR in our sample,
NGC 1700, does not appear in Figure 16, due to the lack of
FUV data.  However, based on optical photometry, the stellar population
age
in NGC 1700 is estimated to be 1 $-$ 2 Gyrs \citep{trancho14} or
$\sim$3 Gyrs \citep{brown00}. 

In Figure 16, 
two of the galaxies in the next highest 
L$_{\rm X}$(gas)/SFR range, 
the stage 5 merger Arp 226 and the stage 6 system Arp 222,
also have somewhat elevated FUV $-$ NUV colors,
implying moderately old luminosity-weighted average stellar ages of 100 $-$ 500 Myrs.
For the rest of the mergers, there is no obvious relationship 
between 
L$_{\rm X}$(gas)/SFR and the UV/IR colors on these plots.
For most of the mergers,
the UV/IR colors imply ages between a few Myrs and 
about 250 Myrs, similar to the expected ages for the normal spirals.
The mergers, on average, have higher implied attenuations than the spirals.
The four mergers with the oldest implied ages have UV/IR colors
similar to those of many of the E/S0 galaxies, except with enhanced
24 $\mu$m flux, thus more dust.

These results indicate that the 
L$_{\rm X}$(gas)/SFR ratio remains relative constant during on-going
star formation, however, 
when star formation diminishes the 
hot gas luminosity decreases on a longer timescale, leading to 
an enhanced
L$_{\rm X}$(gas)/SFR ratio.
This point is discussed further in Section 7.5.

\subsection{Hot Gas Temperature vs.\ SFR}

For 15 systems in our sample, our best-fit models included
a fit to the gas temperature (Table 5).  In Figure 17,
for these 15 systems 
we plot these gas temperatures against the SFR, color-coding by
merger stage.  
In interpreting this plot, one must keep in mind that
a fixed
temperature of kT = 0.3 keV gives an acceptable fit for most of the galaxies
in the sample (see Table 4). In other words, 
for most of the galaxies in the sample, we are not able to reliably constrain the gas temperature,
and we cannot rule out 0.3 keV as the true gas temperature.
To illustrate this point,
these galaxies are included on the plot as small black dots at kT = 0.3 keV.
We have also included the comparison elliptical and spiral samples in Figure 17.
For the spirals and ellipticals, we used the published temperatures; we did not extract
and re-fit the X-ray spectra ourselves.  This introduces additional scatter in the plot because
of different assumptions used in the spectral fitting.

No trend between kT and SFR is seen in Figure 17.  Among the mergers, the temperature is not
strongly dependent upon merger stage or AGN activity, however, this result
is uncertain due to the small number of systems with reliable temperatures.
For a sample of star forming galaxies with moderate SFRs ($\le$ 20 M$_{\sun}$~yr$^{-1}$),
\citet{mineo12b} also did not see any correlation between
hot gas temperature and SFR.
However, \citet{grimes05}
found a tendency for ULIRGs to have higher temperatures.
To better determine the relationship between gas temperature and SFR
for mergers, 
higher S/N X-ray spectra are needed to constrain the temperatures 
for more galaxies.

As shown in Figure 17 and as noted previously 
(e.g., \citealp{goulding16}), elliptical
galaxies tend to have higher hot gas temperatures than spiral galaxies.
In ellipticals, the gas temperatures are thought to be related to the 
depth of the gravitational potential (and therefore the mass of the galaxy),
rather than to star formation processes \citep{goulding16}.   
See Section 7.5 for more
discussion about the hot gas in elliptical galaxies.

\subsection{Large-Scale Environment}

In addition to exploring possible correlations of
L$_{X}$(gas)/SFR with merger stage, we also looked
for correlations with environment on a more global scale.
Using redshifts from the $2M++$ redshift compilation (a 2MASS-based
redshift survey),
\citet{carrick15}
presented a reconstruction of
the density of galaxies within $200$h$^{-1}$ Mpc,
a range which includes our entire sample.  They
have made available a grid of luminosity weighted
density contrast $\delta_g$ smoothed with a Gaussian
of scale Mpc$~h^{-1}$ versus location on a grid
with spacing $1.5625$~Mpc~$h^{-1}$ centered on
the Local Group.  We have associated
the location of
each of our galaxy pairs with the interpolated
luminosity density contrast at that point.
Figure 18
shows a scatter plot of L$_{\rm X}$(gas)/SFR
vs.\ $\log (1 + \delta_g)$.
No obvious correlation is present 
(Spearman
correlation coefficent = 0.07 when AGN and upper limits are excluded).
This
argues against a scenario in which 
the diffuse X-ray emission
from our sample galaxies is dominated 
by intragroup or intracluster gas.
However, 
the variation in L$_{X}$(gas)/SFR
seems to be larger for interacting galaxies in overdense regions,
thus 
it is possible that intragroup or intracluster gas may contribute
to some extent for some of the sample galaxies.
Two of the late-stage mergers with high 
L$_{\rm }$(gas)/SFR ratios, NGC 2865 and NGC 5018, are in dense
regions, but the highest ratio galaxy, NGC 1700, is at average
density. 
NGC 2865 and NGC 5018 have been identified 
by \citet{kourkchi17}
as members of groups.

\subsection{Major Mergers vs.\ Minor Mergers vs.\ Multiple Mergers}

We do not see a trend in the 
L$_{\rm X}$(gas)/SFR ratio with either merger stage or SFR, however, there
is considerable scatter from system to system.
It is possible that some of this scatter is caused by variations in
the mass ratio of the galaxies in the pair.  
Some of the systems in our
sample may be the product of a minor (unequal mass) merger, rather
than a major merger.  Alternatively, they may be the product of
the merger of more than two galaxies, or of a disk galaxy with a spheroid.
After a merger has occurred, it is often difficult to determine the mass
and Hubble type of the progenitor
galaxies.   
In the Appendix to this paper, we describe each system
in detail, and provide a summary of the discussion in the literature 
about the evolutionary history of the system.  

Out of our 49 sample systems, there are 19 for 
which there is some debate in the literature about their classification
as the product of
a merger of two approximately equal mass gas-rich disk galaxies (see Appendix).
To test whether such mis-classification 
contributes to the scatter in the 
L$_{\rm X}$(gas)/SFR ratio, in 
Figure 19, we reproduce our earlier plots of 
L$_{\rm X}$(gas)/SFR vs.\ merger stage and 
L$_{\rm X}$(gas)/SFR vs.\ SFR, but this time identifying the 
19 controversial systems using green squares.
In both panels, the X-ray luminosity has been corrected for internal absorption.

As expected, most of the controversial systems are later stage mergers,
however, even for some of the earlier stage systems 
there is some debate.  However, even when these 
controversial systems are eliminated
from the sample there is still some scatter in the L$_{\rm X}$(gas)/SFR.
This shows that such 
mis-classification is likely not responsible for the observed
scatter.   With the exception of the three stage 7 systems with high 
L$_{\rm X}$(gas)/SFR 
the controversial systems show the same range in
L$_{\rm X}$(gas)/SFR as the other systems.

\section{Discussion}

We have used archival Chandra data to study the
hot gas emission from 49 major mergers
spanning the full merger sequence. 
After removal of the point sources, we have fit the spectrum
of the diffuse X-ray light to two components: a soft
thermal component and a hard power law component.

\subsection{The Diffuse Power Law Component}

The power law component of the diffuse emission can generally be accounted for by 
unresolved HMXBs and LMXBs (Figure 4).   At high SFRs, HMXBs dominate this
component, while at low SFRs LMXBs become more important (Figure 5).
Summing up their expected contributions and comparing to the
observed power law, we find good agreement on average but
with some scatter (Figure 6).
Variations in the HMXB-SFR relation from system to system
may contribute to this scatter.  These variations may be due in part to
different timescales for the formation of HMXBs compared to
the stars responsible for the UV and IR fluxes used to calculate the SFR.
HMXBs typically have ages of 20 $-$ 70 Mpc \citep{antoniou10, williams13},
while the FUV and 24 $\mu$m light is sensitive to stars
with ages from 0 $-$ 100 Myrs \citep{kennicutt12}.
Another factor that may contributes to the scatter is variations
in the LMXB-L$_{\rm K}$ ratio with stellar population age.   
Incompletely-removed
point sources, especially AGN, may also contribute to the scatter.

\subsection{The Relative Constancy of L$_{\rm X}$(gas)/SFR in
Star Forming Galaxies}

As noted above, the thermal component of the diffuse X-ray emission is likely
dominated by radiation from hot gas.
Excluding galaxies
with low SFRs and correcting for internal absorption,
we find no trend between
L$_{\rm X}$(gas)/SFR 
and merger stage, SFR,
AGN activity, or large-scale environment (see Figures 9, 11, and 18).
The lack of a correlation
between L$_{\rm X}$(gas)/SFR and the SFR for star forming systems
is in agreement with previous 
smaller surveys \citep{grimes05, mineo12b}.
These results indicate that the main power source for the hot gas is
star formation.  
The lack of a trend with AGN activity rules out 
feedback from AGNs as the main
source of 
hot gas in our sample galaxies (see Section 7.7 below).
The relatively constant 
L$_{\rm X}$(gas)/SFR ratio and 
the lack of a trend between
L$_{\rm X}$(gas)/SFR and merger stage
also argue against models in
which the hot gas in mergers is primarily
due to shocks from the direct collision of gaseous disks or
gaseous halos
 (e.g., \citealp{cox06b, sinha09}). 
This topic is discussed further below in Section 7.8. 

The relative constancy of 
L$_{\rm X}$(gas)/SFR 
suggest that stellar feedback in star forming
galaxies reaches a quasi-steady state condition on average, with a
constant fraction of the total energy output from supernovae
and stellar winds being converted into X-ray flux.
This simple state of affairs 
disagrees strongly 
with the 
classical \citet{chevalier85} 
theoretical model of hot gas production in starbursts, which
predicts 
L$_{\rm X}$(gas) $\propto$
SFR$^{2}$ \citep{zhang14}.
More recent theoretical models 
of galactic winds including gravitational forces
and more sophisticated radiative cooling calculations 
\citep{bustard16, meiksin16} can reproduce
the observed L$_{\rm X}$(gas) $\propto$ SFR relation,
if 
the mass loading factor (mass outflow
rate/SFR) decreases with increasing SFR.
Observationally, there is evidence from HST UV spectroscopy
that the mass loading factor of galactic winds does decrease
with increasing SFR \citep{chisholm15}.
Both hydrodynamical simulations \citep{muratov15} and 
analytical calculations \citep{hayward17} find that
the mass loading factor of winds should decrease 
as SFR increases.

In an alternative theoretical
treatment, \citet{sarkar16} conclude that at SFRs greater than 
1 M$_{\sun}$~yr$^{-1}$, 
L$_{\rm X}$(gas) $\propto$ SFR$^2$, but at low SFR, 
a hot circumgalactic medium contributes, artificially producing
an overall 
L$_{\rm X}$(gas) $\propto$ SFR relation.
This model is ruled out by our data, however;
even excluding systems with 
SFR $<$ 1 M$_{\sun}$~yr$^{-1}$ 
we do not see a trend of 
L$_{\rm X}$(gas)/SFR with SFR (Figure 11).   

The fraction of the total
mechanical energy of the starburst
converted into diffuse thermal X-ray light is sometimes 
labeled
the X-ray production efficiency. 
In simplistic starburst models in which the X-ray production efficiency
is assumed to be constant,
for an instantaneous burst
the X-ray luminosity from hot gas is expected to rise dramatically
about 3 Myrs after the burst when Type II supernovae
start occurring, and drop off steeply
at an age of about 40 Myrs when SNII activity
decreases
(e.g., \citealp{leitherer99, oskinova05}).
In contrast, if star formation continues at a
constant rate and the X-ray production efficiency is constant, 
L$_{\rm X}$(gas) remains approximately constant 
relative to the absorption-corrected H$\alpha$ 
luminosity (see Figure 12 in \citealp{smith05}).

In realistic models of major mergers, starbursts 
are not instantaneous, but instead prolonged for extended periods
($\ge$100 Myrs; \citealp{dimatteo08, lotz08, bournaud11, fensch17}).
Furthermore,
the hot gas lingers for an extended period.
Using standard
cooling functions \citep{mckee77, mccray87},
we estimated approximate cooling times for the hot gas, based on 
estimates of the angular extent of the diffuse X-ray emission.
These times range from
about 10 to 500 Myrs 
with a median value 
of about 100 Myrs,
similar to the \citet{mineo12b} estimates for disk galaxies.
This cooling timescale is similar to the timescale of
the expected variations in the SFRs.
When the hot gas in a galaxy cools on a timescale approximately
equal to the timescale for the star formation, and/or is
expelled on this timescale, 
the hot gas content will remain 
proportional to the current SFR, not to the integrated 
star formation history of the galaxy. 
In major mergers, star formation occurs in different
locations within a system at different times, with
at least some star formation happening
somewhere in the system throughout most of the merger process.
Another contributing factor
is the fact that the strongest star formation occurs
at different times in different mergers.
The timing of a starburst relative to morphological merger
stage depends upon the parameters of the encounter, with some
mergers having star formation in earlier stages than others 
\citep{dimatteo07, dimatteo08, lotz08}.
These factors conspire to produce no strong trend 
in the absorption-corrected L$_{\rm X}$(gas)/SFR ratio with merger stage.

The similar extended timescales for gas cooling and star formation bursts
means that to first approximation
we can estimate the X-ray production
efficiency by comparing with population synthesis models
assuming continuous star formation.
For galaxies with SFRs greater than 1 M$_{\sun}$, 
the median 
absorption-corrected L$_{\rm X}$(gas)/SFR ratio
for our sample galaxies
is 5.5 $\times$ 10$^{39}$ ((erg~s$^{-1}$)/(M$_{\sun}$~yr$^{-1}$)).
Comparing to 
Starburst99 population synthesis models
with continuous star formation
(e.g., see Figure 12
in \citealp{smith05}), 
the X-ray production efficiency for our sample galaxies is about 2\%.  
That is, about 2\% of the total mechanical energy production due to star formation
(mostly from SNe)
is converted into X-rays.    The system-to-system variation in the 
inferred efficiency is about a factor of two
(see Section 7.8).
Assuming that the total thermal luminosity of the hot gas at all wavelengths is about twice the
luminosity in the 0.3 $-$ 8 keV range \citep{mineo12b}, this implies that about 4\%
of the total mechanical energy from supernovae is thermalized.

\subsection{Comparison to Previous Studies}

For a sample of seven dwarf starbursts, six edge-on starbursts,
and 9 ULIRGs,
\citet{grimes05} found an approximately
constant L$_{\rm X}$(gas)/L$_{\rm FIR}$ ratio
of $\sim$10$^{-4}$.  This is a factor of three times
lower than our median 
L$_{\rm X}$(gas)/L$_{\rm FIR}$ value of 3.0 $\times$ 10$^{-4}$.
For a sample of 21 spiral and irregular galaxies,
\citet{mineo12b} 
find an approximately constant 
L$_{\rm X}$(gas)/SFR after correction for internal absorption
of 7.3 $\times$ 10$^{39}$ ((erg~s$^{-1}$)/(M$_{\sun}$~yr$^{-1}$)).
This value is 30\% higher than our ratio.

Both the \citet{grimes05} and \citet{mineo12b} samples 
have lower stellar masses on average than ours, which may bring down the 
L$_{\rm X}$(gas)/SFR 
ratios 
(see Section 7.6).   
\citet{mineo12b} use a different prescription for their SFRs 
(see \citealp{mineo12a}).  However, when we obtained 
total Spitzer and GALEX
fluxes for their sample galaxies from NED and re-calculated the
SFRs using the \citet{hao11} formula,  we find
little difference in the median, although there is a scatter of 
about 0.15 dex in the ratio of the two SFR calculations.

Our X-ray production 
efficiency of $\sim$2\% is twice that of 
\citet{grimes05}, who quote $\sim$1\%. 
\citet{mineo12b} give an efficiency of 5\%, however, this refers
to the total thermal luminosity from the hot gas at all wavelengths
rather than just that measured within
the 0.3 $-$ 8 keV band that we use.  
Since they use a 
correction factor of 2,
we find good agreement with their results.

\subsection{Low Observed L$_{\rm X}$(gas) at high L$_{\rm FIR}$}

When the X-ray luminosities are not corrected
for internal absorption, 
we see a depression in the median L$_{\rm X}$(gas)/L$_{\rm FIR}$  
and L$_{\rm X}$(gas)/SFR
in the middle of the merger sequence (merger stages 4 and 5), 
compared to other stages
(top and middle panels, Figure 9). 
For stages 4 $-$ 5, with just Galactic absorption the median 
L$_{\rm X}$(gas)/L$_{\rm FIR}$  
is 2.5 $\times$ 10$^{-5}$, 
while for stages 1 $-$ 3 plus stage 6
the median is 1.1 $\times$ 10$^{-4}$, a factor of four times higher.
We exclude stage 7 systems from this comparison, since some stage
7 systems have noticeably higher 
L$_{\rm X}$(gas)/SFR ratios (see Section 7.5).
For stages 4 $-$ 5, with just Galactic absorption the median 
L$_{\rm X}$(gas)/SFR  
is 7.9 $\times$ 10$^{38}$ ((erg~s$^{-1}$)/(M$_{\sun}$~yr$^{-1}$)),
while for stages 1 $-$ 3 plus stage 6
this ratio is 1.6 $\times$ 10$^{39}$ ((erg~s$^{-1}$)/(M$_{\sun}$~yr$^{-1}$)),
twice as large.
Including internal absorption the median 
L$_{\rm X}$(gas)/SFR is 
3.6 $\times$ 10$^{39}$ ((erg~s$^{-1}$)/(M$_{\sun}$~yr$^{-1}$))
for stages 4 $-$ 5, while
stages 1 $-$ 3 plus 6 have a median 1.8 times larger, of 
6.5 $\times$ 10$^{39}$ ((erg~s$^{-1}$)/(M$_{\sun}$~yr$^{-1}$)).
Galaxies in mid-merger tend to have more obscured
star formation.
Correcting for internal absorption and using the SFR rather
than L$_{\rm FIR}$ decreases the drop in the middle
of the sequence (bottom panel, Figure 9).
Given the factor of approximately two in the uncertainties in the absorption-corrected luminosities 
(Section 5.3)
and taking into account possible contributions from AGN to the observed UV and IR fluxes,
we conclude that there is no strong trend between
L$_{\rm X}$(gas)/SFR and merger stage in our sample galaxies.

The systems with the lowest 
Galactic-absorption-corrected 
L$_{\rm X}$(gas)/L$_{\rm FIR}$ 
and 
L$_{\rm X}$(gas)/SFR
ratios
tend to be higher SFR systems.  
The highest SFR systems tend to be in the middle of the merger sequence
(Figure 3).
When only Galactic
absorption is included, 
systems with the highest SFRs appear to be deficient in diffuse
thermal 
X-ray luminosity compared to other galaxies
(top and middle panels, Figure 11).
However, when internal absorption is included,
this discrepancy almost disappears (bottom panel, Figure 11), particularly
if the highest SFR system, Mrk 231, is ignored due to possible contamination
of the UV/IR fluxes by the Seyfert 1 nucleus, and possible AGN contributions
to the high SFR systems Arp 220 and Mrk 273 are taken into consideration.
With only Galactic absorption 
the median L$_{\rm X}$(gas)/L$_{\rm FIR}$ is 3.6 $\times$ 10$^{-5}$
for systems with SFR $>$ 40 M$_{\sun}$~yr$^{-1}$, 
and  
1.1 $\times$ 10$^{-4}$ for 
systems with 1 M$_{\sun}$~yr$^{-1}$ $<$ SFR $<$ 40 M$_{\sun}$~yr$^{-1}$. 
With only Galactic absorption 
the median L$_{\rm X}$(gas)/SFR is 5.4 $\times$ 10$^{38}$
((erg~s$^{-1}$)/(M$_{\sun}$~yr$^{-1}$))
for systems with SFR $>$ 40 M$_{\sun}$~yr$^{-1}$, 
and 
1.5 $\times$ 10$^{39}$
((erg~s$^{-1}$)/(M$_{\sun}$~yr$^{-1}$))
for systems with 1 M$_{\sun}$~yr$^{-1}$ $<$ SFR $<$ 40 M$_{\sun}$~yr$^{-1}$.
Both L$_{\rm X}$(gas)/L$_{\rm FIR}$ and 
L$_{\rm X}$(gas)/SFR are down by a factor of three for high SFR systems
compared to lower SFRs if only Galactic absorption is included.
When internal absorption is included the median
L$_{\rm X}$(gas)/SFR is 3.5 $\times$ 10$^{39}$
((erg~s$^{-1}$)/(M$_{\sun}$~yr$^{-1}$))
for systems with SFR $>$ 40 M$_{\sun}$~yr$^{-1}$
and 1.9 times larger,  
6.5 $\times$ 10$^{39}$
((erg~s$^{-1}$)/(M$_{\sun}$~yr$^{-1}$)),
for systems with 1 M$_{\sun}$~yr$^{-1}$ $<$ SFR $<$ 40 M$_{\sun}$~yr$^{-1}$.
Galaxies with very high SFR and L$_{\rm FIR}$ tend
to have more obscured star formation and are more likely to host AGNs;
when the X-ray luminosity is corrected for internal absorption and AGN contributions
are taken into account,
their L$_{\rm X}$(gas)/SFR ratios generally agree with systems with lower SFRs,
given the uncertainties on the absorption-corrected 
L$_{\rm X}$(gas) values.
Thus we conclude there is no strong correlation 
between L$_{\rm X}$(gas)/SFR and SFR for galaxies with 
SFRs $>$ 1 M$_{\sun}$~yr$^{-1}$.

A deficiency in the global hard (2 $-$ 10 keV) X-ray emission 
from Luminous Infrared Galaxies (LIRGs) and ULIRGs 
relative to L$_{\rm FIR}$ or SFR has been previously noted 
\citep{persic07, iwasawa09, lehmer10}.
In these 
studies, 
the X-ray emission was only corrected for Galactic absorption.
\cite{persic07} suggest that the 2 $-$ 10 keV
L$_{\rm X}$/L$_{\rm FIR}$ is higher
in lower SFR systems since LMXBs contribute substantially
to the total X-ray light, while in higher SFR systems HMXBs dominate.
Alternatively, 
\citet{iwasawa09} suggest that the deficiency at high L$_{\rm FIR}$
may be because 
highly obscured AGN are powering
the FIR luminosity.
A third opinion was given by 
\citet{lehmer10}, who 
argue that the X-ray deficiency at high SFRs is likely due to absorption.

Our current study shows that, after point sources are
removed and the thermal and power law contributions
separated, there is an apparent deficiency in the diffuse
thermal L$_{\rm X}$ from hot gas relative to L$_{\rm FIR}$ and SFR.
We attribute this deficiency
to higher absorption in the most powerful starbursts. 
Interestingly, the number of luminous X-ray point sources relative 
to the far-infrared and SFR also shows a deficiency in LIRGs and ULIRGs 
\citep{smith12, luangtip15}, a result attributed to absorption.
Thus both the extended and 
point source X-ray light, if uncorrected for internal absorption, are lower relative to the SFR in 
LIRGs and ULIRGs. 

Using Chandra data, \citet{brassington07}
found a dip in the diffuse
L$_{\rm X}$/L$_{\rm FIR}$ in the middle of their merger
sequence, near nuclear coalescence. 
They concluded that this drop was caused by the break-out of a galactic
wind.
However, the dip was caused by a single system, Arp 220, 
for which they found a 
diffuse L$_{\rm X}$/L$_{\rm FIR}$ (only correcting for
Galactic absorption) that is a factor of 10 times lower than
for the other systems in their sequence.   
We also 
see a 
low value of Galactic-absorption-corrected
diffuse L$_{\rm X}$/L$_{\rm FIR}$ for Arp 220, 
however, once corrected for internal absorption Arp 220 does not 
stand out relative to the other galaxies in our sample (see Table 6).

Our conclusion that there is no strong trend of 
L$_{\rm X}$(gas)/SFR with SFR is consistent with earlier studies.
In their survey of the hot gas in 22 star forming galaxies,
\citet{grimes05} saw no correlation between the internal-absorption-corrected
L$_{\rm X}$(gas)/L$_{\rm FIR}$ and 
FIR luminosity.  
\citet{mineo12b} also did not see a trend with SFR in their sample.

\subsection{Hot Gas in Post-Starburst Late-Stage Remnants}

In our plot of absorption-corrected
L$_{\rm X}$(gas)/SFR vs. merger stage (Figure 9), 
three low SFR systems in the last
merger stage stand out as having 
higher median 
L$_{\rm X}$(gas)/SFR ratios: 
NGC 1700, NGC 2865, 
and NGC 5018.    
Compared to the median value for the other systems,
the absorption-corrected 
L$_{\rm X}$(gas)/SFR ratio for these three galaxies are
enhanced by a factor of 8 $-$ 47.
As discussed at length in Section 6.5 and in the Appendix,
compared to the other galaxies 
these three systems 
have older stellar populations (500 Myrs $-$ 3 Gyrs, according to 
UV/IR/optical data).

The three late-stage galaxies with enhanced L$_{\rm X}$(gas)/SFR ratios
have lower L$_{\rm FIR}$/L$_{\rm K}$
ratios
and lower [3.6] $-$ [24] colors than the rest of the mergers,
but they are not as extreme in these colors as the majority of the
ellipticals in our comparison samples
(Figures 14 and 15).
This suggests that they are transition objects.
Furthermore, although NGC 2865 and NGC 5018 are as red in FUV $-$ NUV
as some of the ellipticals (implying an older population), 
they are redder in FUV $-$ [24]
and NUV $-$ [3.6] than the ellipticals (implying more extinction)
(Figure 16).
NGC 1700 (which lacks FUV data)
has NUV $-$ [3.6] = 8.55 $\pm$ 0.06, also redder than a typical elliptical.
This also supports the idea that these three galaxies 
are in a transition phase.

The observed increase in L$_{\rm X}$(gas)
in these three systems relative to their SFRs
is less than that expected from 
an aging stellar population 
assuming a constant X-ray production efficiency
and a single instantaneous burst.
By 20 Myrs after an instantaneous burst ends, 
the 
ratio of the 
total mechanical luminosity 
relative to the number of hydrogen-ionizing photons 
is expected to be about a thousand
times
larger than that produced by steady-state star formation
(see Figure 12 in \citealp{smith05}).
This implies that either our SFRs are over-estimates
due to contributions to the UV/IR fluxes from older stars,
or star formation did not completely stop in these systems.
For example, if the SFR in a galaxy with continuous star formation
abruptly dropped to a level 50 times lower than its original
rate, then remained constant at the new level, there would be a time
lag equal to the gas cooling time plus the lifetime of the
most massive stars before the diffuse X-ray luminosity
dropped to match the new SFR.   During this transition period,
L$_{\rm X}$(gas)/SFR will be enhanced by a factor of 50 relative
to the original steady-state value.   
By this same argument, a star formation rate that declines
gradually with time could also lead to an increase 
in the observed L$_{\rm X}$(gas)/SFR.

In classical models of hot gas production in elliptical galaxies,
an early starburst is assumed to quickly use up or clear out
most of the gas, leaving the galaxy deficient in gas until a hot
halo is rebuilt by virialization of mass loss 
from older stars 
and type Ia supernovae
\citep{ciotti91, pellegrini98}.
This rebuilding is expected to be a slow process, with the halo regeneration
taking many Gyrs.
In the \citet{ciotti91} standard model, from its initial
starburst level L$_{\rm X}$(gas)
drops a factor of ten within $\sim$1 Gyrs,
then
drops another factor of 10 to 
a minimum at an age of about 5 Gyrs, then steadily increases.
To study the rebuilding process, 
using ROSAT data 
\citet{osullivan01b}
compared the global L$_{\rm X}$/L$_{\rm B}$ ratios of ellipticals 
with stellar population age.  They found a trend of increasing
L$_{\rm X}$/L$_{\rm B}$ with increasing age.  They concluded
this trend was likely caused by type Ia supernovae and other stellar 
mass loss.
Their relation, however, shows a lot of scatter and includes
many upper limits on the X-ray fluxes, thus is very uncertain.

For a sample of nearby non-cD early-type
galaxies,
\citet{boroson11} used sensitive Chandra images to
remove the bright point sources, and fit the spectrum of the diffuse
emission to multiple components to extract the hot gas
content.   
For the subset of their early-type galaxies
with stellar velocity dispersions $\sigma$$_{*}$
between 170 km~s$^{-1}$ and 
260 km~s$^{-1}$, in Figure 19 we plot the
average stellar age against L$_{\rm X}$(gas) (which is only
corrected for Galactic absorption).
In Figure 19, we color-code the
\citet{boroson11}
galaxies according to $\sigma$$_*$ and gas temperature
(see caption for the ranges).
We have added our three post-starburst systems to Figure 19,
color-coded in the same way but 
also surrounded by large magenta circles\footnote{NGC 
1700, NGC 2865, and NGC 5018
have $\sigma$$_{*}$ of 238 km~s$^{-1}$, 184 km~s$^{-1}$,
and 191 km~s$^{-1}$, respectively \citep{bender94, koprolin00, dressler91},
while their 
stellar
ages have been estimated to be between 1 $-$ 3 Gyrs, 1 $-$ 2 Gyrs, and 3 Gyrs, 
respectively (see Appendix).
The gas temperatures for NGC 1700, NGC 2865, and NGC
5018 have been determined to be 0.47 $\pm$ 0.03 keV
\citep{statler02}, 0.32 $\pm$ $^{0.10}_{0.40}$ keV \citep{fukazawa06},
and 0.41 $\pm$ 0.04 keV \citep{ghosh05}, respectively.}.

As shown in Figure 19 and 
as previously noted by \citet{boroson11}, systems with higher
gas temperatures tend to have higher L$_{\rm X}$(gas), and 
systems with higher dynamical masses (as indicated by $\sigma$$_*$)
also tend to have higher 
L$_{\rm X}$(gas); however, there is a lot of scatter.
As previously pointed out by \citet{boroson11}, 
for the subset of seven ellipticals with 
moderate 
$\sigma$$_*$ (between 202 and 232 km~s$^{-1}$)
and moderate gas temperatures (between 0.32 $-$ 0.36 keV),
there is a trend of increasing L$_{\rm X}$(gas)
with increasing stellar age (blue filled circles in Figure 19).
However, systems with low dynamical masses 
($\sigma$$_*$ between 
170 km~s$^{-1}$ and 200 km~s$^{-1}$)
and moderate to low
gas temperatures 
(kT $\le$ 0.38 keV) 
show a decrease in 
L$_{\rm X}$(gas) with increasing age (red filled squares), although
there are only four galaxies in that subset.
The other subsets show approximately constant 
L$_{\rm X}$(gas) values with age, but only contain
a few galaxies each.
Given the small number of systems in each
subclass, it is difficult to make any strong conclusions  
about the evolution of hot halos in post-mergers from these data.  
We note that 
our three post-starburst systems have L$_{\rm X}$(gas) similar to
systems in the \citet{boroson11} sample with similar velocity
dispersions, gas temperatures, and ages, thus they are not unusual
compared to similar systems.
Observations of additional post-starburst, post-merger
systems would be useful to search for
definitive evolutionary trends with age.

For their sample of early-type
galaxies from the ATLAS$^{\rm 3D}$ survey, \citet{su15} 
calculated the ratio L$_{\rm X}$(gas)/L$_{\rm K}$ and compared
with stellar age.  They found a rough trend, in that galaxies with
older stellar populations 
have higher  
L$_{\rm X}$(gas)/L$_{\rm K}$.  However, there is a very large
amount of scatter in this relation 
(up to a factor of 1000 in the 
L$_{\rm X}$(gas)/L$_{\rm K}$
ratio at a given age).
Furthermore,
their sample only includes a handful of systems with young ages, thus
the evolution of the hot gas 
in the early post-starburst period remains uncertain.

\subsection{Low Mass Galaxies}

As discussed in Section 6.4, 
the galaxies in our sample
with the lowest K band luminosities appear to be deficient
in hot gas relative to the SFR, compared to higher L$_{\rm K}$
galaxies (Figure 13).
Above L$_{\rm K}$ = 10$^{10}$ L$_{\sun}$ no obvious trend is seen.
The absorption-corrected median L$_{\rm X}$(gas)/SFR 
for the systems with L$_{\rm K}$ $<$ 10$^{10}$
L$_{\sun}$ 
is $<$2.3 $\times$ 10$^{39}$ 
(erg~s$^{-1}$)/(M$_{\sun}$~yr$^{-1}$)), while 
systems with 
10$^{10}$ L$_{\sun}$ $<$ L$_{\rm K}$ $<$ 10$^{11}$
L$_{\sun}$ have a median value of 
5.0 $\times$ 10$^{39}$ 
(erg~s$^{-1}$)/(M$_{\sun}$~yr$^{-1}$), and 
systems with 
L$_{\rm K}$ $>$ 10$^{11}$
L$_{\sun}$ have a median of 
6.0 $\times$ 10$^{39}$ 
(erg~s$^{-1}$)/(M$_{\sun}$~yr$^{-1}$)).  
However, our sample only includes a handful of galaxies 
with L$_{\rm K}$ $<$ 10$^{10}$ L$_{\sun}$, so this result is uncertain.

A complicating factor is that 
our sample includes a range of morphological types,
from disk systems to spheroidals, and a range of stellar population ages.
However, inspection of Figure 13 shows that when the merger sample
is divided into three subsets: pre-mergers
(merger stages 1 and 2; green open triangles), 
mid-mergers (merger stages 3, 4, and 5; cyan open diamonds), and 
merger remnants (stages 6 and 7; blue open squares), each subset
shows
lower L$_{\rm X}$(gas)/SFR ratio for low 
L$_{\rm K}$.  
This suggests that morphology is not the only factor, and 
the dynamical mass plays a role.
Each subset, however, only contains a few
galaxies with L$_{\rm K}$ $<$ 10$^{10}$ L$_{\sun}$.  
It would be useful to increase the numbers of galaxies
in each of these classes to get better statistics on the amount of
hot gas as a function of stellar mass in various types of systems.

Lower L$_{\rm X}$(gas)/SFR for galaxies with low L$_{\rm K}$ 
may caused by increased gas loss via winds 
due to weaker gravitational
fields, if galaxies with lower stellar masses also have lower total masses.
From an HST UV spectroscopic study of the wind velocities of starburst galaxies,
\citet{chisholm15} conclude that galaxies with stellar masses greater than
about 3 $\times$ 10$^{10}$ M$_{\sun}$ do not lose interstellar gas via winds, but at lower
masses some gas is removed.   This stellar mass cut-off
for gas lost through winds
is consistent with 
expectations based on binding energy arguments \citep{dekel03}.
Using a typical stellar-mass-to-L$_{\rm K}$ ratio of 1 M$_{\sun}$~L$_{\sun}^{-1}$
(e.g., \citealp{bell03}), this mass limit corresponds to L$_{\rm K}$
$\sim$ 3 $\times$ 10$^{10}$ L$_{\sun}$.  Given the uncertainties, this value
is reasonably consistent
with the L$_{\rm K}$ limit at which we see a 
drop-off in the L$_{\rm X}$(gas)/SFR ratio.
Alternatively, the apparent deficiency in L$_{\rm X}$(gas) at low L$_{\rm K}$
may be due to lower metallicities, as the observed
X-ray luminosity from hot gas increases for increasing
metallicity (e.g., \citealp{silich01}).

As noted in Section 7.3, the \citet{grimes05} 
L$_{\rm X}$(gas)/L$_{\rm FIR}$ is lower on average than ours.
This may be because their sample has lower L$_{\rm K}$ on average.
Half of their galaxies have L$_{\rm K}$ $<$ 10$^{10}$ L$_{\sun}$,
compared to only 8 out of 49 in our sample.
However, 
the \citet{mineo12b} sample also has about half with L$_{\rm K}$
$<$ 10$^{10}$ L$_{\sun}$, but their 
best-fit
L$_{\rm X}$(gas)/SFR ratio is slightly higher than ours.

\subsection{The Importance of AGN Activity}

In none of our plots do we see a systematic difference 
between the properties
of the AGNs and the other galaxies.
The lack of an observed trend with AGN activity
implies that winds and jets from AGN are less important
energetically in major mergers than star formation,
at least as far as the diffuse X-ray-emitting gas is concerned.
Perhaps jets and winds from AGN are
sufficiently 
smothered 
during most major mergers that they do not create
large volumes of hot gas within the galaxies.
Alternatively, perhaps powerful jets and winds from AGN
can be very disruptive of the interstellar medium,
but only for short timescales and only rarely.
In a sample of only 49 systems, maybe we do not have
a large enough subset with AGNs in an outburst phase
for us to detect a trend.

We note that most of the AGNs in our sample are in the middle
merger stages (stages 3, 4, and 5).   This may be in part a selection
effect, since this is an archive-selected sample.  
Galaxies in those stages tend to be more distant (Figure 2)
and more luminous in both the FIR and the K band (Figure 3).

We also note that we excluded radio galaxies from our sample.
This may cause a bias, if 
a radio galaxy phase is responsible for producing 
the hot gas in ellipticals.
Once the cold gas is depleted in a merger and the SFR ends, 
maybe a large radio jet
can be produced (i.e., `smothering' of the
jet by the interstellar medium no longer occurs).  
Such jet activity may be periodic, 
keeping the interstellar gas in ellipticals hot.
In the \citet{ciotti17} hydrodynamical models of AGN feedback
in early-type galaxies, the AGN outbursts are predicted to last
about 30 Myrs, and have a duty cycle of 4\%. 

\subsection{Shocks}

The relatively constant 
L$_{\rm X}$(gas)/SFR ratio 
argues against models in
which the hot gas in mergers is primarily
due to shocks caused by the direct collision of gaseous disks or
gaseous halos
(e.g., \citealp{cox06b, sinha09}). 
The lack of a trend between
L$_{\rm X}$(gas)/SFR and merger stage 
also argues against these models, although
the timing of such shocks depends upon
the impact parameter of the merger.  The \citet{cox06b} simulations
involve nearly radial orbits, which are expected to produce
strong shocks early in the
merger during the first impact, and a second maximum near nuclear
coalescence.   For a larger initial impact
parameter, strong shocks may not be produced 
until 
final coalescence.  Such variations in initial orbital parameters
may smear out any correlation of shock emission with merger stage.

Optical imaging spectroscopy 
\citep{monreal06, monreal10, rich11, rich14, rich15, soto12}
has revealed enhanced
[N~II]/H$\alpha$, [S~II]/H$\alpha$, and [O~I]/H$\alpha$ emission
line ratios over spatially-resolved areas within some LIRGs and ULIRGs, indicative
of excitation by large-scale shocks rather than H~II regions
or AGN.   Further, these systems tend to have larger velocities
dispersions, 100 $-$ 200 km~s$^{-1}$, compared to more
quiescent systems.   Such optical signatures of large-scale shocks tend
to be more frequent in closer pairs and later-stage mergers.
At present, there is some uncertainty 
about what is producing these shocks, with three competing ideas:
1) stellar feedback, including
supernovae and galactic winds, 2)
the direct collision of two gaseous disks, or 
3) large-scale gas flows within the galaxies due to tidal forces.
All of these process directly or indirectly correlate with the SFR; for example,
gas flows within the disks and collisions between
disks likely enhance star formation,
which in turn powers galactic winds. 
Threfore, distinguishing 
between those three scenarios is difficult.

The relationship between the shocks probed by these optical line ratios and the X-ray-emitting
gas is also uncertain, as the optical lines are produced by colder neutral gas compared to the gas detected
in the X-rays.   To better address this question, detailed comparisons of the spatial distribution of the
diffuse X-ray light compared to the location of the shocks traced by the optical emission line maps would
be useful, along with multiphase modeling of the gas.   Such an analysis is beyond the scope of the current paper.

\subsection{The Scatter in L$_{\rm X}$(gas)/SFR}

For moderate and high SFRs,
we see no correlation between the absorption-corrected
L$_{\rm X}$(gas)/SFR with merger stage or SFR, 
however, there is a fair amount of scatter in these 
plots (Figures 9 and 11).
For galaxies with SFR $>$ 1~M$_{\sun}$~yr$^{-1}$, the root mean
square deviation in log L$_{\rm X}$(gas)/SFR is 0.36, or
about a factor of 2.3 in 
L$_{\rm X}$(gas)/SFR.
This variation is similar to that found in earlier studies.
\citet{mineo12b} noted a scatter of 0.34 dex in L$_{\rm X}$(MEKAL)/SFR,
while \citet{grimes05} found a spread of 0.38 dex
in L$_{\rm X}$(gas)/L$_{\rm FIR}$.

One factor that may contribute to the observed
scatter may be strongly variable SFRs.
Depending upon the timing and the duration of the starburst vs.\ the
gas cooling time and/or gas escape time, the 
L$_{\rm X}$(gas)/SFR may vary a great deal (see Section 7.5).
System-to-system variations in the 
parameters of the interaction as well as 
variations in the initial properties 
of the galaxies affect the star formation history of the galaxy.
Also, as noted earlier, system-to-system variations in dynamical mass 
may affect the rate at which the hot gas escapes.
Metallicity variations may also affect the 
L$_{\rm X}$(gas)/SFR ratios.
Shock heating from direct collisions between gas-rich disks
may also heat the gas
\citep{cox06b, sinha09},
adding scatter to the global
L$_{\rm X}$(gas)/SFR ratios.
Differences in the large-scale environments of the galaxies
may also cause some galaxy-to-galaxy 
variations 
in the
L$_{\rm X}$(gas)/SFR ratio, because of infall from intragroup and intracluster
gas.
It is also possible that past AGN activity may 
have affected
the hot gas content in some of these galaxies,
although we see no correlation between L$_{\rm X}$(gas)/SFR and 
current AGN activity in our sample.
Other factors that may contribute to the scatter in 
our L$_{\rm X}$(gas)/SFR values
may be variations in the N$_{\rm H}$/A$_{\rm V}$ ratio
(i.e., the gas-to-dust ratio), variations in the geometry
of the hot gas relative to the stars, temperature gradients across the galaxy, or variations in the absorbing
column across a galaxy.
Also, as noted earlier, different assumptions about the 
absorbing column and gas temperature in the model that is used
can cause the derived luminosities to vary by a factor of two on average.
Further studies are needed to investigate these possibilities. 

\section{Summary and Conclusions}

We have used high resolution X-ray spectral imaging data from
the Chandra telescope to study the properties of the hot interstellar gas 
in a sample of 49 major mergers spanning the full Toomre sequence.
After removal of the bright point sources, we fit the X-ray spectrum
of the diffuse emission to a two component model:  a thermal
component described by a MEKAL function to represent the hot gas,
and a power law component which is dominated by light from
unresolved point sources.
When absorption within
the target galaxy is accounted for, 
L$_{\rm X}$(gas)/SFR does not depend strongly upon SFR.  
The absorption-corrected 
L$_{\rm X}$(gas)/SFR ratio also not depend upon merger stage,
AGN activity, or large-scale environment.
These results indicate
that the hot gas in these mergers is mainly powered by stellar and supernova feedback
associated with a young stellar population. 
This feedback
appears to be regulated such that an approximately constant fraction of about 2\% of
the total energy output
from supernovae and stellar winds is converted into X-ray flux.

Three late-stage merger remnants in our sample have 
L$_{\rm X}$(gas)/SFR ratios that are about 10 $-$ 50 times larger
than the median for the other galaxies.   These three galaxies have
low SFRs ($<$1~M$_{\sun}$~yr$^{-1}$) and UV/optical/IR signatures
of post-starbursts, suggesting that they are in the process of becoming
`red and dead' ellipticals with excess hot gas.

For the merger sample,
L$_{\rm X}$(gas)/SFR does not vary with
L$_{\rm K}$
for galaxies with L$_{\rm K}$ $>$ 10$^{10}$ L$_{\sun}$.
However, galaxies with L$_{\rm K}$ $<$ 10$^{10}$ L$_{\sun}$
show a deficiency
in L$_{\rm X}$(gas) relative to the SFR.   
Since L$_{\rm K}$ approximately traces the stellar mass in galaxies, 
this result suggests that hot gas escapes more easily from these systems due to weaker gravitational
fields, assuming galaxies with lower stellar masses also have lower total masses.

\acknowledgments

We thank the anonymous referee for very helpful suggestions, which greatly improved the paper.
This research was supported by NASA Chandra archive grant AR6-17009X, issued
by the Chandra X-ray Observatory Center, which is operated by the Smithsonian Astrophysical Observatory
for and on behalf of NASA under contract NAS8-03060. 
Support was also provided by National Science Foundation Extragalactic Astronomy Grant ASTR-1311935, as well as 
the NASA Tennessee Space Grant program.
The scientific results reported in this article are based 
on data obtained from the Chandra Data Archive.
This research has also
made use of the NASA/IPAC Extragalatic Database (NED), which is operated by the Jet Propulsion Laboratory, California Institute of Technology,
under contract with NASA.
This work also utilizes on archival data from
the Spitzer Space Telescope, which is operated by
the Jet Propulsion Laboratory (JPL), California Institute
of Technology under a contract with NASA.
This study also uses archival data from the NASA Galaxy
Evolution Explorer (GALEX), which was operated for NASA
by the California Institute of Technology under
NASA contract NAS5-98034.

\vfill
\eject

\appendix{\bf APPENDIX: NOTES ON INDIVIDUAL GALAXIES IN THE SAMPLE}

In this Appendix, we describe each of the galaxies in our sample
and provide references.
In this Appendix, we compare
the results of our analysis of the archival 
Chandra data (Tables 2 $-$ 4) with earlier studies
using the same Chandra data.
See the references listed below for pictures of these galaxies in 
the X-ray and other wavelengths.
In Figure 21, we provide some example Chandra spectra for four of our galaxies:
Arp 293, Arp 295, NGC 5256, and NGC 6240. 

For 19 of the systems in our sample, their classification as major mergers
has been questioned in the literature.  
These include
Arp 157, Arp 160, Arp 163, Arp 178, Arp 186, Arp 217, Arp 220,
Arp 233, Arp 235, Arp 236, Arp 259, Arp 263, NGC 1700, NGC 2865, NGC 3353,
NGC 5018, NGC 7592, UGC 2238, and UGC 5189.
Details on all 49 systems are provided below.

{\bf AM 1146$-$270 (ESO 504-G017):} We classify this as a post-merger stage 6 system, as it is
a single compact source with a short tail.
In NED, it is listed as an Scd/BCD HII galaxy.
Using optical images, \citet{esmith91} found numerous blue knots near the nucleus.
Compared to the other systems in our sample,
it has both a low L$_{\rm K}$ and a low L$_{\rm FIR}$ (10$^{8.98}$L$_{\sun}$ and 10$^{9.49}$L$_{\sun}$,
respectively), as well as a relatively low L$_{\rm FIR}$/L$_{\rm K}$ and a low SFR (0.26 M$_{\sun}$~yr$^{-1}$).
This is likely a low stellar mass system with a low but non-zero specific star formation.
In NED, it is classified as Seyfert 2.
AM 1146$-$270 was included in the \citet{read98} ROSAT study of X-ray emission along a major merger
sequence.  They classify it as `beyond the end of the Toomre sequence', as a `spheroidal
galaxy with a small tail-like structure'.  No X-ray flux was detected by the ROSAT study.
In the current Chandra study, the diffuse X-ray emission is marginally detected.
The Chandra data for AM 1146$-$270 are previously unpublished.

{\bf AM 2055$-$425 (ESO 286-IG 019; IRAS 20551-4250):} 
In the Arp-Madore catalog photograph
of this system \citep{arp87}, two tidal tails are visible, one more pronounced
than the other.
Only one nucleus is visible 
in broadband red images from HST \citep{kim13},
and in archival Spitzer 3.6 $\mu$m images, thus
we classify this system as merger stage 5 (single nucleus with two
strong tails).
The optical surface brightness profile approximates 
an R$^{1/4}$ law \citep{kim13}.
AM 2055$-$425 has a high FIR luminosity 
(10$^{11.72}$L$_{\sun}$), placing it in the `luminous infrared galaxy' (LIRG)
category.  Although its K band luminosity is also quite high
(10$^{11.27}$L$_{\sun}$) compared to many of the other galaxies in our 
sample, its L$_{\rm FIR}$/L$_{\rm K}$ ratio is one of the highest in our sample,
thus it is undergoing a strong starburst, consistent with its
high inferred SFR of 142 M$_{\sun}$~yr$^{-1}$.

Based on XMM-Newton X-ray data, \citet{franceschini03} classify
AM 2055$-$425 as an AGN.  Using optical imaging Fabry-Perot
Spectroscopy, \citet{rich15} find a composite post-starburst 
and LINER spectrum, but 
conclude that the distribution of line strengths
is more consistent with extended shock excitation than with an AGN.

AM 2055$-$425 was included in the 
\citet{grimes05} Chandra survey of hot gas in nearby galaxies,
who found a total thermal 0.3 $-$ 2.0 keV luminosity of 2.8 $\times$ 10$^{41}$
erg~s$^{-1}$ and an absorbing column of 0.1 $\pm$ $^{0.6}_{0.1}$ $\times$
10$^{22}$ cm$^{-2}$.  The large uncertainty on this absorbing column
is consistent with our estimate from the UV/IR data (Table 3), while our
total L$_{\rm X}$ is about twice their value.
The same Chandra data was used earlier by \citet{huo04} to study
the hot gaseous halo, with consistent results.

{\bf 
AM 2312$-$591
(IRAS 23128$-$5919;
ESO 148-IG 002):}
AM 2312$-$591
has two long tidal tails visible on the \citet{arp87} Catalog
photograph.    In archival HST optical and IR images 
two disks are seen in very close contact.  These disks appear
to be just starting
to merge, thus we classify this system as stage 3.
AM 2312$-$591
is infrared-luminous, 
with L$_{\rm FIR}$ = 10$^{11.71}$ L$_{\sun}$
and a SFR of 149 M$_{\sun}$~yr$^{-1}$.

The diffuse X-ray-emitting gas in 
AM 2312$-$591
as seen by Chandra was
previously studied by \citet{grimes05}.  They separated the diffuse
emission into two annuli with a boundary at 3\farcs42, and fit
the spectra of the two parts separately.  For the inner portion,
they used a combination of the VMEKAL function and a power law function,
and fitted for the thermal temperature, the power law index,
and the absorbing column.
For the VMEKAL component for the inner
annulus, they found kT = 0.74 $\pm$ $^{0.10}_{0.09}$ keV,
N$_{\rm H}$ = 7.8 $\pm$ $^{1.3}_{2.9}$ $\times$ 10$^{22}$
cm$^{-2}$, and L$_{\rm X}$(0.3 $-$ 2.0 keV) = 1.7 $\times$ 
10$^{41}$ erg~s$^{-1}$.   
This absorbing column is about 2.4 times larger than our
estimate for the entire system.
For the outer annulus, their fit did 
not require a power law component or internal absorption. For the outer
annulus
using just Galactic absorption, they obtained kT = 0.61 $\pm$ $^{0.07}_{0.05}$
keV and L$_{\rm X}$(0.3 $-$ 2.0 keV) = 1.7 $\times$ 10$^{41}$ erg~s$^{-1}$.
Combining the two annuli, for the thermal component they found 
L$_{\rm X}$(0.3 $-$ 2.0 keV) = 3.4 $\times$ 10$^{41}$ erg~s$^{-1}$.
This is 
40\% less than our total internal-absorption-corrected 0.3 $-$ 8 keV
thermal 
luminosity, calculated using our UV/IR-estimated 
N$_{\rm H}$ = 3.3 $\times$ 10$^{21}$ cm$^{-2}$.

{\bf Arp 91 (NGC 5953/4):} This is a pre-merger pair with a relatively small pair separation but without
strong tidal tails, thus we classify it as merger stage 1.  The more
southern galaxy, NGC 5953, has a Seyfert 2 nucleus and a short tidal tail.
The companion, NGC 5954, is a distorted spiral without tails.
NGC 5954 is classified as SAB(rs)cd? pec in NED, while NGC 5953 is listed as SAa? pec.
The Chandra data for Arp 91 has been exploited earlier in studies of X-ray point
sources \citep{evans10, liu11, smith12}, 
but the diffuse X-ray emission in Arp 91 has not
previously been studied.   

{\bf Arp 147 (IC 298):} The western galaxy in Arp 147 is 
a collisional ring 
\citep{theys76,madore09}.  The likely intruder,
a companion galaxy to the east,
resembles 
a disk galaxy seen edge-on, and is classified as S0? pec
in NED.
The companion is redder in 
the optical and near-infrared than the ring galaxy, 
and has a K band luminosity that is
1/2 that of the ring galaxy \citep{romano08}.
At 3.6 $\mu$m, the companion is 40\% more luminous
than the ring galaxy \citep{rappaport10}.
In HST images, a ring-like structure is seen in the companion
as well \citep{rappaport10}.
Given the lack of prominent tidal tails, 
we classify this system as a stage 1 major merger,
however, the ring morphology indicates that they have
already collided.

The Chandra data for Arp 147 has been previously
used by \citet{rappaport10} to investigate the X-ray point sources.
However, the diffuse X-ray emission in Arp 147 has not previously
been studied.  We find extended X-ray emission along
the northeastern arc of the ring galaxy in Arp 147.

{\bf Arp 148 (VV 32):} The western galaxy in Arp 148 is another 
collisional ring \citep{madore09}
with a disk companion seen edge-on.
In the K band, the luminosity of the edge-on
disk is 2.9 times that of
the ring galaxy \citep{romano08}.
Without strong tidal tails, we classify this as a stage 1 major
merger.
Together, this pair is classified as a LIRG, with 
L$_{\rm FIR}$ $>$ 10$^{11}$ L$_{\sun}$ and a SFR
of 16 M$_{\sun}$~yr$^{-1}$.
In archival Spitzer and GALEX images, the edge-on
galaxy dominates
the total light of the pair.
The Chandra data for Arp 148 has not previously been published.
The Chandra map reveals significant diffuse X-ray emission
in the edge-on disk, and a lower level in the 
ring of the ring galaxy.

{\bf Arp 155 (NGC 3656):} This peculiar elliptical
galaxy has been
classified as a shell galaxy by \citet{balcells97},
who discovered two faint tidal tails.
The interior follows a R$^{1/4}$ law
in both the optical and the near-infrared
\citep{balcells90, balcells97, rothberg04}.
We classify it as a stage 7 merger remnant.
It has a moderate SFR of 1.1 M$_{\sun}$~yr$^{-1}$.

Arp 155 was included in a ROSAT survey 
of early-type galaxies conducted by
\citet{osullivan01a}, who found it was undetected
with an upper limit of L$_{\rm X}$ $<$ 
10$^{40.61}$ erg~s$^{-1}$.  
The archival
Chandra data for Arp 155 has previously been
used for a study of the X-ray point sources
in interacting galaxies \citep{smith12},
however, the diffuse X-ray emission from Arp 155 
as seen by Chandra has not
previously been studied.

{\bf Arp 157 (NGC 520):} This system may be the result of a merger of one gas-rich disk galaxy and one gas-poor disk galaxy since one tail is gas-poor 
\citep{hibbard96}.
We classify it as a stage 4 merger,
as it has two nuclei \citep{stanford90a}
and a long tail \citep{stockton80, stanford90b}.
Arp 157 was included 
in the 
\citet{toomre77}
merger sequence.  

The Chandra X-ray map of Arp 157
was previously presented by 
\citet{read05}, 
who called it a “half-merger.” 
Using Galactic absorption and an absorbed
MEKAL model, they obtained a gas
temperature of 0.58 $\pm$ $^{0.09}_{0.11}$ keV
and an absorption-corrected 0.2 $-$ 10 keV
luminosity of 8.03 $\times$ 10$^{39}$ erg~s$^{-1}$.
This is about half our total (MEKAL + power law)
0.3 $-$ 8 keV luminosity assuming Galactic absorption, but about 3.7 times
our MEKAL luminosity.

{\bf Arp 160 (NGC 4194):} 
In the \citet{arp66} Atlas photograph,
Arp 160 has a short plume-like structure extending
to the north as well as fainter extended arcs
or shells in the south.
It has a long HI tail extending to the south
opposite the optical plume
\citep{manthey08}.
It has a single nucleus and 
a near-infrared K band light profile that approximately
fits an R$^{1/4}$ profile \citep{rothberg04}.
Arp 160 has variously
been classified as a minor merger,
the product of 
a disk galaxy falling into a larger elliptical
\citep{manthey08, weistrop12}, and as a major
merger \citep{rothberg10}.
In the current study, 
we assume it is the result of
an approximately equal mass merger,
and classify it as merger stage 5.

The Chandra X-ray data for 
Arp 160 has been utilized several times before,
both for the point sources \citep{lehmer10,
smith12, mineo12a,
luangtip15} and the diffuse emission \citep{mineo12b}.
\citet{mineo12b} found a diffuse MEKAL component
with L$_{\rm X}$ = 10$^{40.39}$ erg~s$^{-1}$
(with Galactic absorption)
or 10$^{41.1}$ erg~s$^{-1}$
(correcting for internal absorption).
These are consistent with our values.
From fitting the X-ray spectrum, they derived
a hot gas temperature of kT = 0.25 $\pm$ $^{0.04}_{0.03}$ keV
and an absorbing column of 3.1 $\pm$ $^{1.6}_{1.1}$
$\times$ 10$^{21}$ cm$^{-2}$.  This column density
is consistent with the value we obtained from
the UV/IR data. 

{\bf Arp 163 (NGC 4670):}
Arp 163 appears to be a late-stage merger remnant. 
In the \citet{arp66} photograph, the inner region looks
like a warped and distorted lens, while
NED classifies it as
SB0/a(s) pec?.
The K band radial profile of Arp 163 cannot
be fit with either a R$^{1/4}$ profile or an exponential disk \citep{chitre02},
while at V the radial distribution appears a truncated exponential disk
\citep{baggett98}.
Only one nucleus is seen in the K band \citep{chitre02}.
A bar-like structure is detected in the near-infrared \citep{jog06}.
Although it is not an elliptical,
we classify Arp 163 as a stage 7 merger because it has no tails.
Arp 163 has a relatively low K band luminosity of 10$^{9.93}$ L$_{\sun}$,
and is listed as a Liner? in NED.

Arp 163 was included in studies of X-ray point sources as seen
by Chandra 
\citep{swartz09, evans10, liu11, smith12}, however, the diffuse X-ray emission
has not previously been studied.  In the current study,
we do not detect a thermal component
to the diffuse X-ray flux from Arp 163.

{\bf Arp 178:}  
As catalogued by \citet{arp66} and in NED,
Arp 178 consists of three galaxies, NGC 5613/4/5.   
However, NGC 5613 has a redshift that is 4500 km/s higher than the other two, so may not be part of the same system.
We exclude NGC 5613 from our analysis. 
NGC 5614 and 5615 have very similar redshifts.
NGC 5615 is very faint compared to NGC 5614 (the g and 3.6 $\mu$m flux ratios are
1:100 and 1:21, according to NED
and \citealp{brassington15}, respectively).
A strong tail-like structure extends out from NGC 5614.   
NGC 5615 appears to be associated with the tail in some way.
NGC 5614/5 may be an M51-like system, with the tail material pulled out from either NGC 5614 or NGC 5615.
Alternatively, the tail may be tidal debris from NGC 5615 which is being cannibalized by NGC 5614.
Based on the near-infrared light ratio of this pair, NGC 5614/5 would not be classified as a major merger,
unless considerable stripping of NGC 5615 has occurred.
Another possibility is that NGC 5614 is itself a merger remnant, and 
the tail is the result of a past merger rather than an interaction with 
NGC 5615.   
NGC 5614 
is classified as SA(r)ab pec in NED, and has 
a prominent bulge and a distorted disk.
Archival optical HST images show only one nucleus in NGC 5614.
Alternatively, it may be a stage 4 system (two nuclei), or a minor merger.
In the Galaxy Zoo 2 study, 46\% of the participants classified this system as a merger \citep{willett13}.
We tentatively classify NGC 5614 as a stage 5 merger remnant (single nucleus, two long tails), although
only one tail is visible.   
In the current study, we tentatively include NGC 5614/5 as a candidate major 
merger remnant, however, its
exact stage and original stellar mass ratio are uncertain. 

In the Chandra map, only a small amount of diffuse X-ray emission is seen, near the nucleus
of NGC 5614.  The Chandra data were previously unpublished.

{\bf Arp 186 (NGC 1614):}
Arp 186 is a classic stage 5 merger, with two long tidal tails \citep{arp66} but 
only one nucleus on near-infrared images \citep{bushouse92, forbes92, haan11}.
It has a high L$_{\rm FIR}$ of 10$^{11.20}$ L$_{\sun}$ and a high SFR of 67 M$_{\sun}$~yr$^{-1}$.
In NED, Arp 186 is listed as a SB(s)c pec LIRG starburst.
Its red light distribution fits an R$^{1/4}$ law \citep{kim13}.
Its K band light also approximately fits 
the R$^{1/4}$ law except in the inner regions \citep{rothberg04}.
Although Arp 186 has sometimes been classified as a minor merger (e.g., \citealp{konig16}), 
we will follow \citet{haan11, haan13} and assume it is the relic of a major merger.

Based on its optical spectrum, \citet{veilleux95} and \citet{yuan10} classified Arp 186 as a composite
LINER/HII or 
starburst/AGN galaxy;
alternatively, it is catalogued as an H~II region galaxy by \citet{veron06}.
Its near-infrared spectrum is consistent with that of a starburst, not
an AGN \citep{vaisanen12}.
Its mid-infrared neon line ratios from Spitzer also do not indicate an AGN
\citep{inami13}.  However, higher spatial resolution 
mid-infrared spectra show a strong 
mid-infrared continuum but weak PAH features, maybe indicating an obscured
AGN \citep{pereira15}.
Using low signal-to-noise 
archival data from the Advanced Satellite for Cosmology and
Astrophysics (ASCA),
\citet{risaliti00} detected hard X-ray emission
from Arp 186 which they concluded was due to an obscured AGN, however,
later analyses using XMM-Newton observations found that the hard X-ray radiation from
Arp 186 could be accounted for
from HMXBs associated with a starburst \citep{pereira11, vaisanen12}.

The Chandra data for Arp 186 were previously analyzed
by \citet{herrero14} in conjunction with radio, optical, and IR observations.
They provided a map of the diffuse X-ray emission
and fit the nuclear spectrum.   
They concluded that an AGN was not required to account for either
the nuclear X-ray emission or the data at other wavelengths.
For the diffuse X-ray emission, assuming only Galactic absorption,
they obtained their
best fit using the VMEKAL model
(without including an additional power law component).  
Their best-fit temperature 
was kT = 1.7 $\pm$ 0.7 keV, giving a 0.5 $-$ 2 keV luminosity
of 10$^{40.78 \pm ^{0.04}_{0.05}}$
erg~s$^{-1}$.   
This is about four times higher than the luminosity
of the MEKAL component in our MEKAL + power law fit using Galactic absorption, 
assuming a fixed kT = 0.3 keV.

{\bf Arp 217 (NGC 3310):} Arp 217
has been classified as a ``ripple'' system with strong star formation. 
\citet{balick81}
suggested the ripples were caused by a 
smaller galaxy being torn apart by tidal forces. 
However, the system appears to have two HI tails \citep{kregel01}, 
supporting the idea of a major merger.  \citet{wehner06}
conclude that either it is the result of a major merger or 
the result of multiple mergers with smaller companions.
For the current study, we assume it is a major merger and classify
it as stage 6.

The Chandra data for Arp 217 was previously analyzed by 
\citet{lehmer15}, jointly with NuSTAR X-ray data.
They found diffuse emission from hot gas in Arp 217, with a low 
kT $\sim$ 0.2 $-$ 0.3 keV.

{\bf Arp 220:} 
Arp 220 is the prototype of a ULIRG \citep{soifer84}.  
In the \citet{arp66} Atlas photograph, tidal debris is visible around
a compact core.  In near-infrared images, two nuclei are resolved 
\citep{graham90, scoville98}, while the galaxy as a whole obeys
the R$^{1/4}$ law \citep{wright90}.
It has been suggested that Arp 220 might be the product of the merger
of four or more galaxies \citep{taniguchi12}, however, in the current
study we assume it is a single stage 4 major merger.
Although a powerful starburst is clearly on-going in Arp 220
\citep{sturm96, genzel98}, both near-infrared 
\citep{depoy87} and optical \citep{sargsyan11} spectra reveal
broad hydrogen lines, indicating the presence of an obscured AGN 
as well.
However, based on a joint analysis of XMM-Newton
and NuSTAR data, \citet{teng15} concluded that the X-ray spectrum
of Arp 220 could be accounted for by a starburst alone, although
they could not rule out a highly obscured AGN. 

The large-scale diffuse X-ray morphology of Arp 220 as seen by 
Chandra has been discussed in detail by
\citet{mcdowell03}, who note the presence of 
extended emission outside of the optical galaxy in `plumes' and
`lobes'. 
They conclude that at least some of
this light is associated with a starburst-driven superwind. 
These features have a soft X-ray spectrum, with
temperatures corresponding to kT $\sim$ 0.2 $-$ 
1 keV.
Excluding the inner 3$''$ radius core,  
using only Galactic absorption
at our distance they find a total 0.3 $-$ 10 keV
diffuse X-ray luminosity of 
1.1 $\times$ 10$^{41}$ erg~s$^{-1}$.
This is about 60\% higher than our Galactic-absorption-only value.

The Chandra data for Arp 220 was also studied by \citet{huo04},
who analyzed the inner halo separately from the outer
halo, dividing at a radius of 5 kpc ($<$12$''$).
For the outer portion, using Galactic absorption only
and fitting to a pure MEKAL function,
they found kT = 0.62 $\pm$ $^{0.07}_{0.06}$ keV and 
L$_{\rm X}$ = 0.32 $\pm$ $^{0.07}_{0.08}$ $\times$ 10$^{41}$
erg~s$^{-1}$.   For the inner regime, they fit to a MEKAL
function plus two power laws as well as the absorbing column.
They found N$_{\rm H}$ = 0.66 $\pm$ $^{0.54}_{0.41}$ $\times$ 
10$^{21}$ cm$^{-2}$, kT = 0.78 $\pm$ $^{0.09}_{0.12}$ keV,
and 
L$_{\rm X}$(MEKAL) = 0.22 $\pm$ $^{0.03}_{0.03}$
$\times$ 10$^{41}$ erg~s$^{-1}$.  
Their best-fit absorbing column is only 60\% higher than the Galactic
value, and about 1/8th our best-fit and UV/IR-determined values.
Their total MEKAL luminosity is about 2.6 times our Galactic-absorption-only
value, but about 1/4th our value corrected for internal absorption.

A third independent analysis of the Chandra Arp 220 data was
done by \citet{grimes05}, who also studied the inner and outer 
emission separately.   Their inner region was elliptical in shape, with
major and minor axes of 7\farcs72 (3.2 kpc) and 6\farcs23 (2.1 kpc), 
respectively.  The X-ray spectrum of the outer region was successful
fit to a VMEKAL model with Galactic absorption, giving a temperature
of 0.59 $\pm$ $^{0.03}_{0.04}$ keV and a 0.3 $-$ 2 keV luminosity
of 7.3 $\times$ 10$^{40}$ erg~s$^{-1}$.   The inner region
was fit to a VMEKAL+power law model, fitting for both temperature
and absorbing column.   For the inner region they found kT = 
0.85 $\pm$ $^{0.05}_{0.05}$ keV, 
N$_{\rm H}$ = 5 $\pm$ $^{3}_{2}$
$\times$ 10$^{21}$ cm$^{-2}$, 
and a thermal
0.3 $-$ 2 keV 
L$_{\rm X}$ = 1.6 $\times$ 10$^{40}$
erg~s$^{-1}$. 
Their best-fit absorbing column for the inner region agrees well
with our estimate from the UV/IR ratio, while their total thermal
luminosity is about one half our value. 

{\bf Arp 222 (NGC 7727):}
Arp 222 is classified as a late-stage major merger remnant 
by \citet{georgakakis00}.   In optical images,
two extended tidal tails and some inner loops are visible
\citep{arp66, sandage94}.
The K band light fits 
an R$^{1/4}$ profile \citep{chitre02, rothberg04}.
We classify it as a stage 6 merger.
Arp 222 has a very low FIR luminosity (10$^{8.71}$ L$_{\sun}$)
relative to its K band luminosity (10$^{11.12}$ L$_{\sun}$), thus
it is now a relatively quiescent galaxy.
However, high resolution optical and IR images revealed the presence of 
intermediate-age (1 $-$ 2 Gyr) globular clusters in Arp 222 \citep{trancho14},
suggesting a starburst in the past.

Arp 222 was included in the \citet{brassington07} Chandra survey
of major mergers, as the second-to-last in their merger sequence.
Including only Galactic absorption,
they measured a diffuse X-ray luminosity of 6.5 $\times$ 10$^{39}$ erg~s$^{-1}$,
similar to our value.  Their best-fit gas temperature is 0.60 $\pm$ $^{0.07}_{0.06}$
keV.

{\bf Arp 226 (NGC 7252):}  Arp 226 is the classic post-merger
`Atoms for Peace' galaxy with two prominent tidal tails \citep{arp66}.
In the K band, it fits the R$^{1/4}$ law with a single
nucleus \citep{stanford91, rothberg04}.  
Its tails are longer and more prominent than those of Arp 222, thus
we 
classify it as stage 5 rather than stage 6.
Six star clusters in Arp 226 have been age-dated to 
400 $-$ 600 Myrs old
\citep{schweizer98}, 
consistent with its 
location in the `post-starburst' region of Figure 16
(FUV $-$ NUV vs.\ NUV $-$ [3.6]).

The Chandra data for Arp 226 were included in the \citet{brassington07}
survey (see \citealp{nolan04}).  \citet{brassington07}
report an X-ray luminosity from hot gas of 2.4 $\times$ 10$^{40}$
erg~s$^{-1}$, consistent with our value using Galactic absorption.
From ASCA X-ray spectra, \citet{awaki02} derive
a soft (0.5 $-$ 4keV) X-ray luminosity for Arp 226
of 2 $\times$ 10$^{40}$ erg~s$^{-1}$, also consistent with our Chandra results for
Galactic absorption.

{\bf Arp 233 (Mrk 33; Haro 2; UGC 5720):}  The morphological type and evolutionary
state of the peculiar galaxy Arp 233 is uncertain.   In the \citet{arp66}
photograph, a compact central core is surrounded by wispy nebulosity.
In NED, Arp 233 is listed
as Im pec, and it is classified as a blue
compact galaxy by \citet{davidge89}.  
However, 
it
was included in a study of E/S0 galaxies
with dust lanes by
\citet{finkelman10}, and 
\citet{loose86}
concluded that it is a low mass 
elliptical galaxy with a young stellar population.
\citet{buta15} classified Arp 233 as an E3-4 pec galaxy based on Spitzer images.
According to \citet{loose86}, 
its optical radial profile is better fit by a R$^{1/4}$ law than
by an exponential disk, but the best-fit model is in-between these two cases.

Arp 233 clearly has on-going star formation, with blue
optical colors \citep{huchra77, loose86} and a higher 
higher H$\alpha$ equivalent widths than other candidate
dust-lane E/S0 galaxies \citep{finkelman10}.
Arp 233 is likely a low mass system, with 
a somewhat sub-solar metallicity ($\ge$1/3 solar;
\citealp{zhao10}) and 
a moderately low L$_{\rm K}$ (10$^{9.99}$ L$_{\sun}$).
In the current work, we consider it a very late stage (stage 7) 
remnant, perhaps the relic of the merger
of two low mass galaxies.   Its presence in our sample 
helps to anchor the low mass end of the survey.

The Chandra data for Arp 233 were previously used in 
surveys of X-ray point
sources in nearby galaxies 
\citep{evans10, smith12}.
Arp 233 was also included in the \citet{mineo12b} Chandra study of 
the diffuse X-ray emission from star-forming galaxies, who measured
a MEKAL L$_{\rm X}$ similar to our Galactic-absorption-corrected value.
They found a best-fit gas temperature of 0.31 $\pm$ $^{0.09}_{0.05}$ keV.

{\bf Arp 235 (NGC 14; VV 80):} 
In the \citet{arp66} photograph, Arp 235 is a peculiar elongated object
surrounded by nebulosity and a faint outer oval.
Arp 235 is classified as an IB(s)m peculiar galaxy in NED, as Im
based on SDSS images \citep{ann15}, and as IAB(s:)m by \citet{buta15}
based on Spitzer images.
In Galaxy Zoo, 36\% ranked Arp 235 as an elliptical, 33\% as spiral/disk,
15\% as a merger, and 15\% as unknown \citep{lintott11}.
\citet{arkhipova85}
suggested that it could be a former double galaxy which has underwent a merger.
We classify Arp 235 as a possible stage 7 merger remnant.
Like Arp 233, Arp 235 has a low L$_{\rm K}$ (10$^{9.43}$ L$_{\sun}$).
The Chandra point source detections for Arp 235 have previously been studied 
by \citet{swartz09}, \citet{liu11}, and \citet{smith12}.  
No diffuse X-ray emission is detected in the Chandra data for Arp 235.

{\bf Arp 236 (IC 1623; VV 114):} The \citet{arp66} image of Arp 236 shows three galaxies:
IC 1623, a disturbed pair of interacting spirals, and IC 1622,
a third galaxy with a similar redshift and brightness
3$'$ away to the southeast.
In the current study, we only include the two galaxies in the IC 1623 pair in
our analysis.  We classify this pair as a stage 1 merger, since although it
is disturbed with an distorted spiral arm, this feature is not prominent.
This system has a high far-infrared luminosity and a high SFR
(10$^{11.39}$ L$_{\sun}$ and 57 M$_{\sun}$~yr$^{-1}$, respectively).

The Chandra observations of Arp 236 were previously discussed by \citet{grimes06},
along with XMM-Newton and UV spectroscopy.   
Their fit to the X-ray spectrum produced an average
absorbing column of 6 $\pm$ 3 $\times$ 10$^{21}$ cm$^{-2}$, consistent
with our best-fit value and 
slightly higher than
our estimate from the UV/IR data.
They derived a temperature of 0.62 $\pm$ 0.03 keV for the gas and an
absorption-corrected 0.3 $-$ 2 keV luminosity from the hot gas of 2.0 $\times$
10$^{41}$ erg~s$^{-1}$.   This luminosity agrees well with our estimate
using the UV/IR-derived internal absorption, and is about 1/4 our best-fit value.

{\bf Arp 240 (NGC 5257/8):} Arp 240 is a widely-separated pre-merger pair with two tidal tails
and a connecting bridge \citep{arp66}.  We categorize it as a stage 2 merger.
Both galaxies are independently LIRGs.  The total SFR for the pair is high, 40 M$_{\sun}$~yr$^{-1}$.
The diffuse X-ray emission in Arp 240 was previously studied by \citet{smith14},
in conjunction with data at other wavelengths.
Diffuse X-ray
emission is visible along prominent arcs of star formation at the base of the
tidal tails.
Analytical models suggest that the
star formation in these regions has been triggered by intersecting caustics along the base
of the tails,
where a caustic is a narrow pile-up zone produced by orbit-crowding
during a galaxy interaction \citep{struck12}.

{\bf Arp 242 (NGC 4676):} Arp 242 is the famous double-tailed `Mice' pre-merger interacting galaxy pair.
We list it as a stage 2 merger.  
The pair has a moderately high L$_{\rm FIR}$ of 10$^{10.65}$
L$_{\sun}$.   
The diffuse X-ray emission in Arp 242 as seen by Chandra was previously studied by \citet{read03}.
Correcting only for Galactic absorption, they find a total 0.3 $-$ 10 keV thermal luminosity of 
1.9 $\times$ 10$^{40}$ erg~s$^{-1}$, in agreement with our result. 
\citet{read03} finds gas temperatures of 0.50 $\pm$ $^{0.18}_{0.13}$ and 
0.46 $\pm$ $^{0.20}_{0.12}$ for the northern and southern galaxies, respectively.
The \citet{read03} results for Arp 242 were included in the \citet{brassington07} Chandra survey of merging galaxies.

{\bf Arp 243 (NGC 2623):} 
Arp 243 is a single galaxy with two prominent tidal tails \citep{arp66} and a high far-infrared luminosity
(10$^{11.34}$ L$_{\sun}$).  
In HST near-infrared images only one nucleus is visible \citep{haan11},
thus we list it as merger stage 5.
The Chandra data for Arp 243 have previously been utilized in surveys of X-ray point sources \citep{evans10, smith12}.
In the current Chandra study, we do not detect any thermal X-ray emission from Arp 243.
Using ROSAT PSPC data, \citet{read98} claim a detection of soft (kT = 0.20 keV) diffuse X-ray emission with
L$_{\rm X}$ = 4 $\times$ 10$^{40}$ erg~s$^{-1}$ to the west of Arp 243,
outside of the optical
extent of the galaxy.  They suggested this may be a large outflow lobe.  
However, this is a low S/N detection, and emission is only seen on one side of the galaxy.
Furthermore, they do not see this feature with the ROSAT High Resolution Imager (HRI).  
It is also not seen in the Chandra data.
Based on our upper limit to the Chandra flux in the main portion of the galaxy (Table 2), we 
would have expected to
obtain a marginal 
detection ($\sim$3 $-$ 5$\sigma$) of this lobe in our data.

{\bf Arp 244 (NGC 4038/9:}  The stage 3 system Arp 244 (the Antennae) is a pair of
disk galaxies in close contact and two long tidal tails.
The L$_{\rm FIR}$ of Arp 244 is 
10$^{10.61}$
L$_{\sun}$, very similar to that of Arp 242.
The Chandra diffuse X-ray map of Arp 244 has been previously discussed 
by several groups \citep{fabbiano01, fabbiano03, metz04, baldi06a, baldi06b, smith14}.
These data were also included in the \citet{brassington07}
study.
In the spatially-resolved study of \citet{baldi06b}, the diffuse X-ray spectrum for more than a dozen
different regions were analyzed.  The best-fit temperatures for these regions
varied from kT = 0.20 keV to 0.66 keV, with most around 0.62 keV.  Their derived columns
varied from $<$0.12 $\times$ 10$^{20}$ cm$^{-2}$ to 24.6 $\times$ 10$^{20}$ cm$^{-2}$.
For comparison, from the UV/IR data we estimate an average 
N$_{\rm H}$ = 15.4 $\times$ 10$^{20}$ cm$^{-2}$, while our best-fit value for the MEKAL component
is 2.7 times larger.
Summing over all these regions and using our distance, 
the total hot gas L$_{\rm X}$ for Arp 244 from the \citet{baldi06a} analysis
is 2.7 $\times$ 10$^{40}$ erg~s$^{-1}$, about half our best-fit estimate.
Our determination using the UV/IR estimate of absorption
is 7.1 $\times$ 10$^{40}$ erg~s$^{-1}$, 2.6 times larger than the \citet{baldi06a} result.

{\bf Arp 256:} Arp 256 is another stage 2 system, consisting
of a pair of 
widely separated spiral galaxies with tidal features 
\citep{arp66}.
Together, the pair is classified as a LIRG, however,
about 90\% of the FIR light comes from the southern
galaxy \citep{howell10}.
In an earlier study \citep{smith14}, we used the 
Chandra data for Arp 256 to measure
the 
diffuse X-ray flux from a luminous knot of star formation
at the base of the northern tail of the northern
galaxy of Arp 256.  We found that
the X-ray light from this region
is resolved, with L$_{\rm X}$ = 
3.2 $\times$ 10$^{40}$ erg~s$^{-1}$ using
N$_{\rm H}$ = 2.4 $\times$ 10$^{21}$ cm$^{-2}$.
\citet{ricci17} also analyzed the archival Chandra data for both
galaxies of Arp 256,
and concluded that the emission can be accounted for a starburst,
without the need for an AGN.
Neither galaxy was detected by NuSTAR \citep{ricci17}. 
In the current study, we extract the total
diffuse X-ray spectrum for the combined pair,
and derive a luminosity four times larger assuming
a similar absorbing column.  Inspection of the Chandra
map shows that the majority of extended emission in Arp 256
is arising from the southern galaxy.

{\bf Arp 259 (NGC 1741):}
In the \citet{arp66} Atlas photograph,
Arp 259 has a peculiar appearance, consisting
of at least
two edge-on disks surrounded by loops and tidal debris.
The nature and evolutionary state of Arp 259
has been a matter of debate.
\citet{hickson82} classify this system as a compact group
(HCG 31)
and identify four galaxies.
\citet{rubin90} concur, identifying 
two overlapping
strongly interacting edge-on disk galaxies 
(A and C), another edge-on galaxy (B),
and a fainter galaxy (D), as well as other more distant
galaxies.  In contrast, \citet{richer03} conclude
that A and C are a single tidally distorted galaxy,
and some of the more
distant objects are tidal debris rather than pre-existing
galaxies.  
They find that Galaxy B is kinematically distinct from
A+C and is counter-rotating relative to it, thus they
conclude that another more distant galaxy (G) is responsible
for the tidal features of A+C.
In an opposing view, \citet{amram07} and \citet{alfaro15}
conclude that A and C are 
two gas-rich interacting Magellanic-type
irregular galaxies in the process of merging.
In the current study, we will treat Arp 259 as a 
peculiar major merger and classify it as stage 2, although
this classification is very uncertain.
The total K band luminosity of Arp 259 is moderately
low (10$^{10.20}$ L$_{\sun}$), but there are half a dozen
other systems in our sample with lower L$_{\rm K}$ than
Arp 259.  The SFR is also a moderate value, 
6.9 M$_{\sun}$~yr$^{-1}$.

The Chandra X-ray data for Arp 259 were
included in earlier studies of X-ray point sources 
\citep{smith12, tzanavaris14}.  The hot X-ray-emitting
gas in and around Arp 259 was previously
investigated by \citet{fuse13} and \citet{desjardins13}.
\citet{fuse13} found a total
0.3 $-$ 6.0 keV MEKAL 
L$_{\rm X}$ of 10$^{41.6}$ erg~s$^{-1}$
using only Galactic absorption.
In contrast, \citet{desjardins13} measured a factor
of ten times
lower 
L$_{\rm X}$ of 10$^{40.54 \pm ^{0.60}_{0.36}}$ erg~s$^{-1}$,
also with Galactic absorption.   \citet{desjardins13}
also fit for temperature, and find kT = 0.65 $\pm$ $^{0.18}_{0.31}$
keV. 
Our extraction yields a result reasonably consistent with that
of \citet{desjardins13} and much lower than that quoted
by \citet{fuse13}.
Inspection of the Chandra map shows that almost all of the 
diffuse X-ray light lies within A+C.

{\bf Arp 261 (VV 140):} 
The \citet{arp66} picture of Arp 261
shows a close pair of edge-on disk-like galaxies surrounded
by tidal debris, and a smaller irregular 
galaxy (GALEXMSC J144935.10-100444.6) about five 
disk radii (5$'$) away to the north.  
In the current study, we only include the two main galaxies
in our analysis.  We class it as a stage 1 merger, since the
tidal features are not long or prominent.
Arp 261 has a 
low 
L$_{\rm K}$ = 10$^{9.66}$ L$_{\sun}$ and a low SFR of 
0.7 M$_{\sun}$~yr$^{-1}$.
The star cluster population in Arp 261 has been investigated by 
Peterson et al.\ (2017, in preparation),
who found a peak in the age distribution at $\sim$15 Myrs. 
The Chandra data for Arp 261 were included in the point
source study of \citet{evans10}, however, the diffuse 
emission has not previously been studied. 

{\bf Arp 263 (NGC 3239):} Arp 263 is the most nearby of the galaxies
in our sample, at only 9.8 Mpc.
It also has the lowest L$_{\rm K}$ in our sample.
It is classified
as an IB(s)m pec in NED.
In optical images it has a knotty appearance, with two distorted tail-like
structures \citep{arp66}.
Based on its morphology and kinematics,
\citet{krienke90} suggested it was a merger remnant.
\citet{taylor07} also classify it as a merger remnant based on UV/optical images.
We list it as a stage 5 merger assuming only a single nucleus,
although this is uncertain from the archival Spitzer 3.6 $\mu$m image because
of the presence of luminous star forming knots.

The Chandra data for Arp 263 was previously searched for X-ray point sources
\citep{swartz09, swartz11, liu11, smith12}, however, the diffuse X-ray emission has
not previously been studied.
Although we detect low level diffuse emission, 
we do not detect a MEKAL component in the spectrum of the diffuse light.

{\bf Arp 270 (NGC 3395/6):} 
The two galaxies in the Arp 270 pair are relatively
close together, but only short tails are visible
in optical light \citep{arp66} and in the UV \citep{smith10}.  
However, in 21 cm HI
maps a long gaseous tail is visible to the southeast \citep{clemens99}.
In the current study, we classify it as a stage 1 merger
since the optical tails are short.
The Chandra data for Arp 270 were included in the X-ray point
source study of \citet{smith12}, while the diffuse X-ray
light was analyzed by \citet{brassington05}.
They measured a total absorption-corrected MEKAL L$_{\rm X}$ of 
9.5 $\pm$ 0.3 $\times$ 10$^{39}$ erg~s$^{-1}$, in agreement
with our results.
Their best-fit N$_{\rm H}$ was 4 $\pm$ $^{4}_{2}$ $\times$ 10$^{20}$
cm$^{-2}$ for the northeastern galaxy NGC 3395 and 
9 $\pm$ $^{4}_{2}$ $\times$ 10$^{20}$
cm$^{-2}$ for the southwestern galaxy NGC 3396, in reasonable
agreement with our UV/IR-derived absorption for the system.
\citet{brassington05} derived gas temperatures of
0.52 $\pm$ $^{0.05}_{0.10}$ keV and 0.49 $\pm$ $^{0.05}_{0.09}$
keV for NGC 3395 and NGC 3396, respectively. 

{\bf Arp 283 (NGC 2798/9):}
This pre-merger interacting pair only has short tails in the
\citet{arp66} Atlas photograph,
thus we classify it as a stage 1 merger.
A third galaxy at a similar redshift, UGC 4904, lies 5$'$ to the south.
Arp 283 was previously included in the Chandra point source surveys of
\citet{liu11} and 
\citet{smith12}. 
The diffuse X-ray emission of this group as seen by Chandra was previously
investigated by \citet{desjardins14}.
For the group
they obtained an upper
limit for the thermal component of L$_{\rm X}$ $<$  
5 $\times$ 10$^{39}$ erg~s$^{-1}$ assuming Galactic absorption.  This is consistent with
our non-detection of the MEKAL component for Arp 283 assuming Galactic absorption.   
Including internal absorption, however, 
we detect a MEKAL component (Tables 4 and 5).
Inspection of the Chandra image shows diffuse emission 
in the disk of NGC 2798, the western galaxy in the Arp 283 pair.

{\bf Arp 284 (NGC 7714/5):}
The more western galaxy in Arp 284, NGC 7715, has a partial ring
and a long tidal tail stretching to the  west; the companion
NGC 7714 is an edge-on disk connected to NGC 7714 by a bridge
\citep{arp66}.
Numerical simulations suggest that the ring was
produced by an off-center collision between the two
galaxies \citep{smith92, struck03}.
Because of the tidal features we list Arp 284 as a stage 2 merger.
The nucleus of NGC 7714 is the prototypical nuclear starburst
\citep{weedman81}, while NGC 7715 is classified as a post-starburst  
\citep{bernlohr93}.

A complete analysis of the Chandra data for Arp 284 was
conducted by \citet{smith05}.
In the interior of NGC 7714, they noted diffuse hot gas
outside of the compact nucleus.  
They fit the spectrum of this hot gas with a thermal (MEKAL)
component and a hard (power law) component.
For a 9\farcs5 radius region
near the center of NGC 7714 but excluding the nucleus,
they measured a total diffuse MEKAL 0.3 $-$ 8 keV L$_{\rm X}$ 
of 1.1 $\times$ 10$^{40}$
erg~s$^{-1}$. They found a best-fit kT = 0.59 $\pm$ $^{0.05}_{0.06}$ keV
and an absorbing column
of 5 $\pm$ $^{4}_{3}$ $\times$ 10$^{20}$ cm$^{-2}$.
This absorption is consistent with our estimate from the UV/IR
ratio as well as our best-fit value.
Further out in the disk, \citet{smith05} 
detected a total of 1.1 $\times$ 
10$^{40}$ erg~s$^{-1}$ of resolved emission 
associated with four extra-nuclear
H~II regions.
In the current study, our best-fit absorption-corrected
MEKAL luminosity for the entire system 
is about three times the sum of the fluxes calculated
by \citet{smith05}.

{\bf Arp 293 (NGC 6285/6):}
Arp 293 consists of two widely separated disk galaxies
with short tails \citep{arp66}.
We list this system as a stage 2 merger.
Together, their FIR luminosity is 10$^{11.1}$ L$_{\sun}$.
Of this, about 80\% comes from the 
southeastern galaxy NGC 6286 \citep{howell10}.
NGC 6286 
is a peculiar edge-on disk which was classified as a LINER
by \citet{veilleux95} based on optical spectra, and as a composite
AGN/starburst by \citet{yuan10}.
Based on infrared spectra, \citet{dixon11} conclude
that less than half
of the infrared luminosity of NGC 6286 is powered by
an AGN.
\citet{ricci16} analyzed the archival Chandra data
for Arp 293 jointly with XMM-Newton and NuSTAR data,
and concluded that NGC 6286 contains a buried AGN.  This AGN, however,
has a low intrinsic luminosity, and 
contributes less than 1\% of the bolometric luminosity of the system 
\citep{ricci16}.
The companion
nucleus NGC 6285 was undetected by NuSTAR \citep{ricci16}. 
In Figure 21, we display the diffuse X-ray spectrum for Arp 293,
along with the best-fit model.

{\bf Arp 295:}
Arp 295 is a very widely separated pair of spirals connected by
a long bridge \citep{arp66}.
The morphology of this pair was reproduced
in the \citet{toomre72} paper.  Arp 295 was included in
the \citet{hibbard96} survey of HI in tidal tails,
as well as in Spitzer and GALEX imaging studies of
nearby Arp systems \citep{smith07, smith10}.
The southwestern galaxy in the pair, Arp 295A, is listed as a
possible LINER in NED, while the second galaxy, Arp 295B,
is listed as having an H~II region optical spectrum
\citep{corbett03}.
The Chandra data for Arp 295 has not yet been published.
Our best fit for the
X-ray spectrum indicates a highly extincted power law component,
perhaps due to an obscured AGN (see Table 5 
and Figure 21).

{\bf Arp 299 (NGC 3690):}
Arp 299 is a mid-merger system with two disks 
in very close contact \citep{arp66}, thus we classify it as
a stage 3 merger.
Arp 299 is a LIRG with a very high implied SFR of 132 
M$_{\sun}$~yr$^{-1}$.

High energy X-ray observations with the NuStar satellite
reveal a highly absorbed hard source, likely an AGN, 
associated with the western nucleus Arp 299-B \citep{ptak15}.
The Chandra data for Arp 299 were previously
published by \citet{ptak15} and \citet{anastasopoulou16}.  After
removal of the point sources, \citet{anastasopoulou16} fit the spectrum
to an absorbed power law plus thermal spectrum.
After correction for Galactic absorption, 
they found that the 
diffuse emission has a total 0.5 $-$ 8 keV luminosity of 
2.06 $\pm$ 0.02 $\times$ 10$^{41}$ erg~s$^{-1}$,
and the total 0.1 $-$ 2.0 keV luminosity is 
1.70 $\pm$ 0.02 $\times$ 10$^{41}$ erg~s$^{-1}$.
They quote a measured 0.3 $-$ 10 keV
luminosity for the diffuse thermal component
of 1.89 $\times$ 10$^{41}$ erg~s$^{-1}$.
This is about three times larger than our Galactic-absorption-corrected
MEKAL luminosity, and 60\% our value after correction
for internal absorption.
Their best-fit temperature for the thermal component
is kT = 0.72 $\pm$ 0.03 keV.
Inspection of the Chandra map shows considerable diffuse
emission outside of the nuclear regions, however,
this diffuse emission is contained within the $\mu$$_{\rm B}$ = 25.0 
magnitudes~arcsec$^{-2}$ isophotes as seen in the SDSS g
image.

{\bf IRAS 17208$-$0014 (IRAS F17207$-$0014):} 
IRAS 17208$-$0014 is a late-stage merger remnant 
with short tails visible on the SDSS images.
We classify it as a stage 6 merger since
only a single 
nucleus is visible in HST near-infrared images
\citep{haan11}, however, we note that adaptive
optics observations with the Keck telescope may separate 
the two nuclei \citep{medling14}.
It has an R$^{1/4}$ profile in the K band \citep{zenner93}.
IRAS 17208$-$0014 is a ULIRG, with L$_{\rm FIR}$ = 10$^{12.19}$ L$_{\sun}$
and SFR = 149 M$_{\sun}$~yr$^{-1}$.
Although IRAS 17208$-$0014 has sometimes been listed as a LINER
\citep{rupke05}, 
its optical spectrum has also been classified as that
of an H~II region galaxy
\citep{veilleux95, yuan10}.
Its mid-infrared spectrum
is consistent with that of a starburst \citep{farrah07,
stierwalt13}.

The Chandra data for IRAS 17208$-$0014 were previously analyzed
by \citet{iwasawa11}, who found no evidence
for an obscured AGN in this galaxy based on the X-ray data.
From the X-ray spectrum, they derived an absorbing column of 
5.2 $\pm$ 1.0 $\times$ 10$^{21}$ cm$^{-2}$, 
consistent with our estimate of 7.2 $\times$ 10$^{21}$
cm$^{-2}$ based on the UV/IR data.
They obtain a thermal temperature of kT = 0.79 $\pm$ $^{0.12}_{0.10}$
keV, and noted that the X-ray emission extends out to a radius
of 11$''$.  After correction
for Galactic absorption and scaling to our distance,
they found a soft X-ray luminosity (0.5 $-$ 2 keV)
of 9.9 $\times$ 10$^{40}$ erg~s$^{-1}$ and a hard (2 $-$ 10 keV)
luminosity of 2.2 $\times$ 10$^{41}$ erg~s$^{-1}$.
Together, these are 60\% larger than our total 0.3 $-$ 8 keV 
luminosity assuming only Galactic absorption.

{\bf Mrk 231 (IRAS 12540+5708):} 
The SDSS images of Mrk 231 show two moderate-length low surface brightness 
($\mu$$_{\rm B}$ $\sim$ 25 mag~arcsec$^{-2}$) tails and only a single nucleus,
thus we classify it as a stage 5 merger.
Mrk 231 has the highest L$_{\rm K}$ in our sample (10$^{12.36}$ L$_{\sun}$) and 
the second highest L$_{\rm FIR}$ (10$^{12.13}$ L$_{\sun}$).  
Mrk 231 has been classified as a Seyfert 1 galaxy and as a low-ionization broad absorption 
line QSO (e.g., \citealp{boroson92}).   It is included in the \citet{healey07}
catalog of flat-spectrum radio sources.
Millimeter \citep{feruglio10} and FIR \citep{sturm11} 
spectroscopy shows evidence for fast ($\sim$1000 km~s$^{-1}$) 
molecular outflows from Mrk 231, possibly driven by AGN feedback.
Optical imaging spectroscopy shows that neutral gas winds extend 
out to at least 3 kpc from the nucleus \citep{rupke11}.
Outside of the nucleus, a star-forming disk has been observed
\citep{taylor99, davies04}.   This disk is estimated to
contributed 25\%$-$40\% of the bolometric
luminosity of the galaxy \citep{davies04}.
Using multi-wavelength observations and population synthesis, \citet{davies04}
estimate a SFR of $\sim$125 M$_{\sun}$~yr$^{-1}$.  This is about 1/4 the
SFR we get using the standard UV+IR relation (Table 4).  This suggests that the powerful
AGN in this system may contribute significantly to powering the UV and IR measurements,
causing our SFR to be over-estimated.
\citet{veilleux09} and \citet{nardini10} independently estimated 
the fraction of the bolometric luminosity powered by star formation
to be 29\% and 66\%, respectively.

\citet{gallagher02} analyzed the first 40 ksec exposure of the 
Mrk 231 
Chandra data, and noted that the nucleus was strongly variable.
They found that diffuse soft X-ray 
light is visible out $\sim$25$''$ from the nucleus.
Assuming solar metallicity,
a reasonable fit 
to its spectrum could be made
by either two Raymond-Smith plasma models
or by one Raymond-Smith component and a power law component.
In the former case, the best-fit temperatures were 
kT = 0.30 $\pm$ $^{0.07}_{0.05}$ keV and 
1.07 $\pm$ $^{0.22}_{0.18}$ keV, while in the latter case
the best-fit temperature was 0.80 $\pm$ $^{0.07}_{0.11}$ keV.
Using Galactic absorption,
they measured the diffuse 0.5 $-$ 2 keV luminosity 
within a 25$''$ radius to be 1.2 $\times$ 10$^{41}$ erg~s$^{-1}$.
This is in reasonable agreement with our determination.

An early subset (25.3 ksec) of the Chandra data for Mrk 231 
was analyzed by \citet{grimes05}.  
For the diffuse X-ray emission outside of a 3\farcs34 radius
assuming Galactic absorption,
they obtained a temperature of kT = 0.73 $\pm$ $^{0.13}_{0.09}$ keV
and a thermal 0.3 $-$ 2 keV luminosity of 2.6 $\times$ 10$^{41}$
erg~s$^{-1}$.  
For the diffuse emission inside this radius,
they found a thermal L$_{\rm X}$ of 2.4 $\times$ 10$^{41}$ erg~s$^{-1}$,
kT = 0.73 $\pm$ $^{0.13}_{0.09}$ keV, and N$_{\rm H}$ =  
7 $\pm$ $^{3}_{4}$ $\times$ 10$^{21}$ cm$^{-2}$.
This absorbing column agrees with our estimate for the full system
using the UV/IR ratio, but is about 10 times larger than our best-fit
value for the total diffuse X-ray spectrum.  Their total thermal luminosity
agrees with our best-fit value.

The full Chandra dataset for Mrk 231 was analyzed
by \citet{veilleux14}.  They divided the extended light into
four annuli with inner radius of 1 kpc and outer edge of 40 kpc,
and fit the spectrum of each annuli to various functions, including 
a MEKAL plus power law.   Keeping the
temperature fixed to kT = 0.67 keV and adding a second
fixed temperature MEKAL component of kT = 0.27 keV to the outer
two annuli and only including Galactic absorption, 
their total MEKAL luminosity was 2.4 $\times$ 10$^{41}$ erg~s$^{-1}$.
This is about a factor of two larger than our value
assuming Galactic absorption.
\citet{veilleux14} find that a shock model fits the X-ray data equally
well.
A joint Chandra/NuSTAR X-ray analysis by \citet{teng14} found
an intrinsic MEKAL 0.5 $-$ 30 keV luminosity of 2.4 $\times$ 10$^{41}$ erg~s$^{-1}$,
with two temperature components at 0.26 keV and 0.87 keV.

{\bf Mrk 273 (UGC 08696; IRAS 13428+5608):}
In the SDSS optical images of Mrk 273, a long 
straight high surface brightness tail extends 1$'$ to the south, with a weaker
plume extending to the northeast.
In near-infrared images, at least
two nuclei are visible \citep{majewski93, knapen97, scoville00},
thus we classify it as a stage 4 merger.
Mrk 273 has a high L$_{\rm FIR}$ of 10$^{11.90}$ L$_{\sun}$ but
a less extreme L$_{\rm K}$ of 10$^{11.44}$ L$_{\sun}$, and a high inferred
SFR of 130 M$_{\sun}$~yr$^{-1}$.
At least one of the nuclei of Mrk 273 is an AGN, based on its 
optical spectral classification as a Seyfert 2 system \citep{khachikian74, koski78}.
Mrk 273 was detected in the mid-infrared [Ne~V]~14.32 $\mu$m line \citep{farrah07}, confirming
the presence of an AGN.

The Chandra data for Mrk 273 were previously analyzed by 
several groups.  \citet{xia02} identified a compact hard X-ray point source
coincident with the northern nucleus as seen in near-IR maps,
and concluded that its X-ray spectrum
is consistent with that of an obscured AGN.  
The existence of an obscured AGN in Mrk 273 was further confirmed by \citet{teng15} using
Chandra data in conjunction with NuSTAR observations, while \citet{iwasawa17}
used Chandra+NuSTAR data to argue that both nuclei are AGN.
Outside of the nucleus, \citet{xia02} found bright soft X-ray
emission extending to about 10$''$ (7.8 kpc) from the nucleus.  Beyond this,
faint `halo' emission is seen, spanning 
52$''$ $\times$ 33$''$ (108 kpc $\times$ 68 kpc) 
\citep{xia02}.
Most of this `halo' emission lies near the southern tidal tail
and within the $\mu$$_{\rm B}$ = 25.0 mag~arcsec$^{-2}$
contours indicated by the SDSS g image, 
thus it is included in our flux measurements.
Outside of the 
$\mu$$_{\rm B}$ = 25.0 mag~arcsec$^{-2}$ isophote, some faint emission
is visible to the east of the southern tail, however, this does not contribute
much to the total light of the system.
Fitting a MEKAL + power law + 6.7 keV line model to the diffuse
X-ray emission within 10$''$, for the MEKAL component
\citet{xia02} found a best-fit temperature of 0.77 $\pm$ $^{0.09}_{0.04}$ keV
and an absorbing column of 
1.56 $\pm$ $^{1.29}_{1.56}$ $\times$ 10$^{21}$ cm$^{-2}$,
giving an absorption-corrected 0.1 $-$ 10 keV luminosity of 2.6 $\pm$ $^{0.32}_{0.37}$
$\times$ 10$^{41}$ erg~s$^{-1}$.
For this same $\le$10$''$ radius region but only 
correcting for Galactic absorption, 
\citet{iwasawa11} found a thermal temperature of kT = 0.58 $\pm$ 0.04 keV and 
a soft X-ray (0.5 $-$ 2 keV) luminosity
of 5.6 $\times$ 10$^{40}$ erg~s$^{-1}$ for this
emission.  
For the more extended `halo' gas, \citet{xia02} found N$_{\rm H}$ = 
0.30 $\pm$ $^{0.71}_{0.30}$ $\times$ 10$^{21}$ cm$^{-2}$,
kT = 0.62 $\pm$ $^{0.07}_{0.13}$ keV, and 1.9 $\times$ 10$^{41}$ erg~s$^{-1}$.
Another independent analysis of the Chandra data was done by
\citet{grimes05}, who fit the spectrum of the diffuse X-ray light outside of
a central 4\farcs75 $\times$ 3\farcs74 region, and found 
a thermal component with kT = 0.56 $\pm$ $^{0.06}_{0.06}$ keV
and a 0.3 $-$ 2.0 keV luminosity (only including Galactic
absorption) of 3.4 $\times$ 10$^{41}$ erg~s$^{-1}$.
For the diffuse gas at smaller radii, using a VMEKAL+power law
model \citet{grimes05}
found kT = 0.80 $\pm$ $^{0.06}_{0.04}$ keV, 
MEKAL luminosity of 1.5 $\times$ 10$^{41}$ erg~s$^{-1}$,
and column 
N$_{\rm H}$ = 39.2 $\pm$ 
$^{3.3}_{4.6}$ $\times$ 10$^{22}$ cm$^{-2}$.
The \citet{grimes05} absorbing column for the inner region is about 100 times larger than what we 
estimate from the UV/IR data, and 6000 times our best-fit value for
the whole system.  
Our estimate of the 
total absorbing column 
from the global UV/IR fluxes is
2.7$\sigma$ larger
than the \citet{xia02} best-fit value from the X-ray spectrum of the inner diffuse emission,
but our best-fit value of N$_{\rm H}$(MEKAL) is lower, consistent with the Galactic
value and with the \citet{xia02} fit for the extended halo.

Our derived X-ray luminosity for the hot gas is about twice that 
obtained by \citet{iwasawa11}, only including Galactic absorption 
in both cases.
This may be because we included the halo gas as well, 
while they only included
the bright inner region.
In contrast, our estimate with only Galactic absorption is about
a third that of \citet{grimes05}, however, they did not include 
a power law component.
Our best-fit absorption-corrected MEKAL 
0.3 $-$ 8 keV luminosity 
is about 1/4th the \citet{xia02} total MEKAL luminosity (both inner and halo) 
of 4.5 $\times$ 10$^{41}$ erg~s$^{-1}$.

{\bf NGC 34 (NGC 17; VV 850; IRAS 00085$-$1223; Mrk 938):}
NGC 34 is a mid-merger system with a long straight tail extending to the
northeast, and a second fainter tail to the southwest \citep{mazzarella93,
schweizer07}.
Only one nucleus is seen in the near-infrared \citep{haan11},
thus we classify it as a stage 5 merger.
NGC 34 has high FIR and K band luminosities of 10$^{11.18}$ L$_{\sun}$
and 10$^{11.14}$ L$_{\sun}$, respectively.
NGC 34 has been variously listed as a Seyfert 2 galaxy \citep{dahari85, veilleux95, veron06}
or a 
LINER \citep{osterbrock83}.
According to \citet{esquej12}, the far-infrared light is mainly powered
by a starburst.
However, a joint analysis of the Chandra, NuSTAR, and XMM-Newton data for NGC 34
indicates a highly obscured AGN \citep{ricci17}.

{\bf NGC 1700:}  
NGC 1700 is an elliptical-like galaxy (classified as E4 in NED)
with fine structure 
in its outer reaches
variously classified as tidal tails 
\citep{brown00},
shells \citep{forbes92b}, or an outer ring 
\citep{franx89}.
Based on its morphology,
\citet{schweizer92} and \citet{brown00}
conclude that NGC 1700
is the result of a major merger (i.e., two spirals).
However, the existence of a counter-rotating core 
with a younger stellar population led \citet{statler96}
and \citet{kleineberg11}
to conclude that NGC 1700 is the result of
the merger of 3 or more systems:
a minor merger that produced the counter-rotating core, and 
a gas-rich major merger, producing the large scale structure.
In the current study, we assume NGC 1700 is the product of a major
merger, and classify it as a stage 7 system.

Based on optical BVI imaging, \citet{brown00} suggest an age for the younger
stellar population in NGC 1700 of 
about 3 Gyrs.  Using HST and Gemini optical and IR, \citet{trancho14}
concluded that 
NGC 1700 hosts globular clusters with
intermediate ages (1 $-$ 2 Gyr).
The optical radial profile 
provides evidence for a stellar disk in the interior
\citep{franx89,goudfrooij94}.

The Chandra data for NGC 1700 used in the current
study was previously utilized by \citet{statler02}
to study the extended X-ray emission in NGC 1700.
They find a very flattened distribution,
and conclude that the hot gas is in a rotating disk.
They suggest that this gas may have been acquired during a merger.
Their total 0.3 $-$ 7 keV luminosity 
of 8.37 $\pm$ 0.32 $\times$ 10$^{40}$ erg~s$^{-1}$
(obtained only using Galactic absorption)
agrees well with our Galactic-absorption-only value.

{\bf NGC 2207/IC 2163:}  The two disks of this pre-merger pair are
in contact \citep{struck05}, thus we classify it as a stage 3 system, although strong tidal
tails are not visible in optical images.  The western galaxy in the pair,
NGC 2207, hosts a high SFR (1.7 M$_{\sun}$~yr$^{-1}$)
IR-luminous knot of star formation in its outer disk
near the base of a short tail-like structure \citep{elmegreen06, smith14}.
This source contributes 
about 12\% of the total 24 $\mu$m flux of the galaxy \citep{elmegreen06}.

The Chandra data for this system were previously analyzed by
\citet{smith14} and 
\citet{mineo14}.   
\citet{smith14} focused on 
the luminous star forming region in the west, noting
that it is extended in the X-ray with a soft spectrum.  They 
found a total 0.3 $-$ 8 keV luminosity
of 2.5 $\times$ 10$^{40}$ erg~s$^{-1}$ for this region,
after correcting for a total (Galactic plus internal) absorption
of N$_{\rm H}$ = 5.9 $\times$ 10$^{21}$ cm$^{-2}$ 
obtained from the H$\alpha$/24 $\mu$m ratio.
\citet{mineo14} extracted and fit the spectrum of the diffuse X-ray light for the entire
system, and found a thermal component with kT = 0.28 $\pm$ $^{0.05}_{0.04}$ keV,
a total (internal plus Galactic) absorption of N$_{\rm H}$ = 2.0 $\pm$ 1.4 $\times$ 10$^{21}$
cm$^{-2}$,
and an 0.5 $-$ 2 keV L$_{\rm X}$ of 7.9 $\times$ 10$^{40}$ erg~s$^{-1}$.
Their absorbing column 
is about twice the value we obtained from the UV/IR ratio but within the uncertainties;
their luminosity is  a factor of two higher than our determination.

{\bf NGC 2865 (AM 0921-225):} 
NGC 2865 is an elliptical with shell structures
\citep{malin83, fort86} and a short tidal tail or polar ring \citep{whitmore90}.
It was identified as a merger remnant by \citet{hau99},
who noted a kinematically distinct core.
The UV morphology was studied
by \citet{rampazzo07}.
For the current study, we assume NGC 2865 is a stage 7 major merger, although there is some
uncertainty as to the mechanism that produces shell galaxies.
An HST and Gemini study by \citet{trancho14}
concluded that NGC 2865 hosts a 1.8 $\pm$ 0.8 Gyr old
population of globular clusters.
\citet{longhetti00}
estimate
a stellar population age of $\le$1 Gyr, based on H$\beta$ and other optical
line indices.

The archival Chandra data for NGC 2865 
was previously used in several studies.
The X-ray point sources were studied by \citet{liu11}, while the hot gas
as seen by Chandra was studied by several groups 
\citep{sansom06, fukazawa06, diehl05, diehl07, mulchaey10}.
\citet{sansom06} extracted the spectrum of the
diffuse X-ray light out to a radius of 100$''$ (5 effective radii),
and successfully fit 
the spectrum
with a MEKAL plus Bremsstrahlung model.
The temperature of the latter 
component was 
fixed at 7.3 keV and the absorbing column was fixed to the Galactic value.
Their best-fit temperature for the MEKAL component was 0.32 $\pm$ $^{0.10}_{0.04}$ keV,
and their total absorbed diffuse 
0.3 $-$ 7 keV flux was 6.44 $\pm$ 10$^{-14}$ erg~s$^{-1}$~cm$^{-2}$.
This is about twice our value, however, their extraction radius was three
times larger than ours.

In an independent analysis of the same Chandra data, 
\citet{diehl07} extracted the diffuse X-ray light out to three effective radii,
and modeled contributions from unresolved point sources by a power law.
Based on their analysis, unresolved point sources dominate the diffuse
light, and they only obtain an upper limit to the 0.3 $-$ 5 keV X-ray 
luminosity from hot gas of $<$9.9 $\times$ 10$^{40}$ erg~s$^{-1}$.
In our analysis, we also find that the power law component dominates the
total flux, however, we detect the MEKAL component
at the 4$\sigma$ level, at a level well below the upper limit quoted
by \citet{diehl07}.

In a third independent analysis, \citet{fukazawa06} 
extracted the diffuse X-ray emission out a radius of 80$''$ (2.7 $\times$ our
extraction radius).  With Galactic absorption,
they find hot gas temperature of 0.33 $\pm$ 0.10 keV and a 0.2 $-$ 5 keV luminosity of 
2.0 $\times$ 10$^{39}$ erg~s$^{-1}$ and a 2 $-$ 10 keV luminosity of 1.3 $\times$ 10$^{40}$
erg~s$^{-1}$.  Their soft component is similar to ours, although
their high energy flux is about two times larger.

In yet another independent study of the same Chandra
data, \citet{mulchaey10} fit the diffuse X-ray spectrum of NGC 2865 to a 
MEKAL plus power law model with Galactic absorption, and 
find
that the thermal component of the diffuse Chandra emission has an 0.5 $-$ 2 keV
luminosity of 2.2 $\pm$ $^{0.5}_{0.3}$ 
$\times$ 10$^{39}$ erg~s$^{-1}$.  This value is in agreement with ours.
The radius of their extracted region is not given explicitly, however,
they state the radius is set equal to the maximum extent of the X-ray emission, 
which was determined
from radial profiles. 

{\bf NGC 3256 (AM 1025-433):} 
The stage 4 merger NGC 3256 has two long tidal tails
in optical photographs (e.g., \citealp{sandage94}).
Near/mid-infrared images
reveal two nuclei, with the southern being much more obscured
\citep{zenner93, kotilainen96}.
The southern nucleus may be an AGN
\citep{kotilainen96, ohyama15}
although this is uncertain \citep{lira02, lehmer15}.

The 
Chandra data for NGC 3256 has been previously
analyzed by \citet{lira02}.  Fitting to two thermal components and a power law,
they obtained temperatures for the thermal components of 0.60 $\pm$ $^{0.04}_{0.07}$ keV
and 0.91 $\pm$ $^{0.14}_{0.09}$ keV, with the former being subject to only Galactic
absorption, and the latter having N$_{\rm H}$ = 9.2 $\pm$ 2.2 $\times$ 10$^{21}$ cm$^{-2}$.
The total intrinsic 0.5 $-$ 10 keV luminosity was determined to be 6.79 $\times$ 10$^{41}$
erg~s$^{-1}$.  
Our best-fit absorbing column for the MEKAL component
is about 1/3 their value
for their second thermal component, while our total absorption-corrected luminosity
is about half theirs.
\citet{lehmer15} also studied 
the diffuse X-ray light from NGC 3256
as seen by Chandra.  They fit its spectrum
to two thermal components plus a power law.  They obtained
gas temperatures of the two thermal components of kT = 0.30 $\pm$ $^{0.06}_{0.04}$ keV 
and 0.90 $\pm$ $^{0.09}_{0.04}$ keV and an absorbing column of 7.4 $\pm$ $^{0.9}_{0.4}$ $\times$
10$^{21}$ cm$^{-2}$, twice our best-fit value for the MEKAL component.

{\bf NGC 3353 (Mrk 35; Haro 3):} In the RC3, NGC 3353 is classified as Sb? pec, while NED
lists it as a BCD/Irr HII galaxy.  It has
a low K band luminosity (10$^{9.66}$ L$_{\sun}$), a 
low L$_{\rm FIR}$ of 10$^{9.43}$ L$_{\sun}$, and a moderate SFR of 1.0 M$_{\sun}$~yr$^{-1}$.
In HST optical images, the inner region of NGC 3353 shows numerous bright knots 
\citep{malkan98}.
In SDSS images, 
\citet{mezcua14} identify two of these knots 
as galactic nuclei.
These two candidate nuclei are also visible in archival Spitzer 3.6 $\mu$m images,
with the more northern brighter at 3.6 $\mu$m. 
The northern source is also detected in the radio continuum 
spectrum \citep{johnson04}.
Whether these two sources are both galactic nuclei is uncertain;
for example, \citet{johnson04} and \citet{hunt06} 
both assume that the northern source is a luminous
star forming region, not a second nucleus.
In both the HST and the SDSS images, two disks in close contact may be present, though
that is also uncertain.
Based on the HST images, \citet{taylor05} classify NGC 3353 as a merger.
In the HST and SDSS images, a short spiral arm or tail-like 
structure extending to the southwest is
seen.  Two luminous knots
of star formation are visible near the end of this tail/arm.
Another short extension is seen to the northeast.
For the current study, we assume this system is a merger remnant, but the
merger stage is uncertain.  In spite of the possible double nucleus,
we tentatively list it as stage 6 because of the shortness of the 
tail-like structure. 

The Chandra data for NGC 3353 have not previously been published.

{\bf NGC 5018:} 
NGC 5018 is another elliptical-like galaxy with shells \citep{malin83, fort86}. 
It is listed as E3: in the RC3 and in NED, but was classified as an SAB0- by
\citet{buta10} based on Spitzer 3.6 $\mu$m images.
In short wavelength Spitzer images
\citep{kim12} and in optical images \citep{ghosh05},
a tail-like structure extends to the northwest. 
NGC 5018 was identified as 
a very late-stage major merger remnant 
by 
\citet{schweizer90} and \citet{buson04}.
We classify it as merger stage 7.
Population synthesis suggests a typical stellar age of $\sim$3 Gyr
for NGC 5018
\citep{buson04}.

The Chandra X-ray data for NGC 5018 were previously analyzed 
by \citet{ghosh05}.  They note that the nucleus hosts an X-ray point source
which may be a low luminosity AGN (intrinsic L$_{\rm X}$ $\le$ 3.5 $\times$
10$^{39}$ erg~s$^{-1}$).
They found that the 
diffuse X-ray emission is spatially coincident with strong H$\alpha$
emission, suggesting that it is powered by star formation.
\citet{ghosh05} obtained a good fit for the spectrum of the diffuse light with
a MEKAL plus power law model with Galactic absorption.
For the MEKAL component they found kT = 0.41 $\pm$ 0.04 keV.  Their 
total absorption-corrected 0.5 $-$ 8 keV luminosity for the
diffuse light was 13.7 $\pm$ 1.5 $\times$ 10$^{39}$
erg~s$^{-1}$, with about 54\% arising from the MEKAL component.
This is in reasonable agreement with our results.

{\bf NGC 5256 (Markarian 266):}
In the SDSS images,
NGC 5256 is a pair of disks in close contact surrounded by tidal debris, thus
we classify it as a stage 3 merger.
This pair has high FIR and high K luminosities (10$^{11.21}$ L$_{\sun}$ and
10$^{11.62}$ L$_{\sun}$, respectively), as well as a moderately high
SFR of 36 M$_{\sun}$~yr$^{-1}$.  The southwestern galaxy is classified
as a Seyfert 2 nucleus \citep{osterbrock83}.

The Chandra view of the diffuse X-ray light in NGC 5256 was previously investigated 
by \citet{brassington07}.   
Both nuclei are detected as X-ray point sources.  Diffuse emission is seen between
and surrounding the two nuclei.  A separate patch of diffuse emission is
seen 20$''$ to the north, coincident with tidal debris seen in the SDSS images and 
in H$\alpha$ maps \citep{brassington07}.
All of the diffuse X-ray 
emission in NGC 5256 lies within the $\mu$$_{\rm B}$ = 25.0 mag~arcsec$^{-2}$ isophote
as indicated by the SDSS g image, thus is included in
our extraction. \citet{brassington07} divide the diffuse emission
into three zones: between the two nuclei, surrounding the two nuclei, and the northern
region.   
Assuming just Galactic
absorption and fitting the X-ray spectra of the diffuse light to
single
MEKAL functions, \citet{brassington07} find temperatures of 
1.07 $\pm$ $^{0.06}_{0.09}$ keV, 0.52 $\pm$ 0.06 keV,
and 0.30 $\pm$ $^{0.02}_{0.03}$ keV for these three zones, respectively.  
Their total diffuse 0.3 $-$ 6 keV 
luminosity is 27.7 $\pm$ 0.9 $\times$ 10$^{40}$ erg~s$^{-1}$.
Our MEKAL component assuming Galactic absorption is about 25\% lower than theirs,
but we measure an additional power law component with similar luminosity.
In Figure 21, we display the diffuse X-ray spectrum for NGC 5256,
along with our best-fit model.

{\bf NGC 6240 (IRAS 16504+0228):} 
In the SDSS images, NGC 6240 has a distorted structure with a prominent
dust lane and tidal debris.
Two nuclei are seen in both the optical
\citep{fried83} and the near-infrared \citep{eales90},
thus we classify NGC 6240 as a stage 4 merger.
NGC 6240 has both high L$_{\rm FIR}$ and L$_{\rm K}$ (10$^{11.61}$ L$_{\sun}$
and 10$^{11.81}$ L$_{\sun}$, respectively) as well as a high SFR 
(98 M$_{\sun}$~yr$^{-1}$).
Based on its optical spectrum NGC 6240 is sometimes 
classified as a LINER
galaxy 
\citep{veilleux95, veron06}, however, it is sometimes
called a Seyfert galaxy (e.g., \citealp{contini13}.)
Using spatially-resolved spectra,
\citet{rafanelli97} conclude that the northern nucleus
has LINER-like line ratios, while the southern
source shows indications of a Seyfert nucleus.
Mid-infrared spectroscopy reveals the presence of a buried AGN,
however, it may contribute less than half of the total
bolometric luminosity of the system \citep{genzel98, armus06},
although this is uncertain \citep{lutz03, egami06}.
\citet{risaliti06} detected broad Br$\alpha$ emission 
(line width $\sim$ 1800 km~s$^{-1}$)
from the southern nucleus, and a steep 3 $-$ 5 $\mu$m continuum
from the northern nucleus.
They conclude that both nuclei host AGN, but
the AGN contribute little to the bolometric luminosity.

The Chandra data have previously been used to study both nuclear
activity and the diffuse X-ray light.
Using archival Chandra data, \citet{komossa03} 
concluded that both nuclei are AGN.
Hard variable X-rays indicative of an obscured AGN were also detected
by the NuSTAR satellite \citep{puccetti16}.
The diffuse emission in NGC 6240 was studied by \citet{grimes05} using
the initial 35.6 ksec of Chandra data. 
They found that 95\% of the light lies within a radius of 
62$''$ (32 kpc).
They split the emission into two annuli,
and fit the spectra for the two regions with a power law plus MEKAL spectrum.
They found a MEKAL kT = 0.87 $\pm$ $^{0.05}_{0.04}$ keV for the inner
annulus (inside an 8$''$ $\times$ 11$''$ radius), 
and kT = 0.57 $\pm$ $^{0.02}_{0.02}$ keV for the outer region.  For the inner
annulus, they also fit for N$_{\rm H}$, finding 4 $\pm$ 1 $\times$ 10$^{21}$
cm $^{-2}$; for the outer annulus, they used the Galactic value.
Their determination of N$_{\rm H}$ for the inner annulus agrees
well with our estimate from the UV/IR ratio, and is about twice our best-fit value.
They found a total 0.3 $-$ 2 keV thermal emission of 1.3 $\times$ 
10$^{42}$ erg~s$^{-1}$.
Our best-fit absorption-corrected MEKAL luminosity
agrees well with their result.
The diffuse X-ray emission from NGC 6240 was again 
studied by \citet{nardini13} using a much longer Chandra exposure.  
They note diffuse emission spread over a 225$''$ $\times$ 162$''$ diameter
region (110 $\times$ 80 kpc).  They divided the area into six azimuthal
sectors, and fitting to a thermal plus power law model
found gas temperatures between 0.65 $-$ 0.82 keV for these sectors.
They found a total 0.4 $-$ 2.5 keV luminosity 
for the hot gas of 4 $\times$ 10$^{41}$ erg~s$^{-1}$ assuming Galactic absorption
only.  This is about three times our value with only Galactic absorption.
When they included internal absorption in their fit, 
they obtained an absorbing column of
N$_{\rm H}$ = 3 $\times$ 10$^{21}$ cm$^{-2}$, similar to our best-fit estimate.
The observed extent of the diffuse X-ray light extends faintly to the west 
about 1$'$ 
beyond 
the $\mu$$_{\rm B}$ = 25.0 mag~arcsec$^{-2}$ isophote as indicated by the SDSS
emission.  
Thus our nominal aperture misses some of this
faint low level emission.
In Figure 21, we display the diffuse X-ray spectrum for NGC 6240,
along with our best-fit model.

{\bf NGC 7592 (VV 731):}
The SDSS images of NGC 7592 show two disks in close
contact, with at least two faint tidal tails.
One of the tails is very long, extending 2$'$ (58 kpc) to the
south.  
In addition to the two galactic nuclei,
another bright source is visible in the optical SDSS images
at the base of one of the tails.   
This is listed in
NED as a third galaxy (NGC 7592C).  However, 
in the Spitzer 3.6 $\mu$m image, this source is considerably
fainter than the two galactic nuclei.  
Both optical \citep{rafanelli92}
and mid-infrared \citep{haan11} spectra indicate
strong star formation at that position.
We thus argue that this third source 
is likely an extra-nuclear star forming region, perhaps akin
to the `hinge clumps' studied by \citet{smith14}, rather than
a third nucleus.
We thus follow the lead of \citet{haan11}, and classify
NGC 7592 as a major merger, since the near-infrared luminosities
of the two galaxies are similar.   We therefore list
NGC 7592 in Table 1 as 
a
stage 3 merger.
The nucleus of the western galaxy in the pair has
been identified as a Seyfert 2 nucleus,
while the
eastern nucleus has an H~II spectrum
\citep{rafanelli92, veilleux95}.
These classifications have been confirmed by 2.5 $-$ 5 $\mu$m
spectroscopy \citep{imanishi10}.

The X-ray point sources in the Chandra data for NGC 7592 were
included in the \citet{evans10} survey.  The diffuse
X-ray emission has not previously been analyzed.  We find
diffuse emission around both galactic nuclei in NGC 7592.

{\bf UGC 2238 (IRAS 02435+1253):} 
In the Spitzer 3.6 $\mu$m image, UGC 2238
has an edge-on disk-like appearance with a single nucleus
and possible tidal tails.
Only one nucleus is seen in K band images also
\citep{smith96, rothberg04}.
This galaxy was classified as a merger remnant by
\citet{rothberg04} based on its optical appearance, 
although they noted that its K band light
did not fit a
R$^{1/4}$ profile.
The tails are also visible in the \citet{rothberg04} 
K band image.   We classify this system as a stage 5 merger
since only one nucleus is visible.
UGC 2238 has a high FIR luminosity
(10$^{11.04}$ L$_{\sun}$) 
placing it in the LIRG category.
It also has a high L$_{\rm K}$ of 10$^{11.15}$ L$_{\sun}$.  
UGC 2238 is classified as a LINER
by \citet{veilleux95}.
\citet{tateuchi15} note extended Pa$\alpha$ emission over
a 3 kpc region in 
the disk.
The Chandra data for UGC 2238 have not previously been
published.  

{\bf UGC 5101 (IRAS 09320+6134):}
The SDSS images of UGC 5101 show a straight tail extending 
about 50$''$ (40 kpc) to the west, and a fainter
curved structure about the same length in the north.
Only a single nucleus is visible in the SDSS images.
We thus classify it as a stage 5 merger.
It was included in the \citet{rothberg04} catalog of 
merger remnants, who concluded that it has only one nucleus
in the near-infrared, and the K band light fits
an R$^{1/4}$ law.
UGC 5101 has a high L$_{\rm FIR}$ of 
10$^{11.72}$ L$_{\sun}$, a high L$_{\rm K}$ of 10$^{11.51}$
L$_{\sun}$, and a high inferred SFR of 128 M$_{\sun}$~yr$^{-1}$. 
UGC 5101 is classified as a Seyfert 1 galaxy by \citet{veron06},
but as a LINER by \citet{veilleux95}.
UGC 5101 was confirmed as a powerful but obscured
AGN via infrared observations
\citep{imanishi01, armus04}.

The archival Chandra data for UGC 5101 were previously
used by \citet{imanishi03} to confirm the existence
of an obscured AGN.
\citet{imanishi03} also studied the diffuse light
seen in the Chandra image.
They find an absorption-corrected 0.5 $-$ 2
keV luminosity of 1.2 $\times$ 10$^{41}$ erg~s$^{-1}$,
about 1/3rd of our value.
The diffuse hot gas in
UGC 5101 as seen by Chandra was also studied 
by \citet{grimes05}. 
They concluded that 95\% of the diffuse light is
contained within 8.75 kpc (10\farcs5).
They divided the data into two annuli 
(separated at 4\farcs28 = 3.4 kpc)
and
fitted the spectra of each annulus separately to a
VMEKAL plus power law function.
For the inner and outer regions, they obtained
thermal temperatures of 0.65 $\pm$ $^{0.08}_{0.10}$ keV
and 0.69 $\pm$ $^{0.14}_{0.11}$ keV, respectively.
For the outer annulus, they assumed only Galactic
absorption.  For the inner annulus, they fit for the
absorbing column and found N$_{\rm H}$ = 0.9 $\pm$ 
$^{1.9}_{0.7}$ $\times$ 10$^{22}$ cm$^{-2}$.   This
absorbing column agrees within the
uncertainties with our estimate from the UV/IR ratio.
\citet{grimes05}
derived a total
diffuse 
thermal 0.3 $-$ 2 keV luminosity of 1.0 $\times$ 10$^{41}$
erg~s$^{-1}$.
Our total internal-absorption-corrected
thermal L$_{\rm X}$ is about three times the \citet{grimes05}
value.
In a third independent study, \citet{huo04} analyze the diffuse
X-ray light from UGC 5101.
Only including the inner 8.7 kpc
radius (10\farcs9), a fit to the MEKAL
plus power law model gave them N$_{\rm H}$ = 3.37 $\pm$
$^{5.63}_{3.37}$ $\times$ 10$^{21}$ cm$^{-2}$, kT = 0.61 $\pm$
$^{0.16}_{0.44}$ keV, and a total 0.3 $-$ 10 keV 
absorption-corrected luminosity of 1.04 $\pm$ 
$^{0.26}_{0.24}$  $\times$ 10$^{41}$
erg~s$^{-1}$.  Their column density agrees within the
uncertainties with our UV/IR estimate, but their 
total L$_{\rm X}$ is only one quarter our value.

{\bf UGC 5189 (VV 547):} 
In SDSS and other optical images (e.g., \citealp{taylor05}), 
UGC 5189 looks like two disk galaxies connected by a broad tidal bridge.
The more western galaxy (NED01) has higher surface brightness, while 
the more eastern (NED03) is larger but lower surface brightness.   
NED also lists a small point-like object (NED02) as a third galaxy.
NED03 was found to be a low metallicity (log(O/H) + 12 = 8.24)  
star forming region by \citet{izotov06} based on its
SDSS spectrum.
Ignoring the third object, we classify UGC 5189 as a stage 2
merger.   UGC 5189 has low L$_{\rm FIR}$ (10$^{9.48}$ L$_{\sun}$) 
and L$_{\rm K}$ (10$^{9.47}$ L$_{\sun}$).  
\citet{dahari85} classify the optical spectrum
of UGC 5189 as Nuclear Emission Type (NET)
4.0, meaning an H~II region-like spectrum.
In 2010, a bright type IIn supernova, SN 2010jl, was discovered in the
western galaxy of UGC 5189
\citep{newton10, benetti10}.
The Chandra data of this supernovae was analyzed by \citet{chandra12}.
In our analysis, the supernova is detected as a point source.

\vfill
\eject

\begin{figure}
\plotone{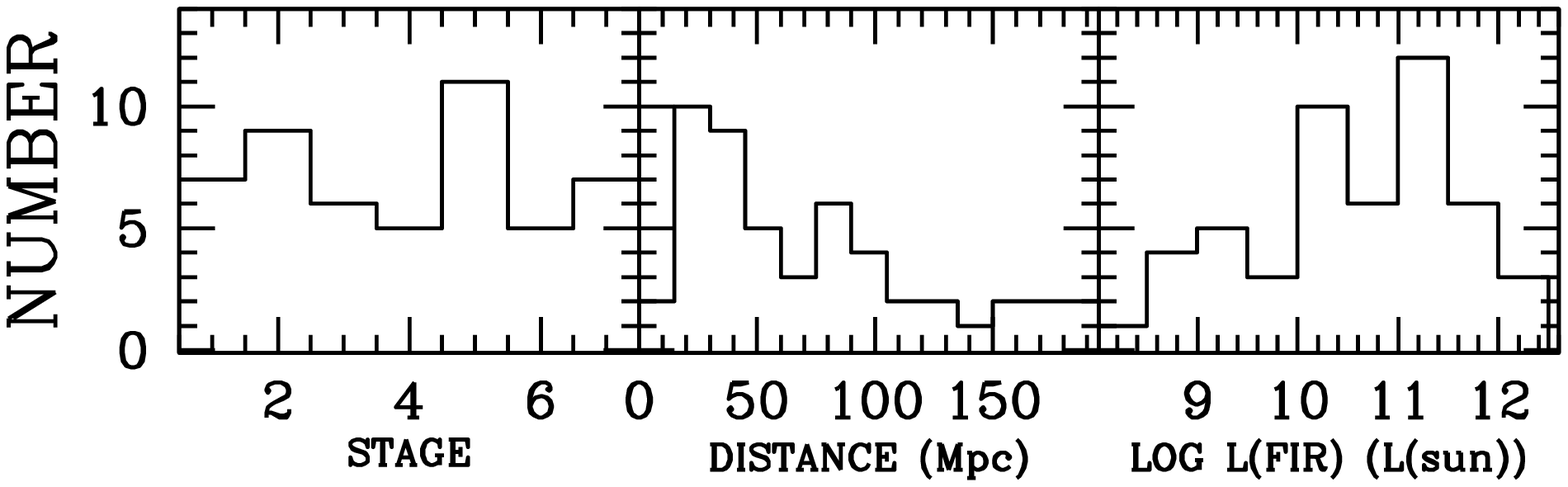}
\caption{Histograms of merger stage, distance, and far-infrared
luminosities for the sample galaxies.}
\end{figure}

\begin{figure}
\plotone{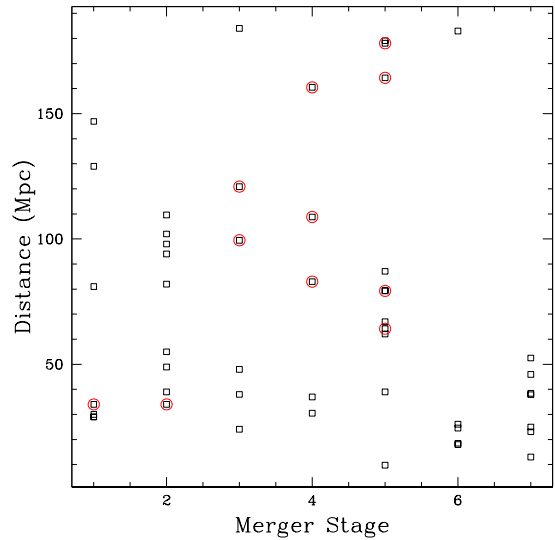}
\caption{Plot of distance vs. interaction stage.
Galaxies containing a Seyfert nucleus are circled in red.
}
\end{figure}

\begin{figure}
\plotone{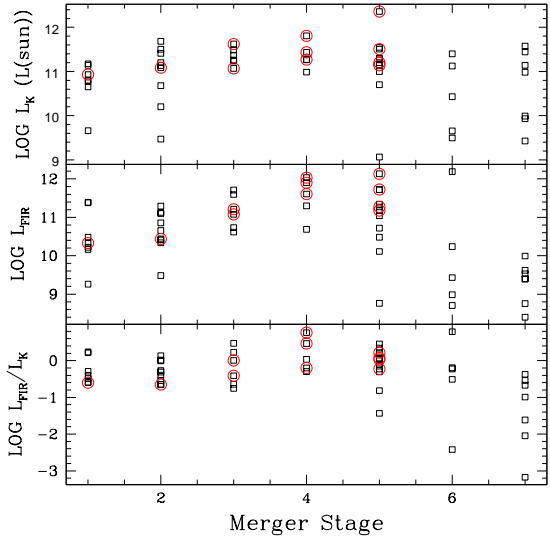}
\caption{
Top panel: plot of log(L$_{\rm K}$) vs. interaction stage.
Middle panel: plot of log(L$_{\rm FIR}$) vs. interaction stage.
Bottom panel: plot of log(L$_{\rm K}$/L$_{\rm FIR}$) vs. interaction stage.
Galaxies containing a Seyfert nucleus are circled in red.
}
\end{figure}

\begin{figure}
\plotone{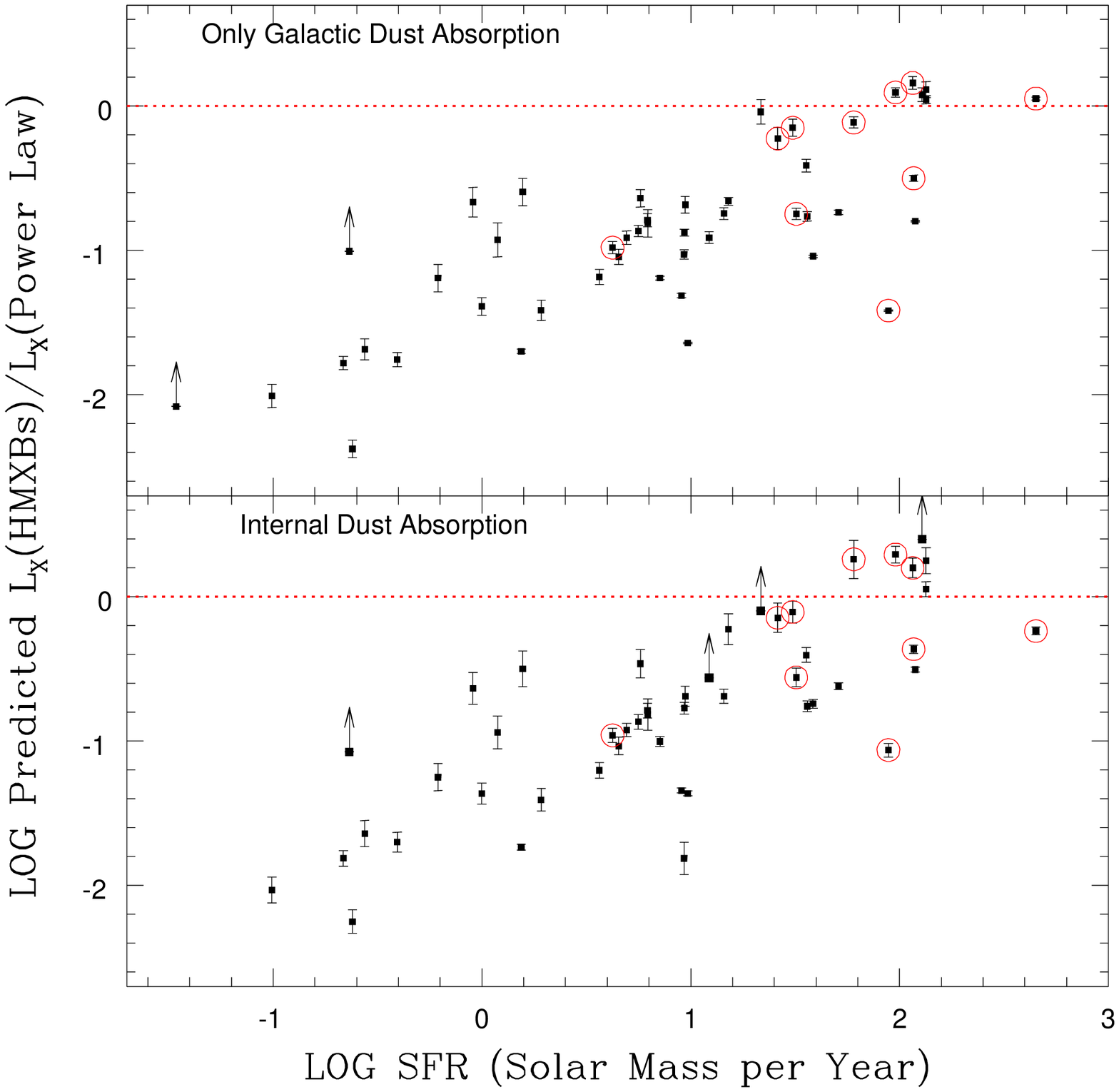}
\caption{
The 
predicted X-ray luminosity from unresolved HMXBs, 
divided by the 
absorption-corrected X-ray luminosity of the power law component, plotted against SFR.
The top panel shows the results assuming only absorption by Galactic absorption, 
while
the bottom panel includes a correction for absorption 
internal to the target galaxy.
The red line shows the expected result if all of the observed
power law flux is due to HMXBs.
The red circles mark the AGNs.
See the text of the paper for more details.
}
\end{figure}

\begin{figure}
\plotone{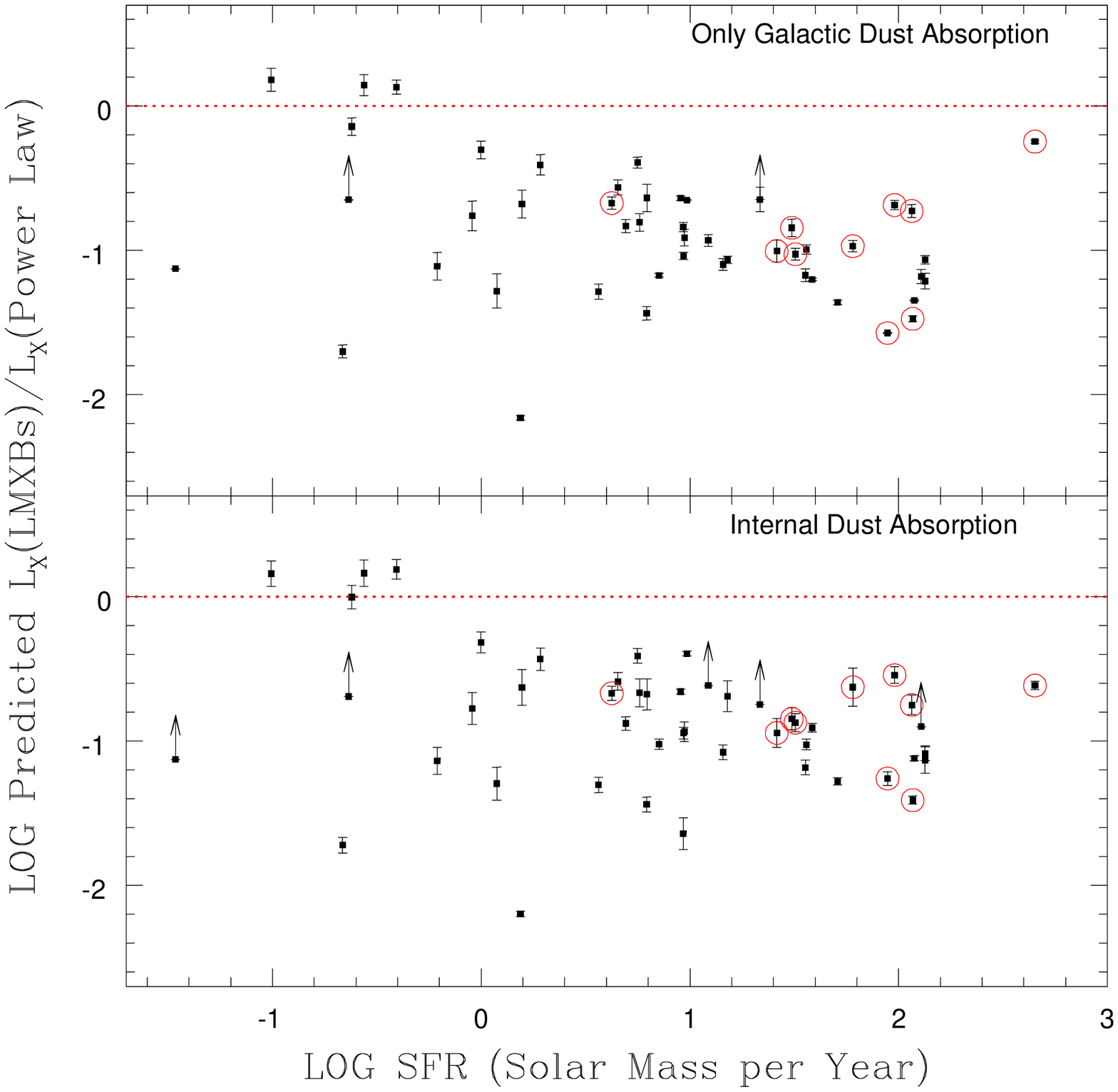}
\caption{
The predicted X-ray luminosity from unresolved LMXBs, divided by 
the absorption-corrected X-ray luminosity of the power law 
component, plotted against SFR.
The top panel shows the results assuming only Galactic absorption,
while
the bottom panel includes a correction for absorption 
internal to the target galaxy.
The red line shows the expected result if all of the observed
power law flux is due to LMXBs.
The red circles mark the AGNs.
See the text of the paper for more details.
}
\end{figure}

\begin{figure}
\plotone{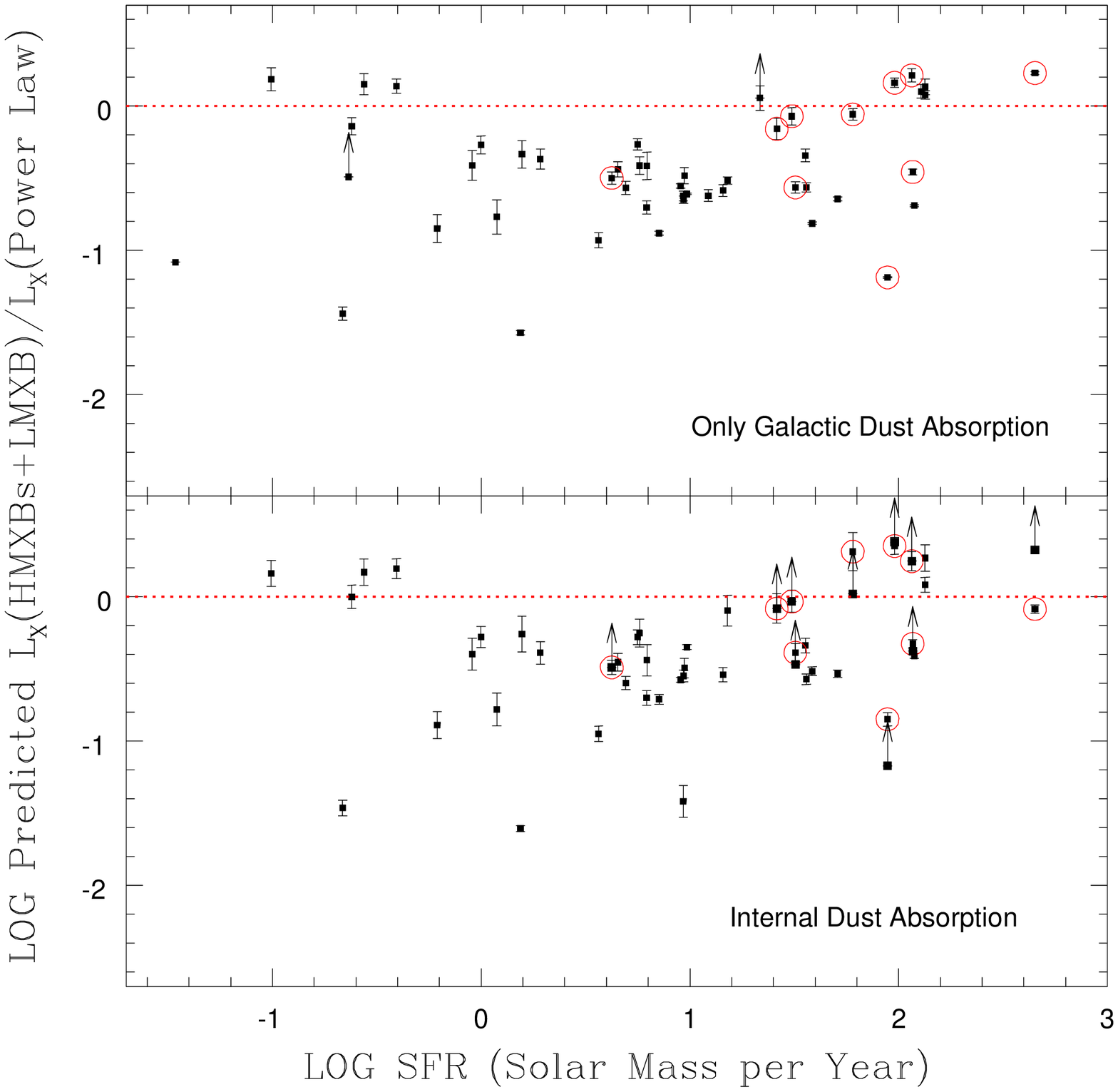}
\caption{
The predicted X-ray luminosity for both unresolved HMXBs and LMXBs, divided by the 
absorption-corrected X-ray luminosity of the power law 
component, plotted against SFR.
The top panel shows the results assuming only Galactic absorption,
while
the bottom panel includes a correction for absorption
internal to the target galaxy.
The red line shows the expected result if all of the observed
power law flux is due to HMXBs and LMXBs.
The red circles mark the AGNs.
See the text of the paper for more details.
}
\end{figure}

\begin{figure}
\plotone{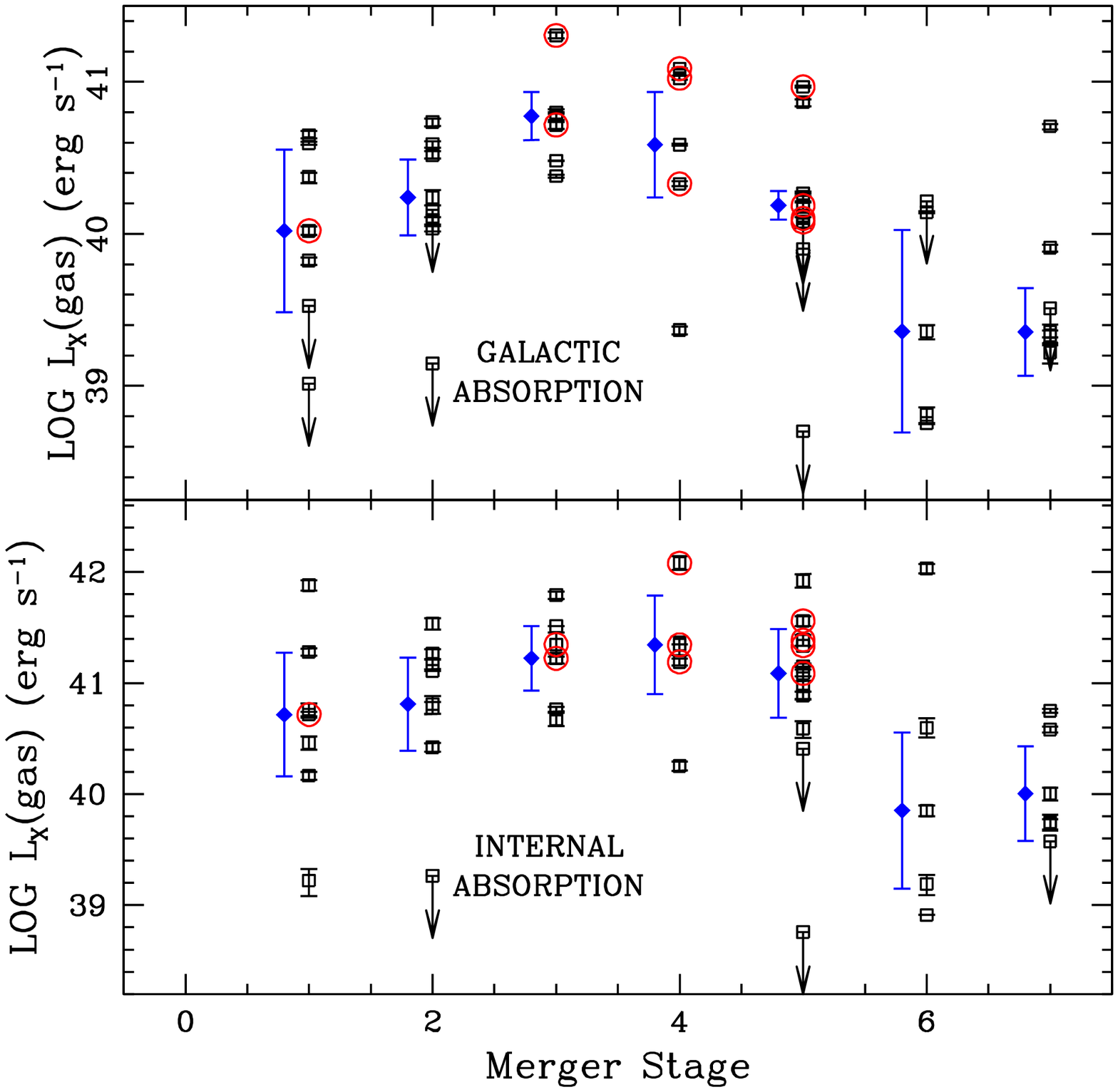}
\caption{Plot of the best-fit 
L$_{\rm X}$(gas) vs. interaction stage (open black squares), 
including the uncertainty in L$_{\rm X}$.
The top panel gives the X-ray luminosities only corrected for Galactic absorption.
The bottom panel includes internal absorption.
The filled blue diamonds are the median values for each stage, slightly
offset to the left.  
The errorbars plotted on the median values are the
semi-interquartile range, equal to half the difference between
the 75th percentile and the 25th percentile.
Upper limits above the median were not included in calculating the median.
Galaxies containing a Seyfert nucleus are circled in red.
The upper limits are 3$\sigma$.
}
\end{figure}

\begin{figure}
\plotone{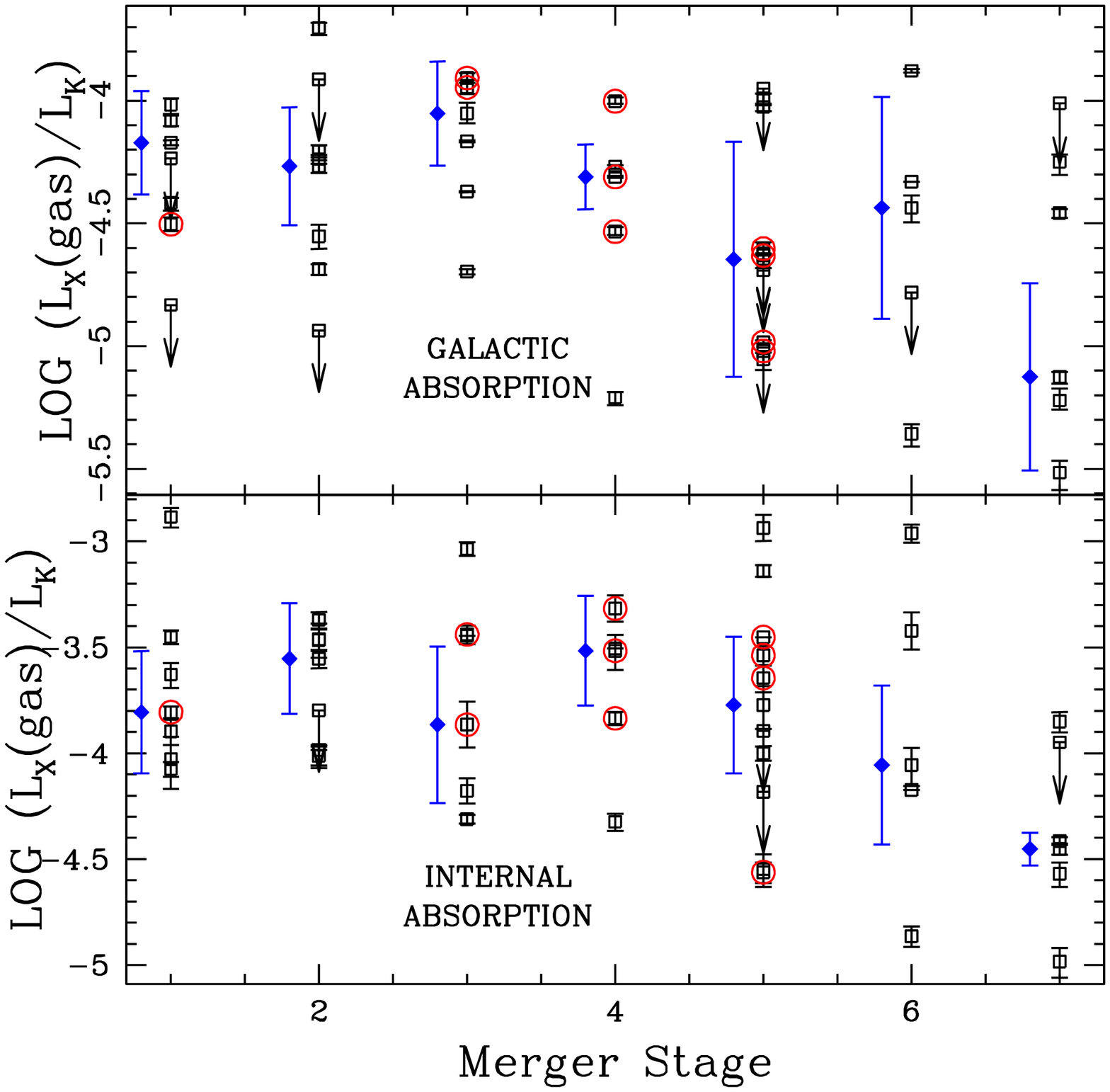}
\caption{Plot of the best-fit diffuse L$_{\rm X}$(gas)/L$_{\rm K}$ vs. interaction stage (open black squares), 
including the uncertainty in L$_{\rm X}$/L$_{\rm K}$.
The top panel gives the X-ray luminosities only corrected for Galactic absorption.
The bottom panel includes internal absorption.
The filled blue diamonds are the median values for each stage, slightly
offset to the left.  The errorbars plotted on the median values are the
semi-interquartile range, equal to half the difference between
the 75th percentile and the 25th percentile.
Upper limits above the median were not included in calculating the median.
Galaxies containing a Seyfert nucleus are circled in red.
}
\end{figure}

\begin{figure}
\plotone{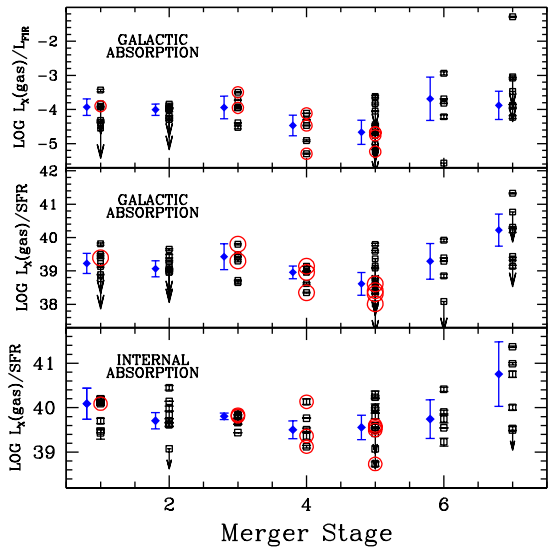}
\caption{
Top panel: plot of 
L$_{\rm X}$(gas)/L$_{\rm FIR}$ vs. interaction stage (open black squares).
The X-ray luminosity in this plot has only been corrected for Galactic 
absorption, not for material within
the host galaxy itself.
Galaxies containing a Seyfert nucleus are circled in red.
The galaxy with the highest 
L$_{\rm X}$/L$_{\rm FIR}$ ratio is NGC 1700, which we classify
as stage 7.
Middle panel: Plot of diffuse L$_{\rm X}$(gas)/{SFR} vs. interaction stage,
where 
the X-ray luminosity has only been corrected for Galactic 
absorption.
Bottom panel: plot of diffuse L$_{\rm X}$(gas)/{SFR} vs. 
interaction stage, where 
the X-ray luminosity has been corrected for absorption within
the host galaxy itself, using the UV/mid-infrared ratio as described in the text.
Note that the galaxy with the highest 
L$_{\rm X}$/L$_{\rm FIR}$
and
L$_{\rm X}$/SFR,  
the stage 7 system
NGC 1700, is not plotted in the bottom panel since no UV data is available.
In all panels, galaxies containing a Seyfert nucleus are circled in red.
The filled blue diamonds are the median values for each stage, slightly
offset to the left.  The errorbars plotted on the median values are the
semi-interquartile range, equal to half the difference between
the 75th percentile and the 25th percentile.
Upper limits above the median were not included in calculating the median.
}
\end{figure}

\begin{figure}
\plotone{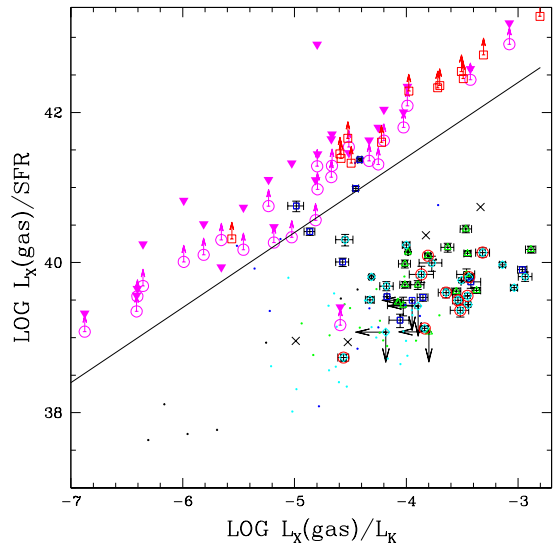}
\caption{
L$_{\rm X}$(gas)/L$_{\rm SFR}$ vs.\ 
L$_{\rm X}$(gas)/L$_{\rm K}$, for the mergers, the normal spirals, and the E/S0 galaxies.
The mergers are color-coded by merger stage, after correction for internal
absorption.
Merger stages 1 and 2
systems are marked as open green triangles.  Merger stages 3, 4, and 5
are open cyan diamonds, and merger stages 6 and 7 are identified by
blue open squares.
Merging galaxies with a Seyfert nucleus are circled in red.
Small dots of the same color represent the mergers with only Galactic absorption.
Large black crosses mark the normal spirals after correction for
internal absorption; small black dots are spirals with only Galactic absorption.
The red open squares represent the \citet{goulding16} massive ellipticals
and the \citet{su15} E/S0 galaxies are marked by magenta open circles, 
calculated 
using
SFRs derived from the UV+IR data with the \citet{hao11}
relation (these SFRs are assumed to be upper limits for the E/S0 galaxies).
The magenta filled upside down triangles mark the \citet{su15} E/S0 galaxies 
using SFRs from SED fitting.	
The black line is the relation expected if the observed 24 $\mu$m flux is powered solely by
the older stellar population (see text for more details).
The E/S0 values were obtained using only Galactic absorption. 
}
\end{figure}

\begin{figure}
\plotone{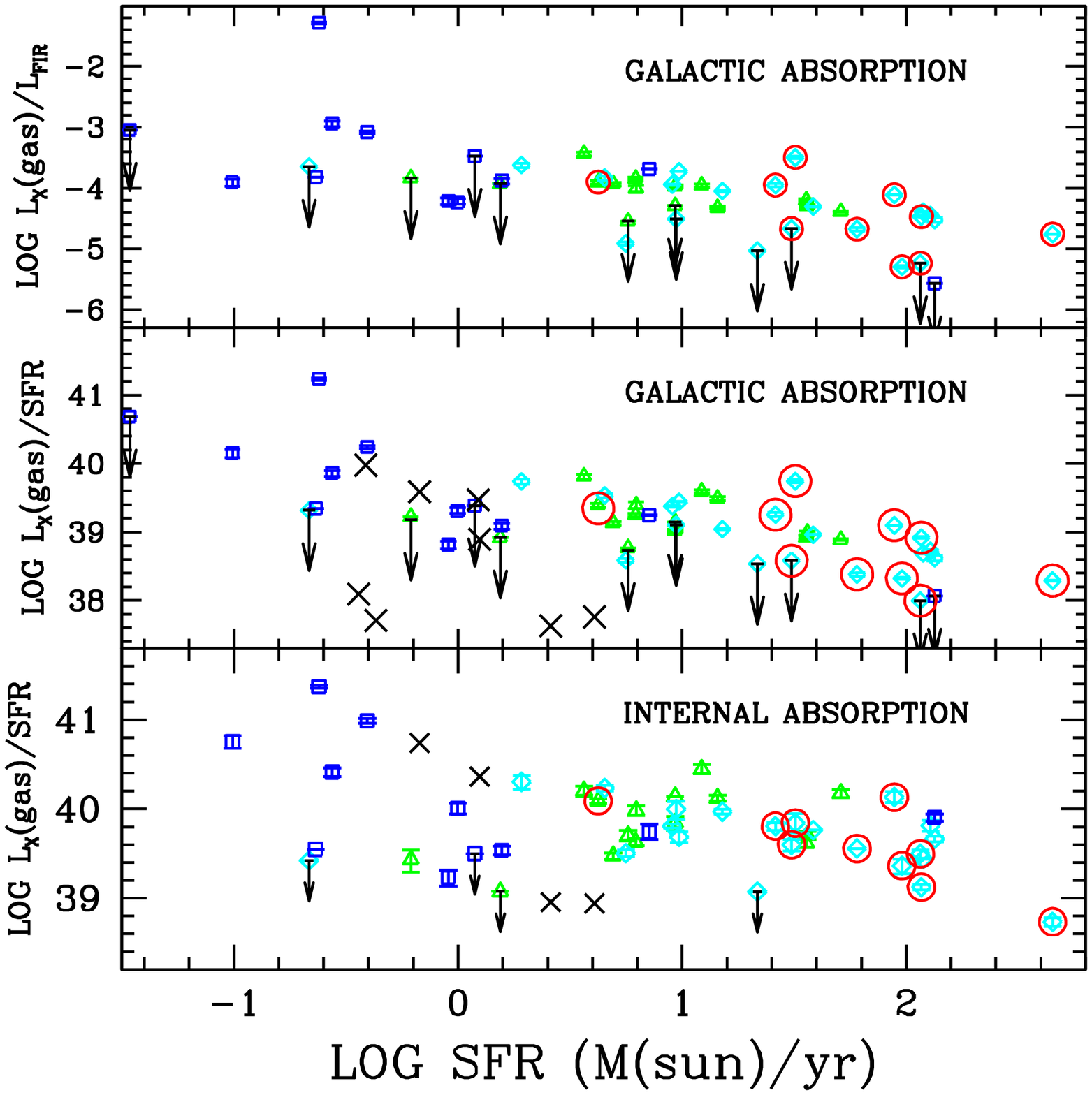}
\caption{
Top panel: Diffuse L$_{\rm X}$(gas)/L$_{\rm FIR}$,
after only correcting for Galactic absorption, plotted vs. SFR.
The datapoints are color-coded by merger stage.  Merger stages 1 and 2
systems are marked as open green triangles.  Merger stages 3, 4, and 5
are open cyan diamonds, and merger stages 6 and 7 are identified by
blue open squares.
Galaxies containing a Seyfert nucleus are circled in red.
Black crosses mark the normal spirals from Mineo et al.\ (2012).
Middle panel: 
plot of diffuse L$_{\rm X}$(gas)/SFR vs. SFR, where the
X-ray is only corrected for Galactic absorption.
Bottom panel: 
plot of diffuse L$_{\rm X}$(gas)/SFR vs. SFR, after correction for internal 
absorption.
}
\end{figure}

\begin{figure}
\plotone{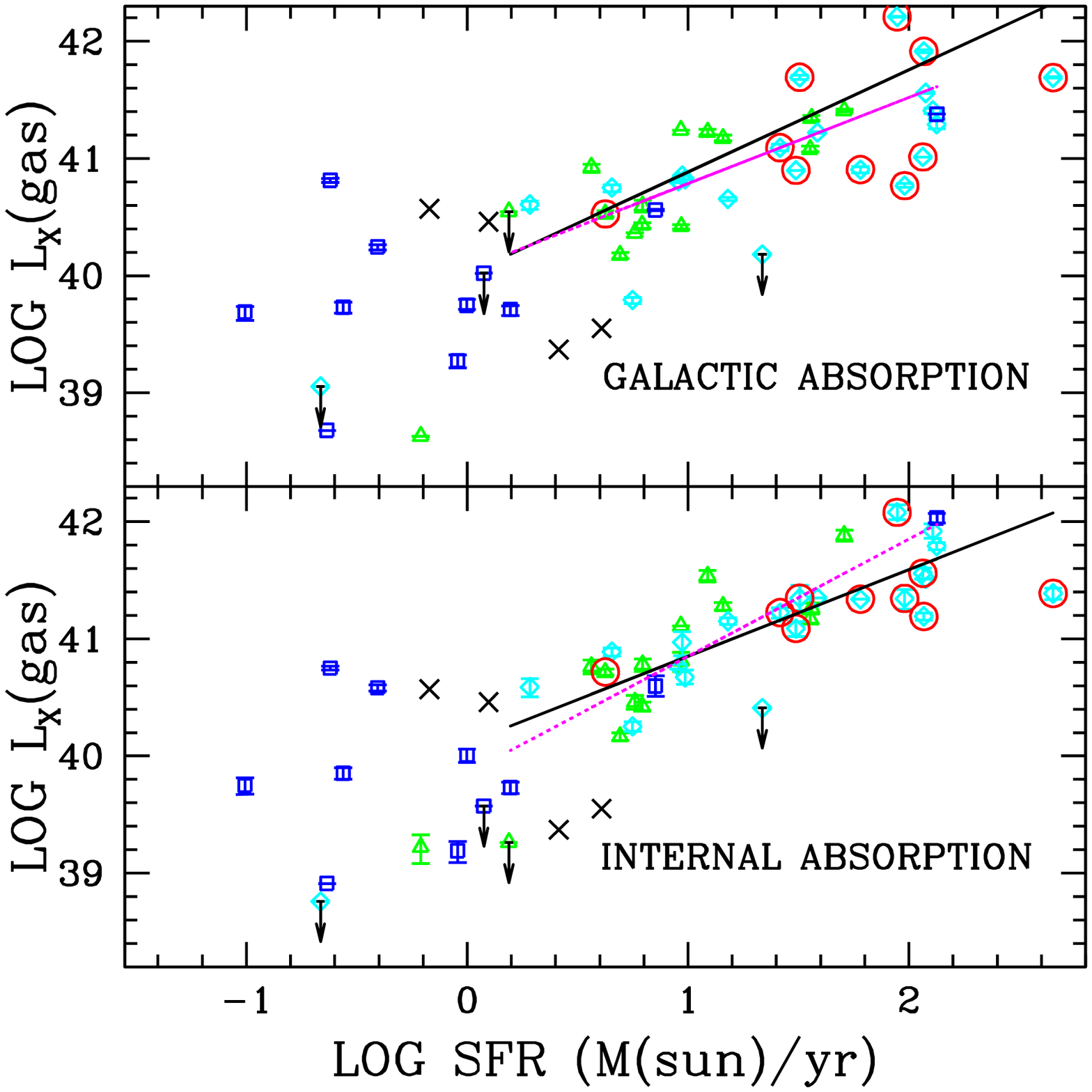}
\caption{
Top panel: Diffuse L$_{\rm X}$(gas) vs.\ SFR,
after only correcting for Galactic absorption.
The datapoints are color-coded by merger stage.  Merger stages 1 and 2
systems are marked as open green triangles.  Merger stages 3, 4, and 5
are open cyan diamonds, and merger stages 6 and 7 are identified by
blue open squares.
Galaxies containing a Seyfert nucleus are circled in red.
Black crosses mark the normal spirals from Mineo et al.\ (2012).
Bottom panel: 
plot of diffuse L$_{\rm X}$(gas) vs.\ SFR, after correction for internal 
absorption.
Best-fit lines for data points above SFR = 1 M$_{\sun}$~yr$^{-1}$ are plotted.  Black solid lines are the
fits including
all of the mergers (slope = 0.87 $\pm$ 0.12 top panel, and 0.74 $\pm$ 0.09 bottom panel), 
and dashed magenta lines when the AGN are omitted (slope = 0.74 $\pm$ 0.05 top panel
and 1.00 $\pm$ 0.10 bottom panel).
}
\end{figure}

\vfill
\eject

\begin{figure}
\plotone{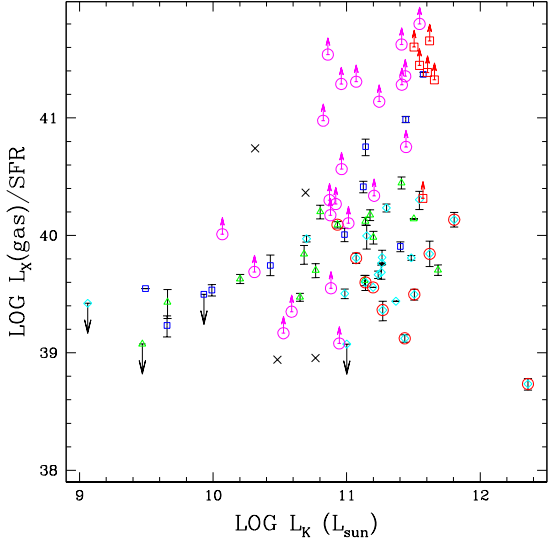}
\caption{Plot of diffuse gaseous internal-absorption-corrected
L$_{\rm X}$/SFR 
vs. L$_{\rm K}$.
The datapoints are color-coded by merger stage.  Merger stages 1 and 2
systems are marked as open green triangles.  Merger stages 3, 4, and 5
are open cyan diamonds, and merger stages 6 and 7 are identified by
blue open squares.
Here, L$_{\sun}$ represents the bolometric luminosity of the Sun, 3.9 $\times$ 10$^{33}$ erg/s.
Black crosses mark the normal spirals from Mineo et al.\ (2012).
The open red squares indicate the \citet{goulding16} ellipticals,
while the \citet{su15} E/S0 galaxies are shown by open magenta circles.
As noted earlier, for the E/S0 galaxies, the SFRs from the UV+IR data using
the \citet{hao11} formula may be upper limits, if the older stellar population is contributing
significantly to the UV or IR data.
Note that Arp 235 is not plotted, since no UV data is available.
}
\end{figure}

\vfill
\eject

\begin{figure}
\plotone{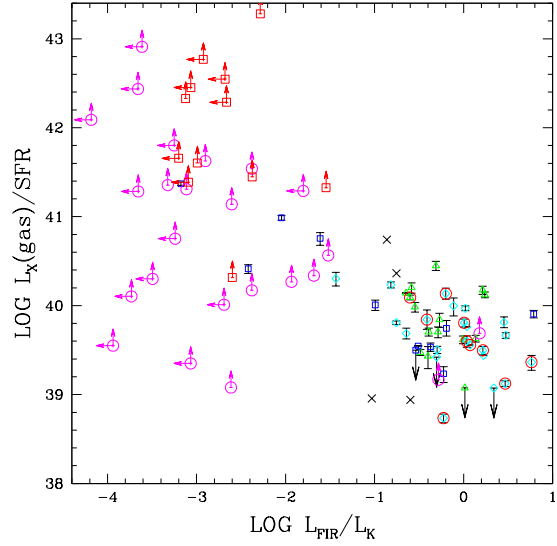}
\caption{
Plot of diffuse gaseous L$_{\rm X}$/SFR after correction for internal 
absorption vs. 
L$_{\rm FIR}$/L$_{\rm K}$.
Merger stages 1 and 2
systems are marked as open green triangles.  Merger stages 3, 4, and 5
are open cyan diamonds, and merger stages 6 and 7 are identified by
blue open squares.
The galaxy with the highest
L$_{\rm FIR}$/L$_{\rm K}$ ratio is IRAS 17208-0018, followed closely
by Arp 220.
Black crosses mark the normal spirals from Mineo et al.\ (2012).
The open red squares indicate the \citet{goulding16} ellipticals,
while the \citet{su15} E/S0 galaxies are shown by open magenta circles.
}
\end{figure}

\eject
\vfill

\begin{figure}
\plotone{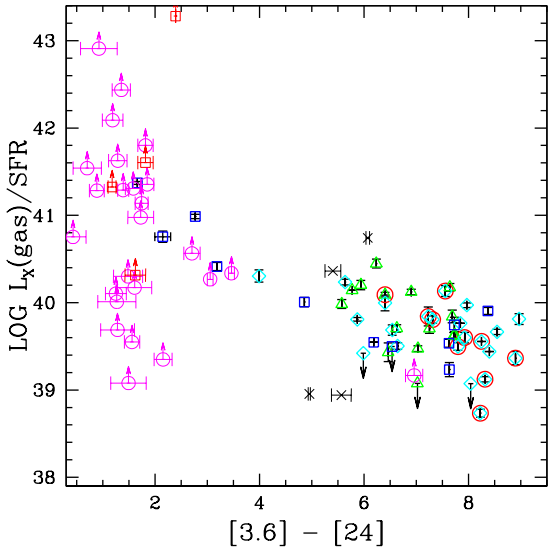}
\caption{Plot of L$_{\rm X}$/SFR vs.
[3.6] $-$ [24].
These X-ray luminosities have been corrected for internal absorption.
Merger stages 1 and 2
systems are marked as open green triangles.  Merger stages 3, 4, and 5
are open cyan diamonds, and merger stages 6 and 7 are identified by
blue open squares.
Black crosses mark the normal spirals from Mineo et al.\ (2012).
The open red squares indicate the \citet{goulding16} ellipticals,
while the \citet{su15} E/S0 galaxies are shown by open magenta circles.
}.
\end{figure}

\begin{figure}
\plotone{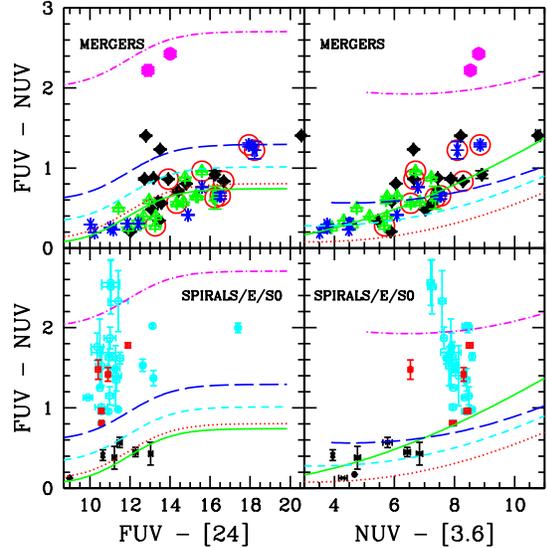}
\caption{
Left: Plot of FUV $-$ NUV vs. FUV $-$ [24].
Right:
FUV $-$ NUV vs. NUV $-$ [3.6].
Top row:  Mergers.  
Bottom row: Spirals, Ellipticals, and S0.
The data points for the mergers are color-coded according to 
the internal-absorption-corrected L$_{\rm X}$(gas)/SFR ratio:
Magenta filled circles: 
L$_{\rm X}$/SFR $>$ 10$^{40.5}$
(in units
of erg~s$^{-1}$)/(M$_{\sun}$~yr$^{-1}$)).
Black filled diamonds:
L$_{\rm X}$/SFR between 10$^{39.9}$ 
and 10$^{40.5}$. 
Green open triangles:
L$_{\rm X}$/SFR between 10$^{39.5}$ and 10$^{39.9}$. 
Blue asterisks: 
L$_{\rm X}$/SFR $\le$ 10$^{39.5}$.
In the bottom row the cyan circles are the \citet{su15} E/S0 galaxies,
red squares are massive E's from \citet{goulding16}, and black
crosses are the \citet{mineo12b} normal spirals.
The curves are Starburst99 models (see text).
}
\end{figure}

\begin{figure}
\plotone{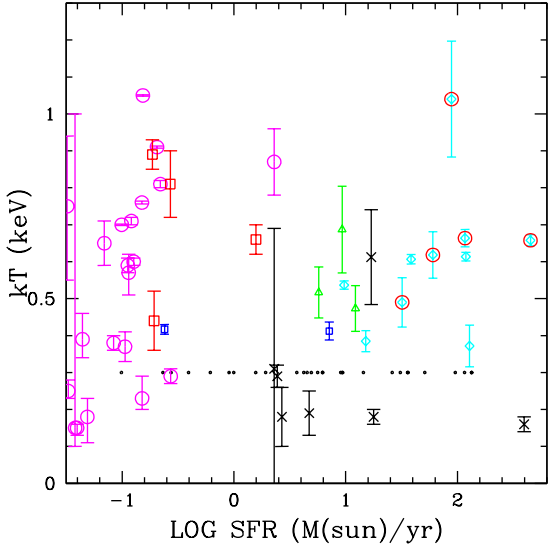}
\caption{
The best-fit hot gas temperature kT plotted against
SFR, for the galaxies in our merger sample for which we were able to obtain a good fit for the gas temperature.
Merger stages 1 and 2
systems are marked as open green triangles.  Merger stages 3, 4, and 5
are open cyan diamonds, and merger stages 6 and 7 are identified by
blue open squares.
AGN are identified by large red circles.
The black dots represent mergers for which a fixed temperature of 0.3 keV provides a good fit to the X-ray spectrum.
The \citet{mineo12b} spirals are marked by black crosses, the \citet{su15} ellipticals
by open magenta circles, and the \citet{goulding16} ellipticals by red open squares.
For systems in which the best model had two temperature
components, the temperature plotted is the luminosity-weighted gas temperature.
}
\end{figure}

\begin{figure}
\plotone{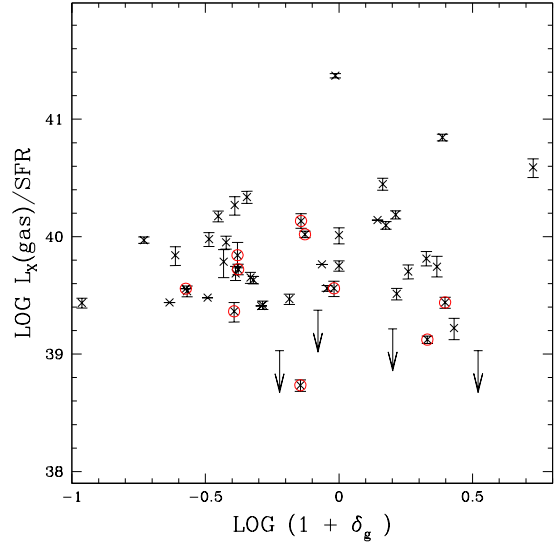}
\caption{
Plot of the internal-absorption-corrected
L$_{\rm X}$(gas)/SFR ratio against the large scale environment luminosity-weighted
density contrast $\delta_g$ (see Section 6.6).
On the x-axis, log(1 + $\delta$$_g$) = 0 corresponds to the average
local density.
}
\end{figure}

\clearpage

\begin{figure}
\plotone{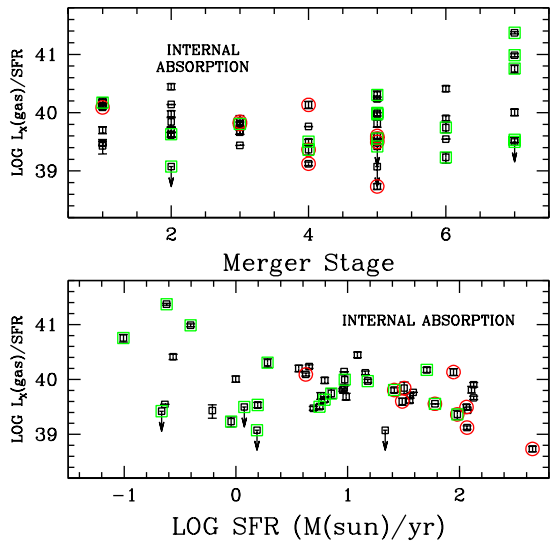}
\caption{
Plots of the internal-absorption-corrected
L$_{\rm X}$(gas)/SFR ratio against merger stage (top panel)
and SFR (bottom panel).
Objects circled in red are classified as Seyferts in NED.
Sources marked by a green square may not be 
classical `wet' major mergers, 
according to the discussion in the literature
(see Appendix).
}
\end{figure}

\clearpage

\begin{figure}
\plotone{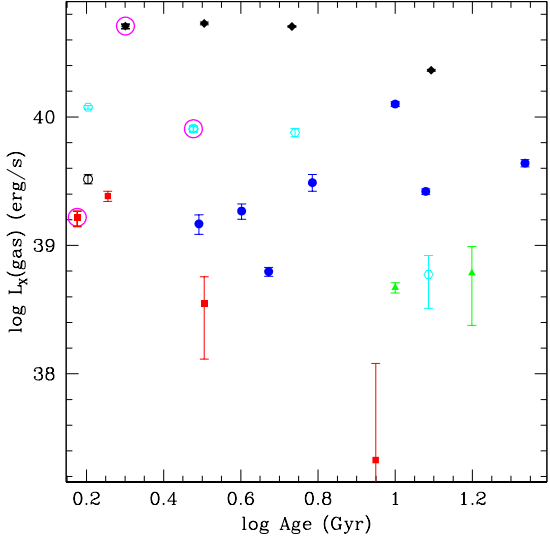}
\caption{Plot of L$_{\rm X}$(gas) vs.\ stellar population age
for our three post-starburst systems (large magenta circles),
compared to the \citet{boroson11} early-type galaxies.
The galaxies are
color-coded with the following ranges: stellar velocity dispersion
between 201 km~s$^{-1}$ and 238 km~s$^{-1}$ and gas temperatures
between 0.32 keV and 0.36 keV (blue filled circles),
velocity dispersion between 223 km~s$^{-1}$ and 260 km~s$^{-1}$
and gas temperature greater than 0.36 keV (black filled diamonds),
velocity dispersion between 170 km~s$^{-1}$ and 200 km~s$^{-1}$
and kT $\le$ 0.38 keV 
(red filled squares), 
velocity
dispersion between 200 km~s$^{-1}$ and 260 km~s$^{-1}$, with kT $<$ 0.32
keV  
(green filled triangles), 
velocity
dispersion between 170 km~s$^{-1}$ and 201 km~s$^{-1}$, with kT $>$ 0.36
keV  
(cyan open hexagons).
velocity
dispersion greater than 232 km~s$^{-1}$ and less than 260 km~s$^{-1}$ with kT $\ge$ 0.32 keV and kT $\le$ 0.36 
keV  
(black open circle).
}
\end{figure}

\begin{figure}
\plotone{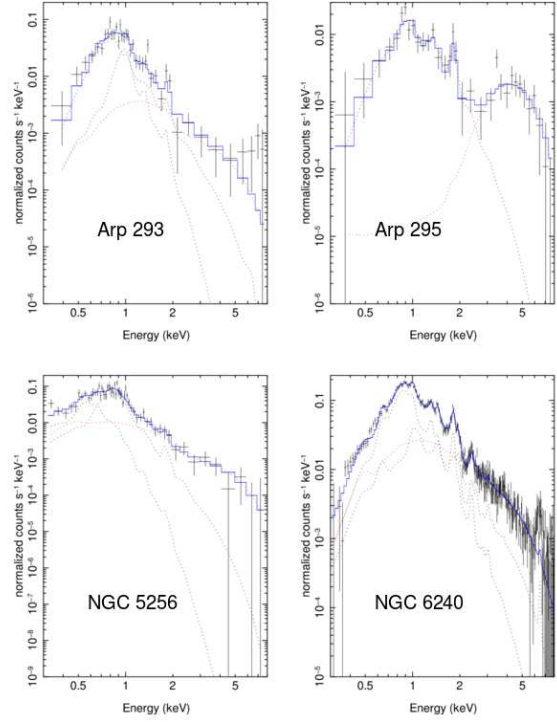}
\caption{
The Chandra diffuse X-ray spectra of four of our sample galaxies, plotted with their best-fit
models.  The x-axis is energy, in units of keV, while the y-axis has units of normalized 
counts~s$^{-1}$~keV$^{-1}$.
All of these models have a power law component plus at least one MEKAL or VMEKAL component.
The individual components of the model are plotted as dotted curves, while the sum of all
of the components is a solid curve (in blue).  
In all cases, the dotted curve highest on the right side of the plot (in red) is the
power law component, while the other dotted curve(s) (in black) are the MEKAL component(s).
Top left: Arp 293.
Top right: Arp 295.
Bottom left: NGC 5256.
Bottom right: NGC 6240.
For details on the models and the fitting process, see Table 5 and Section 5.3.
For more information about the individual galaxies, see the Appendix.
}
\end{figure}

\clearpage

\begin{deluxetable}{ccrrrrrrrrc}
\rotate
\tablecolumns{11}
\tablewidth{0pc}
\tablecaption{Basic Data on Sample Galaxies}
\tablehead{   
\colhead{Name} 
& \colhead{Stage}   
& \colhead{Distance}    
& \colhead{L$_{\rm FUV}$} 
& \colhead{L$_{\rm NUV}$} 
& \colhead{L$_{\rm 24}$} 
& \colhead{log L$_{\rm FIR}$} 
& \colhead{log L$_{\rm K}$} 
& \colhead{log}
& \colhead{SFR} 
& \colhead{AGN?}\\ 
\colhead{} 
& \colhead{}   
& \colhead{(Mpc)}    
& \colhead{(10$^{42}$}
& \colhead{(10$^{42}$}
& \colhead{(10$^{42}$}
& \colhead{(L$_{\sun}$)}  
&\colhead{(L$_{\sun}$)}
& \colhead{L$_{\rm FIR}$/{L$_{\rm K}$}}
&\colhead{(M$_{\sun}$}
&
\\
\colhead{} 
& \colhead{}   
& \colhead{}    
& \colhead{erg~s$^{-1}$)}
& \colhead{erg~s$^{-1}$)}
& \colhead{erg~s$^{-1}$)}
& \colhead{}  
&\colhead{}
& \colhead{}
&\colhead{yr$^{-1}$)}
&
\\
}
\startdata
AM 1146-270 & 6 & 24.6  &  1.80  &  1.92  &  0.86 & 8.98 & 9.49 & -0.50 & 0.23  &  \\
AM 2055-425 & 5 & 179.1  &  40.04  &  48.64  &  722.27 & 11.72 & 11.27 & 0.45 & 128.23  &  \\
AM 2312-591 & 3 & 184  &  45.23  &  55.22  &  752.92 & 11.71 & 11.24 & 0.47 & 133.83  &  \\
Arp 91 & 1 & 34  &  4.85  &  7.21  &  22.87 & 10.33 & 10.93 & -0.59 & 4.22  &  Sy2\\
Arp 147 & 1 & 129  &  19.34  &  15.70  &  15.90 & 10.22 & 10.80 & -0.57 & 3.65  &  \\
Arp 148 & 1 & 146.9  &  25.73  &  29.57  &  75.67 & 11.38 & 11.14 & 0.24 & 14.4  &  \\
Arp 155 & 7 & 46  &  2.16  &  3.24  &  5.14 & 9.99 & 10.98 & -0.98 & 1.00  &  \\
Arp 157 & 4 & 30.5  &  3.07  &  4.65  &  31.27 & 10.69 & 10.99 & -0.29 & 5.61  &  \\
Arp 160 & 5 & 39  &  7.82  &  11.00  &  84.50 & 10.72 & 10.70 & 0.02 & 15.14  &  \\
Arp 163 & 7 & 23.1  &  16.27  &  13.08  &  2.61 & 9.39 & 9.93 & -0.53 & 1.19  &  \\
Arp 178 & 5 & 62.1  &  5.97  &  8.88  &  9.44 & 10.11 & 11.54 & -1.42 & 1.92  &  \\
Arp 186 & 5 & 64.2  &  8.52  &  10.22  &  342.46 & 11.27 & 11.20 & 0.07 & 60.33  &  Sy2:HII\\
Arp 217 & 6 & 18.0  &  28.84  &  26.52  &  33.31 & 10.24 & 10.43 & -0.18 & 7.13  &  \\
Arp 220 & 4 & 83  &  2.21  &  4.57  &  545.82 & 12.03 & 11.27 & 0.76 & 95.65  &  Sy\\
Arp 222 & 6 & 26.1  &  0.81  &  1.98  &  1.36 & 8.71 & 11.12 & -2.41 & 0.27  &  \\
Arp 226 & 5 & 67  &  7.41  &  15.48  &  23.90 & 10.48 & 11.30 & -0.81 & 4.52  &  \\
Arp 233 & 7 & 25  &  5.69  &  5.01  &  7.49 & 9.62 & 9.99 & -0.36 & 1.57  &  \\
Arp 235 & 7 & 13  &  ...  &  ...  &  0.17 & 8.75 & 9.42 & -0.66 & 0.03  &  \\
Arp 236 & 1 & 81  &  87.37  &  80.60  &  269.00 & 11.39 & 11.18 & 0.21 & 51.02  &  \\
Arp 240 & 2 & 102  &  93.68  &  109.18  &  182.60 & 11.29 & 11.68 & -0.38 & 36.18  &  \\
Arp 242 & 2 & 98  &  14.47  &  15.15  &  31.87 & 10.65 & 11.20 & -0.54 & 6.23  &  \\
Arp 243 & 5 & 79.4  &  5.55  &  7.51  &  122.40 & 11.34 & 11.00 & 0.34 & 21.68  &  \\
Arp 244 & 3 & 24.1  &  31.18  &  30.14  &  47.37 & 10.62 & 11.26 & -0.63 & 9.7  &  \\
Arp 256 & 2 & 109.6  &  70.17  &  67.91  &  186.21 & 11.13 & 11.13 & 0.00 & 35.75  &  \\
Arp 259 & 2 & 55  &  35.44  &  30.19  &  26.32 & 10.34 & 10.20 & 0.14 & 6.2  &  \\
Arp 261 & 1 & 29  &  5.55  &  4.60  &  2.09 & 9.26 & 9.66 & -0.39 & 0.62  &  \\
Arp 263 & 5 & 9.8  &  3.18  &  2.79  &  0.42 & 8.76 & 9.06 & -0.29 & 0.22  &  \\
Arp 270 & 1 & 29  &  28.96  &  25.76  &  20.70 & 10.16 & 10.65 & -0.48 & 4.93  &  \\
Arp 283 & 1 & 30  &  3.39  &  3.86  &  31.92 & 10.48 & 10.77 & -0.28 & 5.74  &  \\
Arp 284 & 2 & 39  &  22.10  &  ...  &  47.51 & 10.41 & 10.68 & -0.26 & 9.31  &  \\
Arp 293 & 2 & 82  &  9.10  &  11.77  &  67.61 & 11.10 & 11.41 & -0.30 & 12.24  &  Liner/HII\\
Arp 295 & 2 & 94  &  14.37  &  16.19  &  49.39 & 10.86 & 11.50 & -0.64 & 9.29  &  \\
Arp 299 & 3 & 48  &  57.03  &  56.03  &  665.19 & 11.60 & 11.37 & 0.23 & 119.01  &  \\
IRAS 17208-0014 & 6 & 183  &  0.36  &  0.88  &  765.23 & 12.19 & 11.40 & 0.79 & 133.97  &  Liner\\
Mrk 231 & 5 & 178.1  &  ...  &  37.63  &  2929.16 & 12.13 & 12.36 & -0.22 & 450.05  &  Sy1\\
Mrk 273 & 4 & 160.5  &  12.96  &  15.86  &  664.08 & 11.90 & 11.44 & 0.47 & 116.83  &  Sy2\\
NGC 34 & 5 & 79.3  &  7.89  &  12.84  &  173.81 & 11.18 & 11.14 & 0.04 & 30.78  &  Sy2\\
NGC 1700 & 7 & 52.5  &  ...  &  2.33  &  0.54 & 8.40 & 11.57 & -3.16 & 0.24  &  \\
NGC 2207/IC 2163 & 3 & 38.0  &  ...  &  36.63  &  42.90 & 10.73 & 11.48 & -0.74 & 9.03  &  \\
NGC 2865 & 7 & 37.9  &  0.27  &  1.39  &  0.49 & 9.53 & 11.14 & -1.60 & 0.1  &  \\
NGC 3256 & 4 & 37.0  &  ...  &  13.22  &  246.15 & 11.30 & 11.26 & 0.04 & 38.5  &  \\
NGC 3353 & 6 & 18.5  &  3.58  &  3.18  &  4.26 & 9.43 & 9.66 & -0.21 & 0.91  &  \\
NGC 5018 & 7 & 38.4  &  0.42  &  2.63  &  2.14 & 9.40 & 11.44 & -2.04 & 0.39  &  \\
NGC 5256 & 3 & 120.9  &  25.51  &  28.70  &  176.36 & 11.21 & 11.62 & -0.40 & 32.02  &  Sy2\\
NGC 6240 & 4 & 108.8  &  8.40  &  12.17  &  503.17 & 11.61 & 11.81 & -0.19 & 88.46  &  Sy2/Liner\\
NGC 7592 & 3 & 99.5  &  52.41  &  45.44  &  135.50 & 11.08 & 11.07 & 0.01 & 26.08  &  Sy2\\
UGC 2238 & 5 & 87.1  &  1.34  &  2.10  &  53.38 & 11.04 & 11.15 & -0.10 & 9.41  &  Liner\\
UGC 5101 & 5 & 164.3  &  3.44  &  7.58  &  659.66 & 11.72 & 11.51 & 0.22 & 115.63  &  Sy1\\
UGC 5189 & 2 & 48.9  &  15.01  &  12.72  &  4.96 & 9.48 & 9.47 & 0.01 & 1.54  &  \\
\enddata
\end{deluxetable}

\vfill
\eject
                                                                                
\begin{deluxetable}{crcrrr}
\tablecolumns{5}
\tablewidth{0pc}
\tablecaption{X-Ray Data }
\tablehead{   
\colhead{Name} & \colhead{Exp} & \colhead{Dataset(s)}& \colhead{Point} 
& \colhead{Unresolved$^{\dagger}$}
& \colhead{Unresolved$^{\dagger}$}\\
\colhead{} & \colhead{Time} &  & \colhead{Source} 
& \colhead{HMXB}
& \colhead{LMXB}
\\
\colhead{} & \colhead{(Sec)} &                    & \colhead{Limit}  & 
\colhead{L$_{\rm X}$} 
& \colhead{L$_{\rm X}$} 
\\
\colhead{} & \colhead{} &                      & \colhead{(10$^{39}$} 
& \colhead{(10$^{39}$}
& \colhead{(10$^{39}$}
\\
\colhead{} & \colhead{} &                       
& \colhead{erg~s$^{-1}$)}  
& \colhead{erg~s$^{-1}$)}  
& \colhead{erg~s$^{-1}$)}  
\\
\\
}
\startdata
AM 1146-270 & 36.6  &  10540 & 0.07  &  0.08 &  0.19 \\
AM 2055-425 & 35.9  &  2036 & 2.82  &  224.28 &  11.25 \\
AM 2312-591 & 31.5  &  2037 & 3.28  &  253.23 &  10.5 \\
Arp 91 & 14.1  &  2930,4023 & 0.26  &  2.67 &  5.21 \\
Arp 147 & 42.5  &  11280,11887 & 1.73  &  4.88 &  3.87 \\
Arp 148 & 51.8  &  12977 & 1.86  &  20.64 &  8.44 \\
Arp 155 & 53.8  &  10541 & 0.16  &  0.52 &  5.88 \\
Arp 157 & 42.5  &  2924 & 0.07  &  2.07 &  5.91 \\
Arp 160 & 35.5  &  7071 & 0.18  &  8.88 &  3.06 \\
Arp 163 & 2.6  &  7117 & 0.91  &  1.17 &  0.52 \\
Arp 178 & 14.6  &  11679 & 1.15  &  2.27 &  21.38 \\
Arp 186 & 15.8  &  15050 & 1.19  &  74.02 &  9.63 \\
Arp 217 & 42.4  &  2939 & 0.02  &  1.71 &  1.64 \\
Arp 220 & 56.5  &  869 & 0.39  &  77.7 &  11.36 \\
Arp 222 & 18.0  &  2045 & 0.12  &  0.13 &  8.13 \\
Arp 226 & 20.1  &  2980 & 0.69  &  4.32 &  12.13 \\
Arp 233 & 17.4  &  9519 & 0.14  &  0.8 &  0.6 \\
Arp 235 & 4.0  &  7127 & 0.18  &  0.02 &  0.16 \\
Arp 236 & 59.1  &  7063 & 0.47  &  41.83 &  9.15 \\
Arp 240 & 19.9  &  10565 & 2.09  &  54.28 &  29.49 \\
Arp 242 & 28.6  &  2043 & 1.02  &  7.00 &  9.67 \\
Arp 243 & 17.8  &  4059 & 1.24  &  27.22 &  6.12 \\
Arp 244 & 327.4  &  3040,315,3041,3042,3043,3044 & 0.01  &  1.19 &  11.09 \\
Arp 256 & 29.6  &  13823 & 1.82  &  49.85 &  8.29 \\
Arp 259 & 35.6  &  9405 & 0.35  &  4.35 &  0.97 \\
Arp 261 & 54.5  &  5191 & 0.07  &  0.21 &  0.28 \\
Arp 263 & 11.8  &  7094,13764 & 0.04  &  0.06 &  0.07 \\
Arp 270 & 18.9  &  2042 & 0.14  &  2.45 &  2.73 \\
Arp 283 & 5.1  &  10567 & 0.75  &  5.69 &  3.58 \\
Arp 284 & 57.2  &  4800 & 0.13  &  4.36 &  2.93 \\
Arp 293 & 14.0  &  10566 & 1.92  &  17.75 &  15.72 \\
Arp 295 & 18.9  &  10570 & 1.88  &  13.12 &  19.44 \\
Arp 299 & 89.8  &  15077,15619 & 0.12  &  58.92 &  14.29 \\
IRAS 17208-0014 & 56.2  &  2035,4114 & 2.12  &  211.8 &  15.42 \\
Mrk 231 & 455.8  &  4028,4029,4030,13947,13948,13949 & 0.29  &  333.00 &  139.28 \\
Mrk 273 & 39.3  &  809 & 1.97  &  185.18 &  16.62 \\
NGC 34 & 14.8  &  15061 & 1.92  &  46.5 &  8.44 \\
NGC 1700 & 41.5  &  2069 & 0.21  &  0.13 &  22.91 \\
NGC 2207/IC 2163 & 62.2  &  11228,14799,14914,14915 & 0.11  &  3.84 &  18.63 \\
NGC 2865 & 24.6  &  2020 & 0.19  &  0.05 &  8.46 \\
NGC 3256 & 54.0  &  3569,835 & 0.1  &  16.43 &  11.2 \\
NGC 3353 & 17.5  &  13927 & 0.09  &  0.38 &  0.28 \\
NGC 5018 & 25.0  &  2070 & 0.2  &  0.22 &  16.92 \\
NGC 5256 & 17.1  &  2044 & 2.62  &  52.61 &  25.54 \\
NGC 6240 & 181.0  &  12713,1590 & 0.28  &  61.34 &  39.01 \\
NGC 7592 & 13.5  &  6860 & 3.13  &  45.17 &  7.19 \\
UGC 2238 & 14.4  &  15068 & 2.42  &  15.19 &  8.63 \\
UGC 5101 & 48.0  &  2033 & 1.76  &  174.32 &  19.59 \\
UGC 5189 & 129.2  &  11122,11237,13199,13781,13782,15869 & 0.06  &  0.52 &  0.18 \\
\enddata
\tablenotetext{\dagger}{0.3 $-$ 8.0 keV}
\end{deluxetable}

\begin{deluxetable}{crrrrrr}
\tablecolumns{7}
\tablewidth{0pc}
\tablecaption{X-Ray Parameters and Results for Galactic Absorption Correction Only}
\tablehead{   
\colhead{Name} &  F(0.3$-$8 keV) 
& \colhead{N$_{\rm H}$}  
& \colhead{$\chi_{\nu}^2$} 
& \colhead{$\chi^2$ / dof} 
& \colhead{MEKAL L$_{\rm X}$$^{1}$ }   
& \colhead{Power Law L$_{\rm X}$$^{1}$ }   
\\ 
\colhead{} &  (observed) 
&
\colhead{(10$^{20}$ } &  
&
& \colhead{(10$^{39}$ erg~s$^{-1}$)} 
& \colhead{(10$^{39}$ erg~s$^{-1}$)} \\
\colhead{} & \colhead{(10$^{-14}$} & 
\colhead{cm$^{-2}$)} &  
& \colhead{} \\
& \colhead{\rm erg~s$^{-1}$} 
&\colhead{} 
&\colhead{} 
& \colhead{} 
& \colhead{} 
& \colhead{} \\
\colhead{}  & 
\colhead{cm$^{-2}$)} & 
\colhead{}  & 
\colhead{} &  
\colhead{} \\
\\
}
\startdata
AM 1146-270 & 0.5 $\pm$ $^{0.32}_{0.05}$ &  7.4 & 1 & 15.6/15 & 0.6 $\pm$ 0.1 & $<$0.1    \\
AM 2055-425 & 5.6 $\pm$ $^{0.25}_{0.34}$ &  3.3 & 3.2 & 97.2/30 & 73.6 $\pm$ 9.1 & 170.3 $\pm$ 18.8  \\
AM 2312-591 & 5.35 $\pm$ $^{0.58}_{0.46}$ &  1.6 & 2.4 & 71.2/30 & 59.5 $\pm$ 9.5 & 171.1 $\pm$ 21.8  \\
Arp 91 & 22.42 $\pm$ $^{1.5}_{1.36}$ &  3.2 & 1.9 & 88.5/47 & 10.5 $\pm$ 1.1 & 24.5 $\pm$ 2.4  \\
Arp 147 & 4.04 $\pm$ $^{0.24}_{0.39}$ &  6.2 & 1 & 20/21 & 23.8 $\pm$ 4.6 & 74.5 $\pm$ 8.9  \\
Arp 148 & 5.57 $\pm$ $^{0.37}_{0.27}$ &  1.0 & 2 & 59.4/29 & 44.6 $\pm$ 5.5 & 105.4 $\pm$ 9.9  \\
Arp 155 & 5.35 $\pm$ $^{0.65}_{0.43}$ &  1.0 & 0.9 & 85.9/91 & 2.3 $\pm$ 0.6 & 11.8 $\pm$ 1.7  \\
Arp 157 & 13.72 $\pm$ $^{0.75}_{0.92}$ &  3.2 & 0.9 & 136.1/148 & 2.3 $\pm$ 0.5 & 14.5 $\pm$ 1.3  \\
Arp 160 & 27.83 $\pm$ $^{0.85}_{1.01}$ &  1.5 & 2.9 & 288.8/101 & 18.4 $\pm$ 1 & 35.5 $\pm$ 2.1  \\
Arp 163 & 17.33 $\pm$ $^{3.32}_{2.63}$ &  1.1 & 0.9 & 4.6/5 & $<$3.2   & 9.9 $\pm$ 2.7  \\
Arp 178 & 13.73 $\pm$ $^{0.92}_{1.27}$ &  1.4 & 1.2 & 55.2/47 & 12.1 $\pm$ 3.6 & 54.4 $\pm$ 8.8  \\
Arp 186 & 17.81 $\pm$ $^{1.1}_{1.09}$ &  6.3 & 3.1 & 93.8/30 & 15.4 $\pm$ 3.9 & 89.9 $\pm$ 8.1  \\
Arp 217 & 93.11 $\pm$ $^{1.4}_{1.35}$ &  1.4 & 3.5 & 763/215 & 13.9 $\pm$ 0.4 & 24.4 $\pm$ 0.8  \\
Arp 220 & 8.04 $\pm$ $^{0.34}_{0.27}$ &  3.9 & 2.2 & 166.6/75 & 21.3 $\pm$ 1.9 & 55 $\pm$ 4  \\
Arp 222 & 8.91 $\pm$ $^{0.87}_{1}$ &  2.9 & 1.4 & 43.4/32 & 2.3 $\pm$ 0.4 & 5.8 $\pm$ 1  \\
Arp 226 & 10.65 $\pm$ $^{0.58}_{0.86}$ &  2.0 & 1.6 & 58.8/36 & 17.5 $\pm$ 2.4 & 44.3 $\pm$ 5.3  \\
Arp 233 & 6.4 $\pm$ $^{0.47}_{0.74}$ &  1.0 & 1.6 & 28.8/18 & 2.1 $\pm$ 0.3 & 2.9 $\pm$ 0.6  \\
Arp 235 & $<$0.11  $^{}_{}$ &  4.3 & 0.8 & 3.2/4 & $<$2   & $<$2.2    \\
Arp 236 & 30.23 $\pm$ $^{0.77}_{0.67}$ &  1.4 & 4.3 & 587.4/138 & 39.5 $\pm$ 3.1 & 209.6 $\pm$ 7.1  \\
Arp 240 & 24.94 $\pm$ $^{1.21}_{1.34}$ &  1.9 & 1.1 & 80.7/75 & 38.9 $\pm$ 9 & 291 $\pm$ 22  \\
Arp 242 & 4.9 $\pm$ $^{0.56}_{0.55}$ &  1.3 & 1.2 & 53.2/45 & 17.4 $\pm$ 3.4 & 41.8 $\pm$ 9.1  \\
Arp 243 & 3.31 $\pm$ $^{0.99}_{0.74}$ &  3.1 & 1.8 & 19.5/11 & $<$8   & 27.1 $\pm$ 5.3  \\
Arp 244 & 101.24 $\pm$ $^{0.46}_{0.74}$ &  3.2 & 11 & 5751.7/524 & 30.3 $\pm$ 0.3 & 49.7 $\pm$ 0.6  \\
Arp 256 & 9.74 $\pm$ $^{0.62}_{0.51}$ &  3.1 & 1.8 & 62.4/35 & 32.9 $\pm$ 6.2 & 123.1 $\pm$ 12.4  \\
Arp 259 & 8.74 $\pm$ $^{0.43}_{0.54}$ &  5.7 & 0.9 & 46.1/51 & 12.3 $\pm$ 1.5 & 26.5 $\pm$ 2.8  \\
Arp 261 & 3.57 $\pm$ $^{0.54}_{0.45}$ &  7.8 & 0.9 & 93.8/100 & $<$1   & 3.6 $\pm$ 0.8  \\
Arp 263 & 28.42 $\pm$ $^{3.15}_{2.9}$ &  2.8 & 0.9 & 57.7/66 & $<$0.5   & 3.5 $\pm$ 0.4  \\
Arp 270 & 23.17 $\pm$ $^{1.06}_{1.61}$ &  2.0 & 1.1 & 110.1/100 & 6.7 $\pm$ 0.7 & 18.4 $\pm$ 2  \\
Arp 283 & 21.82 $\pm$ $^{1.99}_{1.97}$ &  1.5 & 4.8 & 57.2/12 & $<$3.4   & 22.8 $\pm$ 3.2  \\
Arp 284 & 19.82 $\pm$ $^{0.81}_{0.86}$ &  5.0 & 1.6 & 219.2/139 & 10.8 $\pm$ 0.8 & 31.9 $\pm$ 1.8  \\
Arp 293 & 21.8 $\pm$ $^{1.24}_{1.34}$ &  1.8 & 3.3 & 120.6/37 & 54.4 $\pm$ 6.3 & 133.6 $\pm$ 12.5  \\
Arp 295 & 11.53 $\pm$ $^{0.94}_{0.9}$ &  3.6 & 2.1 & 67.6/32 & $<$14.4   & 133.8 $\pm$ 10.2  \\
Arp 299 & 133.28 $\pm$ $^{0.92}_{1.38}$ &  0.9 & 10.4 & 2924.5/282 & 62.5 $\pm$ 2.1 & 317 $\pm$ 4.2  \\
IRAS 17208-0014 & 3.63 $\pm$ $^{0.25}_{0.29}$ &  9.7 & 2.9 & 55.1/19 & $<$16.3   & 178.4 $\pm$ 11.7  \\
Mrk 231 & 8.56 $\pm$ $^{0.15}_{0.11}$ &  1.0 & 3.6 & 830.8/228 & 92.7 $\pm$ 3.2 & 245.1 $\pm$ 5.9  \\
Mrk 273 & 18.83 $\pm$ $^{0.67}_{0.46}$ &  0.9 & 2.6 & 240.6/94 & 105.7 $\pm$ 9.9 & 493.9 $\pm$ 24.3  \\
NGC 34 & 8.8 $\pm$ $^{0.81}_{0.94}$ &  2.1 & 1.1 & 12.8/12 & $<$12.6   & 58.8 $\pm$ 8.1  \\
NGC 1700 & 21.05 $\pm$ $^{0.77}_{0.95}$ &  3.6 & 2 & 378.4/188 & 51 $\pm$ 2 & 31.7 $\pm$ 4.4  \\
NGC 2207/IC 2163 & 46.79 $\pm$ $^{0.97}_{1.29}$ &  8.8 & 1.1 & 291/262 & 24 $\pm$ 1.6 & 80.9 $\pm$ 2.9  \\
NGC 2865 & 3.44 $\pm$ $^{0.4}_{0.52}$ &  6.2 & 1.3 & 19.4/15 & 1.7 $\pm$ 0.4 & 5.5 $\pm$ 1  \\
NGC 3256 & 102.93 $\pm$ $^{1.37}_{1.73}$ &  9.1 & 10.8 & 2611.9/242 & 38.6 $\pm$ 1.5 & 178 $\pm$ 3.4  \\
NGC 3353 & 5.35 $\pm$ $^{0.65}_{0.68}$ &  0.5 & 0.9 & 11.8/13 & 0.6 $\pm$ 0.2 & 1.6 $\pm$ 0.4  \\
NGC 5018 & 9.15 $\pm$ $^{0.42}_{0.54}$ &  6.4 & 2.2 & 77.7/35 & 8.1 $\pm$ 0.8 & 12.5 $\pm$ 1.4  \\
NGC 5256 & 25.08 $\pm$ $^{1.17}_{1.06}$ &  1.7 & 1.7 & 95.5/55 & 202.1 $\pm$ 13.1 & 271 $\pm$ 24.8  \\
NGC 6240 & 97.33 $\pm$ $^{0.69}_{0.68}$ &  4.9 & 13.5 & 4557.4/337 & 122.1 $\pm$ 6.3 & 1455.9 $\pm$ 13.4  \\
NGC 7592 & 8.99 $\pm$ $^{0.67}_{0.56}$ &  3.8 & 1.3 & 20.3/16 & 52.1 $\pm$ 7 & 72.5 $\pm$ 12.5  \\
UGC 2238 & 6.48 $\pm$ $^{0.72}_{0.67}$ &  8.9 & 1.1 & 6.4/6 & $<$13.2   & 70.4 $\pm$ 9.3  \\
UGC 5101 & 3.04 $\pm$ $^{0.22}_{0.16}$ &  3.0 & 1.6 & 32.4/20 & $<$12   & 104.4 $\pm$ 10.8  \\
UGC 5189 & 8.47 $\pm$ $^{0.3}_{0.26}$ &  2.6 & 1.9 & 484.5/260 & $<$1.4   & 26 $\pm$ 0.9  \\
\enddata
\tablenotetext{1}{0.3 $-$ 8.0 keV.  Calculated assuming
kT is fixed at 0.3 keV.}
\end{deluxetable}

\begin{deluxetable}{crrrrr}
\tablecolumns{6}
\tablewidth{0pc}
\tablecaption{X-Ray Luminosities including Internal Absorption Correction from UV/IR}
\tablehead{   
\colhead{Name} & 
\colhead{N$_{\rm H}$}  & 
\colhead{$\chi_{\nu}^2$} & 
\colhead{$\chi^2$ / dof} & 
\colhead{MEKAL L$_{\rm X}$$^{1}$ }    & 
\colhead{Power Law L$_{\rm X}$$^{1}$ }    \\
\colhead{}  & 
\colhead{(10$^{20}$ } & 
\colhead{}    & 
\colhead{}    & 
\colhead{(10$^{39}$ erg~s$^{-1}$)} & 
\colhead{(10$^{39}$ erg~s$^{-1}$)} \\
\colhead{} & 
\colhead{cm$^{-2}$)} & 
\colhead{} & 
\colhead{} & 
\colhead{} & 
\colhead{} \\
\\
}
\startdata
AM 1146-270 & 8.4 & 1.1  &  16.3/15 & 0.8 $\pm$ 0.3 & $<$0.9   \\
AM 2055-425 & 34.1 & 1.5  &  44.6/30 & 745.7 $\pm$ 76.3 & 142 $\pm$ 24.2 \\
AM 2312-591 & 33.4 & 1.5  &  43.5/30 & 619.2 $\pm$ 74.5 & 142 $\pm$ 29.5 \\
Arp 91 & 23.6 & 1.2  &  54.2/47 & 52.1 $\pm$ 5.3 & 24.3 $\pm$ 2.8 \\
Arp 147 & 11.4 & 0.7  &  13.7/21 & 58.2 $\pm$ 12.7 & 77.8 $\pm$ 9.6 \\
Arp 148 & 20.1 & 1.3  &  36.8/29 & 190.7 $\pm$ 23.5 & 100.8 $\pm$ 11.6 \\
Arp 155 & 18.6 & 0.9  &  78.1/91 & 10.1 $\pm$ 2.2 & 12.2 $\pm$ 2 \\
Arp 157 & 29.6 & 1.1  &  170.2/148 & 17.9 $\pm$ 2.7 & 15.2 $\pm$ 1.8 \\
Arp 160 & 30.0 & 1.5  &  150.4/101 & 122.3 $\pm$ 6.5 & 30 $\pm$ 2.6 \\
Arp 163 & 3.8 & 0.8  &  4/5 & $<$2.5   & 10.2 $\pm$ 2.7 \\
Arp 178 & 15.7 & 1.2  &  54.9/47 & 38.8 $\pm$ 11.2 & 57.8 $\pm$ 10.3 \\
Arp 186 & 40.4 & 1.9  &  58.4/30 & 271.9 $\pm$ 36.8 & 79.7 $\pm$ 9.8 \\
Arp 217 & 13.6 & 2  &  426.7/215 & 35.6 $\pm$ 1.1 & 26.2 $\pm$ 0.9 \\
Arp 220 & 54.9 & 2  &  149.4/75 & 433.4 $\pm$ 30.2 & 37.3 $\pm$ 6.3 \\
Arp 222 & 16.1 & 1.4  &  44.7/32 & 7.1 $\pm$ 1.3 & 5.6 $\pm$ 1.2 \\
Arp 226 & 20.8 & 1.1  &  38.9/36 & 77.6 $\pm$ 10 & 46.7 $\pm$ 6.6 \\
Arp 233 & 14.4 & 1.5  &  26.1/18 & 5.4 $\pm$ 1 & 2.5 $\pm$ 0.7 \\
Arp 235 & ... & ...  &  .../... & ...   & ...   \\
Arp 236 & 20.4 & 1.8  &  251.6/138 & 253.7 $\pm$ 13.3 & 229.4 $\pm$ 8.4 \\
Arp 240 & 17.1 & 1.1  &  79.1/75 & 183.8 $\pm$ 31 & 311.4 $\pm$ 26.6 \\
Arp 242 & 18.0 & 1.1  &  47.9/45 & 60.1 $\pm$ 12.2 & 45.8 $\pm$ 11.4 \\
Arp 243 & 35.6 & 1.6  &  17.2/11 & $<$19.4   & $<$34   \\
Arp 244 & 15.4 & 5.2  &  2718.4/524 & 86.5 $\pm$ 0.5 & 49.4 $\pm$ 0.7 \\
Arp 256 & 19.4 & 1.1  &  40.1/35 & 148.3 $\pm$ 24.3 & 126.2 $\pm$ 14.8 \\
Arp 259 & 10.8 & 0.7  &  38/51 & 26.5 $\pm$ 3.9 & 26.7 $\pm$ 3.1 \\
Arp 261 & 7.2 & 0.9  &  91.2/100 & 1.7 $\pm$ 0.8 & 3.8 $\pm$ 0.8 \\
Arp 263 & 3.3 & 0.8  &  55.7/66 & $<$0.6   & 3.7 $\pm$ 0.5 \\
Arp 270 & 10.6 & 1  &  101.3/100 & 14.7 $\pm$ 1.9 & 20.5 $\pm$ 2.2 \\
Arp 283 & 29.0 & 2.7  &  32.9/12 & 36.1 $\pm$ 4.5 & 21.6 $\pm$ 3.4 \\
Arp 284 & 17.8 & 1.2  &  164.4/139 & 38.9 $\pm$ 2.9 & 32.3 $\pm$ 2.1 \\
Arp 293 & 27.1 & 1.6  &  58.7/37 & 364.9 $\pm$ 36.8 & 120.6 $\pm$ 15.3 \\
Arp 295 & 21.3 & 1.8  &  56.6/32 & 48.8 $\pm$ 19.9 & 155 $\pm$ 15.7 \\
Arp 299 & 30.6 & 4.1  &  1147.6/282 & 661.6 $\pm$ 8.4 & 317.7 $\pm$ 5.2 \\
IRAS 17208-0014 & 72.1 & 1.1  &  21.6/19 & 1077.4 $\pm$ 170.7 & 187.2 $\pm$ 22 \\
Mrk 231 & 51.2 & 3.6  &  827.1/228 & 1514 $\pm$ 28.3 & 224.1 $\pm$ 8.6 \\
Mrk 273 & 42.3 & 3  &  286.7/94 & 1886.4 $\pm$ 67.3 & 479.6 $\pm$ 35.6 \\
NGC 34 & 35.6 & 1  &  11.5/12 & 122.5 $\pm$ 29.6 & 59.2 $\pm$ 10.4 \\
NGC 1700 & 4.1 & 1.7  &  322.3/188 & 68.3 $\pm$ 3.9 & 31.9 $\pm$ 4.5 \\
NGC 2207/IC 2163 & 12.8 & 1  &  264/262 & 58.2 $\pm$ 4.5 & 84.5 $\pm$ 3.3 \\
NGC 2865 & 16.7 & 1.2  &  18/15 & 5.6 $\pm$ 1.5 & 5.8 $\pm$ 1.2 \\
NGC 3256 & 37.2 & 3.4  &  828.7/242 & 589.7 $\pm$ 8 & 173 $\pm$ 4.4 \\
NGC 3353 & 13.8 & 0.9  &  11.5/13 & 1.6 $\pm$ 0.5 & 1.6 $\pm$ 0.4 \\
NGC 5018 & 24.3 & 1.1  &  38.6/35 & 38.2 $\pm$ 3.8 & 10.9 $\pm$ 1.7 \\
NGC 5256 & 26.6 & 2  &  110.7/55 & 993 $\pm$ 44.6 & 230.5 $\pm$ 31.2 \\
NGC 6240 & 43.6 & 6.2  &  2100.2/337 & 4249.6 $\pm$ 42.3 & 1488.4 $\pm$ 17.8 \\
NGC 7592 & 19.2 & 1.2  &  19.6/16 & 167.2 $\pm$ 27.5 & 63.1 $\pm$ 14.7 \\
UGC 2238 & 40.3 & 1.6  &  9.5/6 & 93.4 $\pm$ 35.3 & 74.1 $\pm$ 11.6 \\
UGC 5101 & 52.8 & 1.2  &  24.3/20 & 363.5 $\pm$ 62.8 & 109.8 $\pm$ 16.9 \\
UGC 5189 & 6.6 & 1.8  &  467.7/260 & $<$0.9   & 28.3 $\pm$ 1.2 \\
\enddata
\tablenotetext{1}{0.3 $-$ 8.0 keV.  Calculated assuming kT is fixed at 0.3 keV.}
\end{deluxetable}

\begin{deluxetable}{ccrrcccrr}
\rotate
\tablecolumns{9}
\tablewidth{0pc}
\tablecaption{X-Ray Fits for More Complex Models}
\tablehead{   
\colhead{Name} & 
\colhead{Best$^1$} & 
\colhead{$\chi_{\nu}^2$} & 
\colhead{$\chi^2$ / dof} & 
\colhead{N$_{\rm H}$}  & 
\colhead{N$_{\rm H}$}  & 
\colhead{kT} & 
\colhead{MEKAL L$_{\rm X}$$^{2}$ }    & 
\colhead{Power Law L$_{\rm X}$$^{2}$ }    \\
\colhead{}  & 
\colhead{Model}    & 
\colhead{}    & 
\colhead{}    & 
\colhead{(MEKAL) } & 
\colhead{(P.L.) } & 
\colhead{(keV)}    & 
\colhead{(10$^{39}$ erg~s$^{-1}$)} & 
\colhead{(10$^{39}$ erg~s$^{-1}$)} \\
\colhead{} & 
\colhead{} & 
\colhead{} & 
\colhead{} & 
\colhead{(10$^{20}$ } & 
\colhead{(10$^{20}$ } & 
\colhead{} & 
\colhead{} & 
\colhead{} \\
\colhead{} & 
\colhead{} & 
\colhead{} & 
\colhead{} & 
\colhead{cm$^{-2}$)} & 
\colhead{cm$^{-2}$)} & 
\colhead{} & 
\colhead{} & 
\colhead{} \\
}
\startdata
AM 2055-425  &  2T  &  1  &  26.4/27  &  34.1F  &  34.1F  &  0.23$\pm$0.03/0.77$\pm$0.1  &  567.5$\pm$184.6/222.9$\pm$106.5  &  74.6$\pm$29.8 \\
Arp 160  &  2T  &  1  &  94.6/98  &  30.1F  &  30.1F  &  0.25$\pm$0.01/1.03$\pm$0.06  &  116.7$\pm$12/25$\pm$5.2  &  14.9$\pm$3.6 \\
Arp 186  &  1T2N  &  1.2  &  31.6/27  &  61.6$\pm$10.3  &  2.9$\pm$7  &  0.62$\pm$0.06    &  213.7$\pm$102.9    &  40.7$\pm$12.4 \\
Arp 217  &  2TV  &  1.3  &  276.8/210  &  13.6F  &  13.6F  &  0.24$\pm$0.02/0.58$\pm$0.03  &  19.1$\pm$5.1/20.5$\pm$4  &  17.2$\pm$1.4 \\
Arp 220  &  2N  &  1.1  &  77.2/73  &  47.8$\pm$4.6  &  1.3$\pm$1.9  &  0.3F    &  221.5$\pm$63.5    &  39.6$\pm$5.3 \\
Arp 236  &  2N  &  1  &  129.6/136  &  49.5$\pm$2.8  &  6.8$\pm$1.7  &  0.3F    &  762.9$\pm$123.7    &  174$\pm$9.7 \\
Arp 244  &  2TV1N  &  1.6  &  842.1/518  &  4.1$\pm$0.7  &  tied  &  0.2$\pm$0.03/0.58$\pm$0  &  5.6$\pm$1.1/38.4$\pm$2.5  &  27.4$\pm$1 \\
Arp 283  &  1T  &  1.7  &  18.7/11  &  29F  &  29F  &  0.52$\pm$0.07    &  29$\pm$5.7    &  16.5$\pm$3.7 \\
Arp 284  &  2N  &  1  &  140.8/137  &  38.9$\pm$4.5  &  4.8$\pm$2.5  &  0.3F    &  64.7$\pm$17.4    &  25.7$\pm$2.4 \\
Arp 293  &  2T  &  1.1  &  36.9/34  &  27.2F  &  27.2F  &  0.3$\pm$0.03/1.04$\pm$0.12  &  260.8$\pm$50.3/82.1$\pm$32.7  &  63.6$\pm$21.6 \\
Arp 295  &  1TV2N  &  0.8  &  21/27  &  28.6$\pm$11.5  &  1268.9$\pm$330.2  &  0.69$\pm$0.12    &  129.1$\pm$41.2    &  852$\pm$217.6 \\
Arp 299  &  1TV2N  &  1.1  &  292.3/277  &  15.8$\pm$2.8  &  19.8$\pm$16.9  &  0.61$\pm$0.01    &  328.5$\pm$29.5    &  188$\pm$7.3 \\
Mrk 231  &  1TV2N  &  1.1  &  251.4/223  &  3.9$\pm$1.1  &  509.1$\pm$49.4  &  0.66$\pm$0.02    &  244.5$\pm$19.1    &  573.8$\pm$37.5 \\
Mrk 273  &  1T1N  &  1.3  &  121.3/92  &  0.6$\pm$0.8  &  tied  &  0.66$\pm$0.02    &  155.3$\pm$18    &  426.1$\pm$28.2 \\
NGC 1700  &  1T  &  1.1  &  205.8/187  &  4.2F  &  4.2F  &  0.42$\pm$0.01    &  56.4$\pm$3.3    &  23$\pm$4.3 \\
NGC 3256  &  1TV2N  &  1.2  &  280/237  &  29.3$\pm$2  &  12.6$\pm$1.9  &  0.61$\pm$0.01    &  223.8$\pm$21.5    &  90.5$\pm$6.2 \\
NGC 5256  &  2T1N  &  1  &  52.9/51  &  3.1$\pm$2.8  &  tied  &  0.2$\pm$0.04/0.78$\pm$0.09  &  112.5$\pm$54.7/175.2$\pm$63.7  &  190.2$\pm$27.7 \\
NGC 6240  &  2TV1N  &  1.4  &  463.9/331  &  18.1$\pm$0.7  &  tied  &  0.65$\pm$0.01/1.85$\pm$0.28  &  815$\pm$62.8/387.2$\pm$56.8  &  708.3$\pm$76.8 \\
\enddata
\tablenotetext{1}
{
1T: one temperature component, absorbing column fixed;
2T: two temperature components, with absorbing column fixed;
1T1N: one temperature component, one absorbing column (absorbing columns for MEKAL and 
power law components tied together); 
2N: fitting for MEKAL and power law absorbing columns separately, with fixed single-component MEKAL temperature. 
1T2N: one temperature component; MEKAL and power law absorption fit separately;
2T1N: two MEKAL temperature components; MEKAL and power law absorbing columns tied together;
1TV2N: one temperature component, VMEKAL, two absorbing columns;
2TV: two temperature components, VMEKAL, absorbing column fixed;
2TV1N: two temperature components, VMEKAL, absorbing columns for MEKAL and power law components
tied together.
For the VMEKAL models, the $\alpha$/Fe ratios are:
Arp 217:  1.97 $\pm$ 0.83
;
Arp 244:  1.96 $\pm$ 0.56
;
Arp 295:  6.56 $\pm$ 14.86
;
Arp 299:  3.25 $\pm$ 2.48
;
NGC 3256:  3 $\pm$ 2.05
;
NGC 6240:  3.61 $\pm$ 0.97
;
Mrk 231:  2.86 $\pm$ 0.58
.
}
\tablenotetext{2}{0.3 $-$ 8.0 keV}
\end{deluxetable}

\begin{deluxetable}{c|rrr|rrr}
\tablecolumns{7}
\tablewidth{0pc}
\tablecaption{Ratios Involving the Diffuse MEKAL Component of the X-Ray Luminosity}
\tablehead{   
\colhead{Name} & 
\multicolumn{3}{|c|}{Galactic Absorption}&
\multicolumn{3}{|c|}{Internal Absorption}\\
\colhead{} & 
\colhead{log($\frac{\rm L_{\rm X}}{\rm L_{\rm FIR}}$)} & 
\colhead{log($\frac{\rm L_{\rm X}}{\rm L_{\rm K}}$)} &
\colhead{log($\frac{\rm L_{\rm X}}{\rm SFR}$)} & 

\colhead{log($\frac{\rm L_{\rm X}}{\rm L_{\rm FIR}}$)} & 
\colhead{log($\frac{\rm L_{\rm X}}{\rm L_{\rm K}}$)} &
\colhead{log($\frac{\rm L_{\rm X}}{\rm SFR}$)}  \\
\colhead{} & 
\colhead{} & 
\colhead{} & 
\colhead{$(\frac{erg~s^{-1}}{M_{\sun}~yr^{-1}})$} & 

\colhead{} & 
\colhead{} & 
\colhead{$(\frac{erg~s^{-1}}{M_{\sun}~yr^{-1}})$} 
}
\startdata
AM 1146-270 & -3.81 & -4.32 & 39.39 & -3.65 & -4.16 & 39.55 \\
AM 2055-425 & -4.43 & -3.98 & 38.76 & -3.38 & -2.93 & 39.81 \\
AM 2312-591 & -4.52 & -4.04 & 38.65 & -3.5 & -3.03 & 39.67 \\
Arp 91 & -3.89 & -4.49 & 39.39 & -3.2 & -3.8 & 40.09 \\
Arp 147 & -3.42 & -4.01 & 39.81 & -3.03 & -3.62 & 40.2 \\
Arp 148 & -4.31 & -4.07 & 39.49 & -3.68 & -3.44 & 40.12 \\
Arp 155 & -4.22 & -5.21 & 39.36 & -3.57 & -4.56 & 40.01 \\
Arp 157 & -4.9 & -5.2 & 38.62 & -4.02 & -4.31 & 39.5 \\
Arp 160 & -4.04 & -4.02 & 39.09 & -3.15 & -3.13 & 39.97 \\
Arp 163 & $<$-3.46 & $<$-4 & $<$39.44 & $<$-3.4 & $<$-3.94 & $<$39.5 \\
Arp 178 & -3.61 & -5.04 & 39.8 & -3.1 & -4.54 & 40.3 \\
Arp 186 & -4.67 & -4.59 & 38.41 & -3.51 & -3.44 & 39.56 \\
Arp 217 & -3.68 & -3.87 & 39.29 & -3.22 & -3.41 & 39.74 \\
Arp 220 & -5.29 & -4.52 & 38.35 & -4.27 & -3.51 & 39.36 \\
Arp 222 & -2.93 & -5.35 & 39.92 & -2.43 & -4.85 & 40.41 \\
Arp 226 & -3.82 & -4.64 & 39.59 & -3.17 & -3.99 & 40.23 \\
Arp 233 & -3.87 & -4.24 & 39.14 & -3.47 & -3.84 & 39.54 \\
Arp 235 & $<$-3.03 & $<$-3.71 & $<$40.77 &  &  &  \\
Arp 236 & -4.38 & -4.16 & 38.89 & -3.09 & -2.87 & 40.17 \\
Arp 240 & -4.28 & -4.68 & 39.03 & -3.61 & -4 & 39.71 \\
Arp 242 & -4 & -4.54 & 39.44 & -3.46 & -4 & 39.98 \\
Arp 243 & $<$-5.02 & $<$-4.68 & $<$38.57 & $<$-4.51 & $<$-4.17 & $<$39.07 \\
Arp 244 & -3.72 & -4.36 & 39.49 & -3.52 & -4.17 & 39.69 \\
Arp 256 & -4.19 & -4.2 & 38.96 & -3.54 & -3.54 & 39.62 \\
Arp 259 & -3.83 & -3.69 & 39.3 & -3.5 & -3.36 & 39.63 \\
Arp 261 & $<$-3.82 & $<$-4.23 & $<$39.22 & -3.62 & -4.02 & 39.43 \\
Arp 263 & $<$-3.64 & $<$-3.94 & $<$39.37 & $<$-3.58 & $<$-3.88 & $<$39.42 \\
Arp 270 & -3.92 & -4.41 & 39.13 & -3.58 & -4.06 & 39.47 \\
Arp 283 & $<$-4.53 & $<$-4.82 & $<$38.77 & -3.6 & -3.89 & 39.7 \\
Arp 284 & -3.96 & -4.23 & 39.07 & -3.18 & -3.45 & 39.84 \\
Arp 293 & -3.95 & -4.26 & 39.65 & -3.15 & -3.46 & 40.45 \\
Arp 295 & $<$-4.28 & $<$-4.93 & $<$39.19 & -3.33 & -3.97 & 40.14 \\
Arp 299 & -4.38 & -4.15 & 38.72 & -3.66 & -3.44 & 39.44 \\
IRAS 17208-0014 & $<$-5.56 & $<$-4.77 & $<$38.09 & -3.74 & -2.95 & 39.91 \\
Mrk 231 & -4.75 & -4.97 & 38.31 & -4.33 & -4.55 & 38.73 \\
Mrk 273 & -4.46 & -3.99 & 38.96 & -4.29 & -3.83 & 39.12 \\
NGC 34 & $<$-4.66 & $<$-4.62 & $<$38.61 & -3.67 & -3.63 & 39.6 \\
NGC 1700 & -1.27 & -4.45 & 41.33 & -1.23 & -4.4 & 41.37 \\
NGC 2207/IC 2163 & -3.93 & -4.69 & 39.42 & -3.55 & -4.3 & 39.81 \\
NGC 2865 & -3.89 & -5.5 & 40.22 & -3.36 & -4.97 & 40.76 \\
NGC 3256 & -4.3 & -4.26 & 39 & -3.53 & -3.5 & 39.76 \\
NGC 3353 & -4.2 & -4.43 & 38.85 & -3.82 & -4.05 & 39.23 \\
NGC 5018 & -3.07 & -5.12 & 40.31 & -2.4 & -4.44 & 40.99 \\
NGC 5256 & -3.49 & -3.9 & 39.8 & -3.44 & -3.86 & 39.84 \\
NGC 6240 & -4.1 & -4.3 & 39.14 & -3.11 & -3.31 & 40.13 \\
NGC 7592 & -3.94 & -3.94 & 39.3 & -3.44 & -3.43 & 39.81 \\
UGC 2238 & $<$-4.5 & $<$-4.61 & $<$39.15 & -3.65 & -3.76 & 40 \\
UGC 5101 & $<$-5.23 & $<$-5.01 & $<$38.02 & -3.74 & -3.53 & 39.5 \\
UGC 5189 & $<$-3.92 & $<$-3.9 & $<$38.96 & $<$-3.8 & $<$-3.79 & $<$39.08 \\
\enddata
\end{deluxetable}

\end{document}